\documentclass[pra,aps,notitlepage,superscriptaddress,noshowpacs,floatfix]{revtex4}

\pdfoutput=1

\usepackage{amssymb}
\usepackage{amsmath}
\usepackage{amsfonts}
\usepackage{mathrsfs}
\usepackage{color}
\usepackage{graphicx}
\usepackage{epstopdf,epsfig}
\usepackage{subfigure}
\usepackage{latexsym}
\usepackage{tikz}
\usetikzlibrary{positioning,arrows}
\usetikzlibrary{decorations.pathmorphing}
\usetikzlibrary{decorations.markings}
\usetikzlibrary{calc}

\newcommand{\EQ}{\begin{equation}}
\newcommand{\EN}{\end{equation}}

\newcommand{\ud}[1][\!]{\mathrm{d}#1\,}
\newcommand{\ket}[1]{%
 \left|\, #1 \,\right\rangle}
\newcommand{\bra}[1]{%
 \left\langle\, #1 \,\right|}

\newcommand{\braopket}[3]{%
 \left\langle\, #1 \,\middle|\, #2 \,\middle|\, #3 \,\right\rangle}
\newcommand{\ave}[1]{%
 \left\langle #1 \right\rangle}

\graphicspath{{./img/}}

\begin{document}

\title{\LARGE{Truncated Conformal Space Approach \\
for 2D Landau-Ginzburg Theories}}

\author{A. Coser}
\affiliation{SISSA$-$International School for Advanced Studies and INFN, Sezione di Trieste, Via Bonomea 265, I-34136 Trieste, Italy, EU}

\author{M. Beria}
\affiliation{SISSA$-$International School for Advanced Studies and INFN, Sezione di Trieste, Via Bonomea 265, I-34136 Trieste, Italy, EU}

\author{G. P. Brandino}
\affiliation{Institute for Theoretical Physics, University of Amsterdam, Science Park 904, Postbus 94485, 1090 GL Amsterdam, The Netherlands.}

\author{R. M. Konik}
\affiliation{CMPMS Dept., Bldg. 734, Brookhaven National Laboratory, Upton, NY 11973-5000, USA.}

\author{G. Mussardo}
\affiliation{SISSA$-$International School for Advanced Studies and INFN, Sezione di Trieste, Via Bonomea 265, I-34136 Trieste, Italy, EU}
\affiliation{The Abdus Salam International Centre of Theoretical Physics, 34100 Trieste, Italy}

\begin{abstract}
We study the spectrum of Landau-Ginzburg theories in $1+1$
dimensions using the truncated conformal space approach employing a compactified boson. 
We study these theories both in their broken and unbroken phases.  We first demonstrate that we can reproduce the
expected spectrum of a $\Phi^2$ theory (i.e.\ a free massive boson) in this framework.  We then turn
to $\Phi^4$ in its unbroken phase and compare our numerical results 
with the predictions of two-loop perturbation theory, finding excellent agreement.
We then analyze the broken phase of $\Phi^4$ where kink excitations together with their bound states are present.  We
confirm the semiclassical predictions for this model on the number of stable kink-antikink bound states.
We also test the semiclassics in the double well phase of $\Phi^6$ Landau-Ginzburg theory, again finding
agreement.
\end{abstract}

\maketitle

\section{Introduction}

A key datum in a massive quantum field theory consists of the spectrum of its stable excitations: this is the basic information one needs to build up its Hilbert space, so that one can proceed further in the analysis of the theory's scattering processes, matrix elements,  correlation functions, and so on. The determination of this spectrum is however a dynamical problem in and of itself: typically
the elementary fields explicitly present in the Lagrangian of a model do not in fact exhaust the spectrum of stable excitations.  Beyond the particles created by these fields, there could also be for instance kink-like excitations, where a field interpolates between two
different minima of the potential, and additional kink-antikink bound states. The analysis of the entire spectrum can be daunting.  However in certain cases even a partial solution of the dynamics of the theory can lead to the derivation of the entire spectrum by imposing certain self-consistency constraints on the poles in the physical strip of the various $S$-matrix amplitudes. 

Such an approach, termed the bootstrap, works efficiently for $1+1$ dimensional integrable quantum field theories (IQFT)~\cite{ZamZam} 
and has led to an exact solution of many models, among which the sine-Gordon, the Gross-Neveu models, and the Ising model in 
a magnetic field~\cite{Zam} stand out (for an overview of IQFT and an extensive set of references, see Ref.~\cite{mussardo-book}). 
In the first two examples, the spectrum of the stable excitations of the models is far richer than what one would infer from their 
Lagrangians~\cite{ZamZam,Zam}.  We also mention that, as a particular feature of integrable theories, there can be stable excitations with 
mass $m$ higher than the natural energy threshold $E= 2 m_1$ dictated by the lowest mass excitation $m_1$: the stability of 
these particles is ensured in this case by the infinite number of conservation laws which characterize the integrable dynamics.     

In quantum field theories which are not integrable, determining their full spectrum is less elementary.  
Here the dynamics are considerably more complicated because of the presence of inelastic processes such as particle 
production and resonances.  All of these phenomena strongly effect the analytic structure of the $S$-matrix 
amplitudes and make it difficult to determine the location of their poles in the physical strip.  Relatively few analytic methods 
are available to make progress on these aspects of non-integrable models. In the case of two-dimensional quantum 
field theories, there are essentially two techniques (beyond standard Feynman diagram perturbation theory): 
\begin{itemize}
\item 
Form factor perturbation theory \cite{DMS,DM}: this is a method particularly well suited to the study of non-integrable field theories 
obtained as weak deformations of integrable models.  A prototypical example here may be considered the multi-frequency 
sine-Gordon model, 
whose Lagrangian contains two or more cosines of different frequencies \cite{DM,bajnok-2001}.
\item Semi-classics \cite{DHN1,DHN2,GJ,FFvolume,mussardo-2007}: this approach can be used to analyze in simple terms 
non-integrable field theories with heavy topological (kink-like) excitations.
\end{itemize}

In addition to these analytic techniques, there is an efficient numerical method that allows one to extract the basic features of 
2D non-integrable field theories, the so-called truncated conformal space approach (TCSA). 
With this method, one gets direct 
access to the energy eigenvalues (and eigenvectors) of a theory defined in a cylindrical spacetime that can be thought of
as a perturbation of a conformal field theory (CFT).  Originally introduced by V.P. Yurov and Al.B. Zamolodchikov \cite{yurov-1990,yurov-1991} 
and used later by several authors, the TCSA has been significantly improved in its accuracy courtesy of renormalization group 
considerations \cite{feverati2008renormalization,konik-2007,konik-2009,brandino-2010,Watts1,Watts2,konik-2011,caux-2012,konik-2014,takacsRG}. 
As discussed in more 
detail later, the key idea of this algorithm consists in studying the numerical spectrum of the off-critical Hamiltonian on an 
infinite cylinder of circumference $R$, using the basis of the unperturbed conformal field theory as a computational basis.  
Introducing a finite
energy cutoff into this spectrum then reduces the problem to a numerical diagonalization of a finite dimensional Hamiltonian. 

In this paper we extend the TCSA to study the important class of quantum field theories given by the Landau-Ginzburg (LG) models
(Contemporaneously with our own work, M. Hogervorst, S. Rychkov,  and
B. C. van Rees \cite{tcsa_d>2} have studied LG theories using TCSA
in dimensions greater than 2.)
The most familiar example of these models is $\Phi^4$ theory with $\phi(x)$ a real scalar field:
\EQ
{\cal L} = \frac{1}{8\pi}\bigg[(\partial_\mu\phi)^2 +  g_2 \phi^2 + g_4 \phi^4 \bigg] \,, \quad g_4 > 0 \,.
\label{LPhi4}
\EN 
Understanding this action to be normal-ordered, for $g_2>0$ this
theory has a unique vacuum while for $g_2 < 0$, 
the theory possesses two degenerate vacua, connected by kink excitations.  One of our goals of this work is to confirm a conjecture put forward by one of us (GM) in \cite{mussardo-2007}, i.e.\ that a non-integrable quantum field theory with degenerate vacua (connected by kink excitations) cannot have more than two stable topological neutral excitations over a given vacuum. 

The paper is organized as follows: in Section \ref{sec:semi} we briefly review the main steps of the semi-classical approach
that leads to the prediction of the spectrum of neutral bound states in theories with kink excitations.
In Section \ref{sec:num met} we examine the obstacles one faces in implementing the TCSA for the Landau-Ginzburg theories 
and the way we overcome them.  In Section \ref{sec:phi2} we demonstrate that we can reproduce the expected spectrum
of a $\Phi^2$ theory (a free massive boson), a non-trivial check given our starting point is a perturbed massless compactified boson.
In Section \ref{sec:posg2g4} we study the unbroken phase of $\Phi^4$ LG theories, 
comparing our result with standard perturbation theory.
Section \ref{sec:double_well} is devoted to the analysis of the stable neutral
excitations present in the broken phase of the $\Phi^4$ LG theories 
where we show that the number of such excitations is never higher than two, as conjectured in \cite{mussardo-2007}.
We also check this conjecture for higher order LG theories with two degenerate vacua.
We then draw our conclusions in Section \ref{conclusions}.

\section{Semi-classical predictions in kink-like theories}
\label{sec:semi}

\begin{figure}[b]
\centerline{\scalebox{0.5}
{\includegraphics{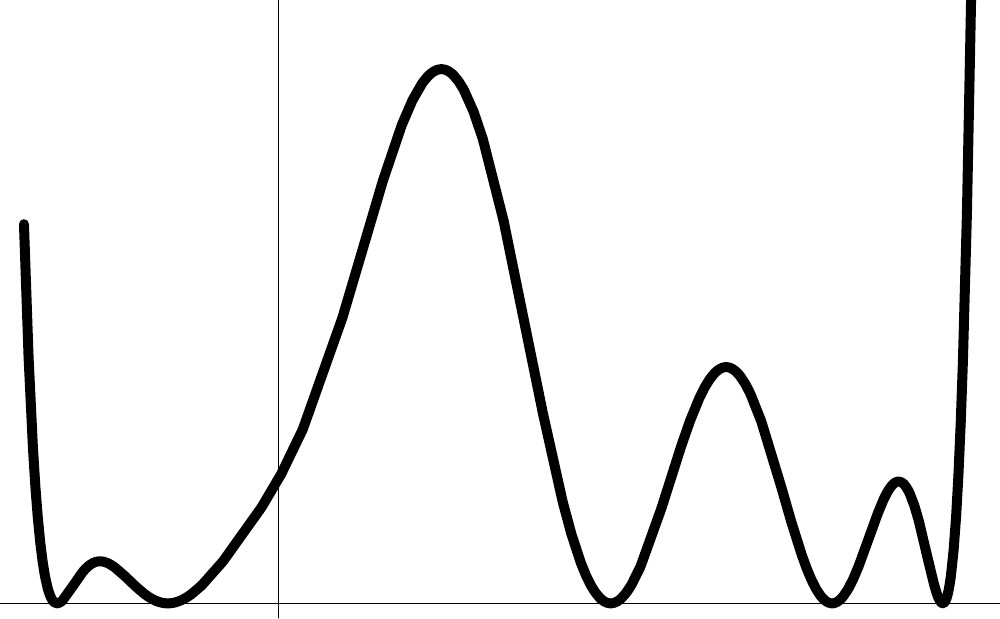}}} 
\caption{ Potential $U(\phi)$ of a quantum field theory with kink 
excitations.}
\label{potential}
\end{figure}

The spectrum of non-integrable quantum field theories with topological excitations (kink-like) can be studied by means of semiclassical methods \cite{DHN1,DHN2,GJ,FFvolume,mussardo-2007}. 
The simplest examples of these theories involve a scalar real field $\phi(x)$  
and a Lagrangian density 
\EQ
{\cal L} \,=\,\frac{1}{4\pi}\bigg[ \frac{1}{2}(\partial_{\mu} \phi)^2 + U(\phi)\bigg]\,, 
\label{Lagrangian}
\EN 
where the potential $U(\phi)$ has several degenerate minima at $\phi_a^{(0)}$ ($a =1,2,\ldots,n$), as sketched in 
Fig. \ref{potential}. These minima correspond to the different vacua $\ket{a}$ of 
the associated quantum field theory. 

\vspace{1mm}

\noindent
{\bf Basic Excitations:} The basic excitations of these kinds of models are the kinks and anti-kinks which interpolate between two neighboring vacua. Semi-classically they are described by the static solutions of the equation of motion, i.e.
\EQ
\partial^2_x \,\phi(x) \,=\,U'[\phi(x)] \,, 
\label{static}
\EN
with boundary conditions $\phi(-\infty) = \phi_a^{(0)}$ and 
$\phi(+\infty)= \phi_{b}^{(0)}$, where $b = a \pm 1$. 
Denoting by $\phi_{ab}(x)$ the solutions of this equation and by $\epsilon_{ab}(x)$ their 
classical energy density,
\EQ
\epsilon_{ab}(x) \,=\,\frac{1}{8\pi}\bigg[ \left(\frac{d\phi_{ab}}{d x}\right)^2 + 2U(\phi_{ab}(x))\bigg]  \,,
\EN 
the classical expression of the kink masses is given by  
\EQ
M_{ab} \,=\,\int_{-\infty}^{\infty} \epsilon_{ab}(x) \,.
\label{integralmass}
\EN   
The boundary values assumed by the field at $x = \pm \infty$ is preserved by the time evolution, hence we can characterize all finite-energy and 
regular solutions in terms of sectors labelled by two indices, namely $\phi(x=\infty, t)$ and $\phi(x=-\infty, t)$. These sectors are topologically 
disconnected. A simple way to keep track of these sectors is to introduce a topological charge given by 
\begin{equation}
{\mathcal T} \,=\, \int^{\infty}_{-\infty} dx \,t^0 \,,
\label{topological charge}
\end{equation}
where 
\begin{equation}
t^{\mu} = \epsilon^{\mu \nu} \partial_{\nu} \phi \,,
\qquad
\partial_{\mu} t^{\mu} = 0 \,.
\end{equation}
If properly normalised by the vacuum ``jump'' 
$\Delta_a \phi \equiv \phi_a^{(0)} - \phi_{a+1}^{(0)}$ in each sector,  the topological charge can take values ${\mathcal T} = \pm 1$ in all sectors. 
If boosted by a Lorentz transformation, i.e.\ $\phi_{ab}(x) \rightarrow \phi_{ab}\left[(x \pm v t)/\sqrt{1-v^2}\right]$, these field configurations 
describe in the quantum theory the kink states $\ket{K_{ab}(\theta)}$, where $a$ and $b$ are the indices of the initial and final vacuum, respectively. 
The quantity $\theta$ is the rapidity variable which parametrizes 
the relativistic dispersion relation of these excitations, i.e. 
\EQ
E = M_{ab}\,\cosh\theta \,, \qquad
P = M_{ab} \,\sinh\theta\,.
\label{rapidity}
\EN 
Conventionally $\ket{K_{a,a+1}(\theta)}$ denotes the {\it kink} between the pair of vacua $\left\{\ket{a\,\rangle ,\mid a+1}\right\}$ 
with topological charge ${\mathcal T} = 1$ while $\ket{K_{a+1,a}(\theta)}$ is the corresponding {\it anti-kink}, with 
topological charge ${\mathcal T} = -1$:  the multi-particle states are then given by strings of these 
excitations satisfying the adjacency condition of the consecutive indices for the continuity of the field configuration, i.e.
\EQ
\ket{ K_{a_1,a_2}(\theta_1) \,K_{a_2,a_3}(\theta_2)\,K_{a_3,a_4}(\theta_3) \ldots } \,, \quad
(a_{i+1} = a_i \pm 1)\,.
\EN 

\vspace{1mm}

\begin{figure}[t]
\centerline{\scalebox{0.4}
{\includegraphics{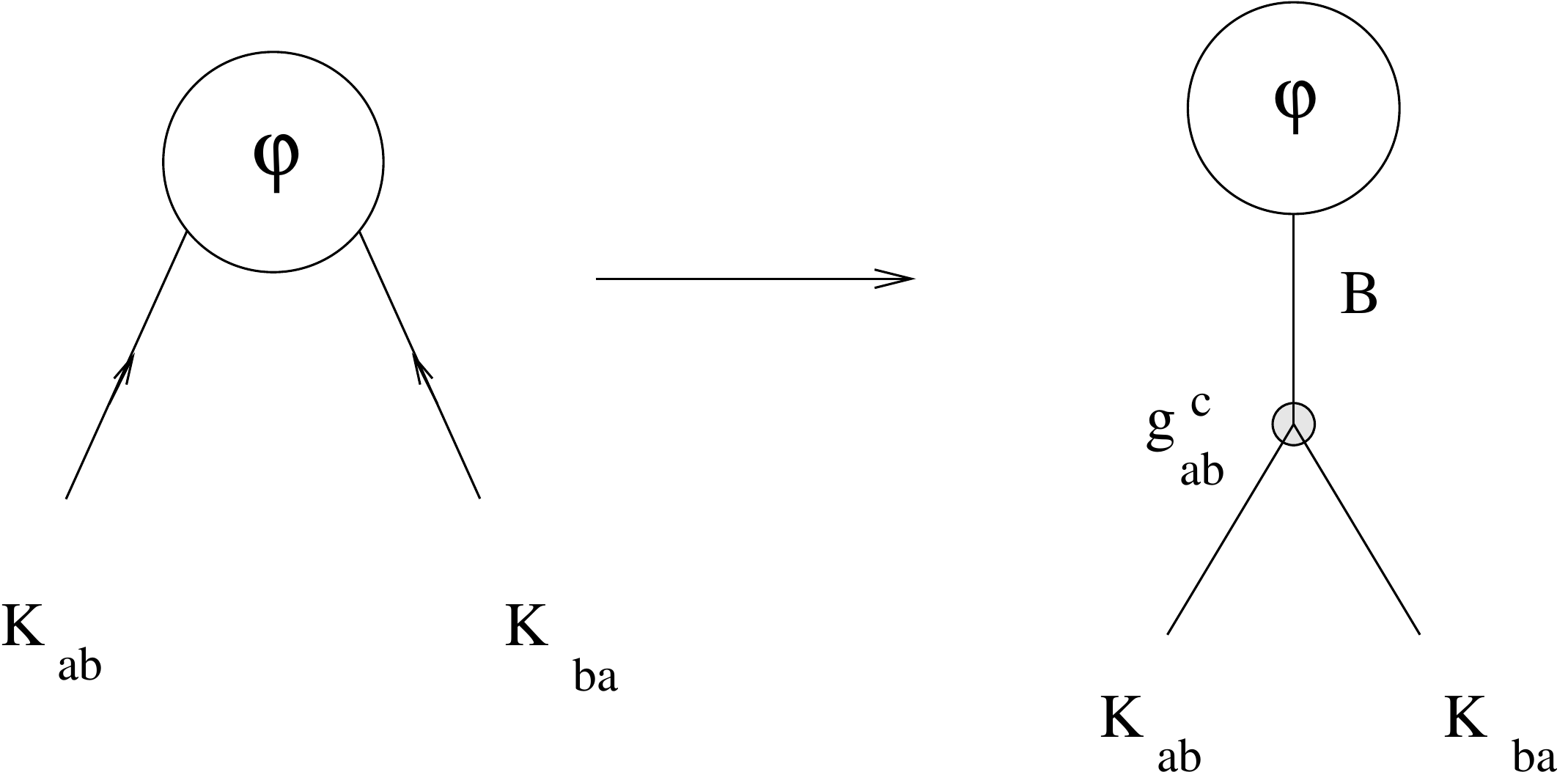}}} 
\caption{ Residue equation for the matrix element on the kink states.}
\label{residuef}
\end{figure}

\noindent {\bf Neutral Bound States:}
Beyond the kinks, in the quantum theory there may also exist kink-antikink bound states with topological charge ${\mathcal T} = 0$. 
These are topological neutral excitations, $\ket{B_c(\theta)}_a$ ($c=1,2,\ldots$), over each of the 
vacua, $\ket{a}$, namely the classical field configuration takes the boundary value, $\phi_a^{(0)}$, both at $ x = \pm \infty$.  For a theory 
based on a Lagrangian of a single real field, these states are all non-degenerate: there are no 
additional operators which commute with the Hamiltonian that would give rise to a multiplicity of such states. The neutral particles 
must be identified as the bound states of the kink-antikink configurations that start and end at the same vacuum, $\ket{a}$, 
i.e.\ $\ket{K_{ab}(\theta_1) \,K_{ba}(\theta_2)}$. If such two-kink states have a pole at an imaginary value, $i \,u_{a b}^c$, within 
the physical strip, $0 < {\rm Im}\, \theta < \pi$, of their rapidity difference, $\theta = \theta_1 - \theta_2$, then their bound 
states are defined through the factorization formula which holds in the vicinity of this singularity:
\EQ
\ket{K_{ab}(\theta_1) \,K_{ba}(\theta_2)} \,\simeq \,i\,\frac{g_{ab}^c}{\theta - i u_{ab}^c}\,\ket{B_c}_a \,.
\label{factorization}
\EN 
In this expression $g_{ab}^c$ is the on-shell 3-particle coupling between the kinks and the neutral particle. 
Moreover, the mass of the bound states is simply obtained by substituting the resonance value $i \,u_{ab}^c$ 
within the expression for the Mandelstam variable $s$ of the two-kink channel:
\EQ
s = 4 M^2_{ab} \,\cosh^2\frac{\theta}{2} 
\quad \longrightarrow \quad
m_c \,=\,2 M_{ab} \cos\frac{u_{ab}^c}{2} \,.
\label{massboundstate}
\EN 
In order to determine the resonance values $u_{ab}^c$, one can make use of a remarkably simple formula due to Goldstone-Jackiw \cite{GJ}, that applies in a semiclassical approximation, i.e.\ when the coupling constant goes to zero and the mass of the kinks becomes correspondingly very large with respect to any other mass scale. In its refined version, given in \cite{FFvolume}, this formula reads as follows 
\EQ
f_{ab}^{\phi}(\theta) \,=\,
\braopket{K_{ab}(\theta_1)}{\phi(0)}{K_{ab}(\theta_2)}
\,\simeq \,
\int_{-\infty}^{\infty} dx \,e^{i M_{ab} \,\theta\,x} \,\,
\phi_{ab}(x) \,,
\label{remarkable1}
\EN  
where $\theta = \theta_1 - \theta_2$. Substituting in this formula $\theta \rightarrow i \pi - \theta$, 
the corresponding expression may be interpreted as the following form factor: 
\EQ
F_{ab}^{\phi}(\theta) \,=\, f(i \pi - \theta) \,=\,
\braopket{a}{\phi(0)}{K_{ab}(\theta_1) \, K_{ba}(\theta_2)} \,,
\label{remarkable2}
\EN    
in which appears the two-particle kink state about the vacuum $\ket{a}$ of interest.
Eq.\,(\ref{remarkable1}) deserves several comments: 
\begin{itemize}
\item 
One appealing aspect of the formula (\ref{remarkable1}) consists of the relation between the 
Fourier transform of the {\em classical} configuration of the kink, -- i.e.\ the solution $\phi_{ab}(x)$ of the 
differential equation (\ref{static}) -- and the {\em quantum} matrix element of the field $\phi(0)$ between 
the vacuum $\ket{a}$ and the 2-particle kink state $\ket{K_{ab}(\theta_1) \,K_{ba}(\theta_2)}$. 
\item Given the solution of Eq.\,(\ref{static}) and its Fourier transform, the poles of $F_{ab}(\theta)$ within the physical strip of $\theta$ identify the neutral bound states which couple to $\phi$. The mass of the neutral particles can be extracted by using Eq.\,(\ref{massboundstate}), while the on-shell 3-particle coupling $g_{ab}^c$ can be obtained from the residue at these poles (see Fig. \ref{residuef}): 
\EQ
\lim_{\theta \rightarrow i \,u_{ab}^c} (\theta - i u_{ab}^c)\, F_{ab}(\theta)
\,=\,i \,g_{ab}^c \,\braopket{a}{\phi(0)}{B_c} \,.
\label{residue}
\EN  
\end{itemize}

The simplest application of this analysis is to the broken phase of the $\Phi^4$ LG theory, whose potential can be chosen as   
\EQ
U(\phi) = \frac{g_2}{2}\phi^2 + \frac{g_4}{2}\phi^4 .
\EN 
Denoting by $\ket{\pm 1}$ the vacua relative to the classical minima $\phi_{\pm}^{(0)} \,=\,\pm \sqrt{-g_2/(2g_4)}$ and 
expanding $U(\phi )$ around them via $\phi \,=\,\phi_{\pm}^{(0)} + \eta$, we have 
\EQ
U(\phi_{\pm}^{(0)} + \eta) \,=\, -g_2\eta^2 \pm \sqrt{-2g_2g_4}\eta^3 + \frac{g_4}{2}\eta^4.
\label{potentialphi4}
\EN 
Hence, ordinary perturbation theory predicts the existence of a neutral particle for each of the two vacua, with a bare mass given by $m_0 = \sqrt{-2g_2}$, irrespective of the value of the coupling $g_4$. Let us see, instead, what is the conclusion reached by 
the semiclassical analysis. The classical kink solutions are 
\EQ
\phi_{-a,a}(x) \,=\,a \,\frac{m_0}{2\sqrt{g_4}} \,\tanh\left[\frac{m_0 x}{{2}}\right]
\,, \qquad a = \pm 1 
\label{kinksolphi4}
\EN 
and their classical mass is 
\EQ
M_0\,=\,\int_{-\infty}^{\infty} \epsilon(x) \,dx \,=\, -\frac{m_0 g_2}{12\pi g_4}.
\EN 
By taking into account the contribution of the small oscillations around the classical static configurations, 
the kink mass gets corrected to \cite{DHN1,DHN2}
\EQ
M \,=\, -\frac{m_0 g_2}{12\pi g_4} - m_0\bigg(\frac{3}{2\pi}-\frac{1}{4\sqrt{3}}\bigg)+ {\cal O}(g_4) \,.
\label{mass1phi4}
\EN 
So, introducing   
\EQ
c =  \left(\frac{3}{2\pi} - \frac{1}{4 \sqrt{3}}\right) > 0 \,,
\EN
and the dimensionless quantities
\EQ
g = -12\frac{g_4}{g_2}
\quad \text{and} \quad
\xi \,=\,\frac{g}{1 - \pi c g} \,, 
\label{definitiong}
\EN 
the mass of the kink can be expressed as 
\EQ
M \,=\,\frac{m_0}{\pi \,\xi}\,.
\label{newmassphi4}
\EN
Since the kink and the anti-kink solutions are equal functions (up to a sign), their Fourier transforms have the same poles and therefore the spectrum of the neutral particles will be the same on both vacua, in agreement with the $Z_2$ symmetry of the model. We have  
\EQ
f_{-a,a}(\theta) \,= \, \int_{-\infty}^{\infty} 
dx \,e^{i M  \theta\,x} \phi_{-a,a}(x)  \, =\, 
 i a \sqrt{\frac{1}{g_4}}\,\frac{1}{\sinh\left(\frac{\pi M}{m_0} \theta\right)}\,.
\EN
By making now the analytical continuation $\theta \rightarrow i \pi - \theta$ and using the above relation (\ref{newmassphi4}), we have  
\EQ
F_{-a,a}(\theta) \,= \, \braopket{a}{\phi(0)}{K_{-a,a}(\theta_1) K_{a,-a}(\theta_2)}
\,\propto \,
\frac{1}{\sinh\left(\frac{i \pi - \theta}{\xi}\right)} \,.
\label{FFphi4}
\EN 
whose poles are placed at 
\begin{equation}
\theta_{n}\,=\,i \pi \left(1 - \xi \,n\right)
\,, \qquad
n = 0,\pm 1,\pm 2,\ldots
\label{polesphi4}
\end{equation}
Notice that if 
\EQ
\xi \geq 1 \,,
\label{conditionphi4}
\EN 
none of these poles is in the physical strip $0 < {\rm Im}\,\theta < \pi$. Consequently, in the 
range of the coupling constant  
\EQ
\frac{g_4}{|g_2|} \geq \frac{1}{12} \,\frac{1}{1 + \pi c} \,=\,0.040719\ldots
\label{criticalg}
\EN 
the theory does not have any neutral bound states, neither above the vacuum to the right nor above the one to the left. 
Conversely, if $\xi < 1$, there are $n = \left[\frac{1}{\xi}\right]$ neutral bound states, where $[ x ]$ denotes the integer 
part of the number $x$. Their semiclassical masses are given by
\begin{equation}
m_{\text{sc},n} \,=\, 2 M\,\sin\left(n \frac{\pi \xi}{2}\right)
\,=\,
n\,\,m_0\left[1 - 6\pi^2\bigg(\frac{g_4}{g_2}\bigg)^2n^2 + \ldots \right] .
\label{massphi4}
\end{equation}
It can be shown \cite{mussardo-2007} that for general two-well potentials the semiclassical analysis predicts the same universal formula for the bound state masses $m_{\text{sc},n}^{(a)}$ above each vacuum $\ket{a}$, with $\xi^{(a)} = m_0^{(a)}/(\pi M)$, as in Eq.~\eqref{newmassphi4}.
Here $m_0^{(a)}$ is the perturbative mass around the corresponding minimum of the potential (its curvature) and $M$ is the mass of the kink interpolating between the two vacua.
It is easy to see that the leading term of this expression is given by multiples of the mass of the elementary boson $\ket{B_1}$.
This leads to the interpretation of the $n$-th neutral excitation as a loosely bound state of $n$ $\ket{B_1}$'s, 
with the binding energy provided by the remaining terms of the above expansion. 
However, as a consequence of the non-integrability of the theory, all particles with mass $m_{\text{sc},n} > 2 m_{\text{sc},1}$ decay. 

We can further see that if there are at most two particles in 
the spectrum, the inequality $m_{\text{sc},2} < 2 m_{\text{sc},1}$ is always valid. However, if $\xi < \frac{1}{3}$, one always has 
\EQ
m_{\text{sc},k} > 2 m_{\text{sc},1} \,,\qquad
\text{for}\,\, k=3,4,\ldots n \,.
\EN 
Hence, according to the semiclassical analysis, the spectrum of neutral particles of $\phi^4$ theory is, for each vacuum, as 
follows \cite{mussardo-2007}:
\begin{itemize}
\item if $\xi > 1$, there are no neutral particles; 
\item if $\frac{1}{2} < \xi < 1$, there is one particle; 
\item if $\frac{1}{3} < \xi < \frac{1}{2}$ there are two particles; 
\item if $\xi < \frac{1}{3}$ there are $\left[\frac{1}{\xi}\right]$ particles, with the first two stable and the latter as resonances.  
\end{itemize}


\section{Numerical methods}
\label{sec:num met}

In this section we discuss the implementation of the truncated conformal space approach (TCSA), originally 
proposed by V. Yurov and Al. Zamolodchikov \cite{yurov-1990}, to Landau-Ginzburg theories. TCSA is a numerical method suitable 
to study theories which can be written as a conformal field theory (CFT) perturbed by a relevant or marginal operator. 
The method amounts to building the Hamiltonian of a quantum field theory defined on a cylinder of radius $R$ using 
as a computational basis the eigenstates of the unperturbed CFT.  This basis is truncated, typically, by throwing away states
above some energy scale and then diagonalizing the resulting finite dimensional matrix in order to extract the approximate energy levels.

The computational basis is provided by the irreducible representations of the unperturbed CFT.  Each irreducible representation 
of the CFT is associated with a highest weight state and its associated tower of descendant states formed by acting on the highest
weight state with the operators of the Virasoro algebra -- taken all together this set of states is known as a Verma module.  

The Hamiltonian associated to the perturbed CFT is always made of two pieces 
\begin{equation}
H\,=\,H_{\text{CFT}} + V \,\,\,,
\label{pertham}
\end{equation}
where the first term $H_{\text{CFT}}$ corresponds to the Hamiltonian of the unperturbed CFT and is purely diagonal, while the second term $V$ 
corresponds to the perturbation.   Its matrix elements on the computational basis are computed in terms of the three-point functions on 
the cylinder involving the perturbing operators and the towers of conformal states. 

The TCSA has been successfully applied to perturbed minimal models of CFT (with central charge $c<1$)~
\cite{yurov-1991,lassing-1991,lepori-2008,lepori-2009}, to compactified 
bosonic theories ($c=1$)~\cite{feverati-1998,konik-2011,konik-2014}, 
to the sine-Gordon model~\cite{feverati-1998b,bajnok-2001}, and to boundary conformal field theories~\cite{dorey-2000}. 
Recently~\cite{beria-2013} it has been also applied to perturbed Wess-Zumino-Witten models with $SU(2)$ symmetry ($c=1$). 
In all these cases, the number of representations involved is {\em finite}, and the truncation can be easily implemented as a 
UV cutoff within each representation.  If the CFT has a countable number of representations, whose highest-weight states are 
ordered in energy, the truncation can still be performed by keeping those representations (at least partially)
where the energy of the highest-weight state 
is below the UV cutoff.

Serious problems arise instead when the CFT has an uncountable set of representations, as happens in the Landau-Ginzburg theories 
described by an Euclidean action of the form
\begin{equation}
\mathcal S=\frac{1}{8\pi}\int \ud^2x \left( :\partial_\mu\phi\,\partial^\mu\phi: + \sum_n g_n :\phi^n: \right).
\label{ss}
\end{equation}
The boson here is uncompactified.  It has a mode expansion given by \cite{DiFrancesco} 
\begin{equation}
\phi(z,\overline z) \,=\,\phi_0 - \frac{i}{4\pi} \pi_0 \,\log(z \overline z) + \frac{i}{\sqrt{4\pi}} \,
\sum_{n \neq 0} \frac{1}{n} \left(a_n \, z^{-n}  + \bar a_n \bar z^{-n}\right) \,\,\,,
\label{expansionphi}
\end{equation}
where $z/\bar{z} = e^{\frac{2\pi}{R}(\tau \mp ix)}$ and $\phi_0$ is the zero-mode of the boson.
The modes $a_n,\bar{a}_n$ satisfy the following set of commutation relations
\begin{eqnarray} \label{eq:algebra}
\left[ a_n,a_m \right]=n\,\delta_{n+m,0}\,,
\qquad
\left[ \bar a_n, a_m \right]=0\,,
\qquad
\left[ \bar a_n,\bar a_m \right]=n\,\delta_{n+m,0}.
\end{eqnarray} 
The normal ordering prescription $: \phi^k:$ for the various powers of the field and its derivatives that we use is that of the CFT. 
Normal ordering any operators means we will move all the modes of the algebra~\eqref{eq:algebra} with $n<0$ to the left of all 
modes with $n>0$. A more detailed discussion on the effects of this normal ordering is given in Section~\ref{subsec:normal order}.
The Virasoro algebra admits here a {\em continuum} of representations with conformal dimension $h = \bar{h}=\alpha^2/2$, 
for all $\alpha\in \mathbb R$.
There is an uncountable set of primary states in the conformal basis: therefore introducing an energy cutoff still leaves
the conformal basis infinite.

In the light of this, to apply the TCSA we abandon its description in terms of an uncompactified boson and turn to a more
numerically amenable one.

\subsection{The (conformal) computational basis}

With the aim of finding a discrete computational basis, let us consider a free compactified boson~\cite{DiFrancesco}
taking values on a circle of length $2\pi/\beta$ so that 
\begin{equation}
\phi(x+R)\equiv \phi(x)+\frac{2\pi}{\beta}w\,,\qquad w\in\mathbb{Z}\,.
\end{equation}
where here $R$ is the system size.
The integer $w$ counts the winding number of the field when the spatial coordinate runs once around the cylinder.
Unlike its uncompactified counterpart, the compactified boson has a countable set of Verma modules $V_{n,w}$, with conformal weights
\begin{equation}
h_{n,w}=\frac{1}{2}\left(n\beta+\frac{w}{2\beta}\right)^2\,,\qquad
\bar h_{n,w}=\frac{1}{2}\left(n\beta-\frac{w}{2\beta}\right)^2 \,,
\end{equation}
where $n$ and $w$ are integer numbers.   The energies and momenta of the highest weight states in these Verma modules are given by
\begin{equation} \label{eq:full E P}
E_{n,w}=\frac{2\pi}{R}\left[\left(n\beta\right)^2+\left(\frac{w}{2\beta}\right)^2 -\frac{1}{12} \right] \,,\qquad P_{n,w} = \frac{2\pi}{R}\;n\cdot w \,.
\end{equation}
These highest weight states can be represented by the action of a vertex operator on the vacuum, i.e.\ $\ket{n,w}\equiv V_{n,w}(0)\ket{0}$.
The Verma module is created by 
acting with the modes $a_n,\bar{a}_n$ with $n<0$ on the highest weight state.
Verma modules with finite winding number correspond to U(1) charged sectors of the theory -- as these have no counterpart
in the Hilbert space of the uncompactified boson, we will always work at $w=0$.

The uncompactified boson can always be recovered in the limit $1/\beta \rightarrow \infty$.   For finite values of $\beta$ the potential
of the boson becomes periodic and its potential $U(\phi)$ will typically be bounded from above.  However provided $\beta$ is small
enough, the low energy states of the full theory will not feel this bound and we expect to obtain a low energy spectrum appropriate
for an uncompactified boson.  We will demonstrate this explicitly in
Section \ref{sec:phi2}.

More concretely, the $w=0$ highest weight states, denoted by $\ket{n}$, are generated from the vacuum $\ket{0}$ 
by applying the primary fields as follows:
\begin{equation}
 \ket{n} \equiv\,\,  : e^{i\beta n \phi(0)} : \ket{0} \mbox{\quad \,\,\,  with \quad} \phi(0)=\phi_0+\phi_L(0)+\phi_R(0)\,. 
\end{equation}
Here $\phi_0$ is the zero mode, while $\phi_L$ and $\phi_R$ contain respectively the holomorphic and antiholomorphic modes, $a_j$'s and $\bar a_j$'s.
The energies and momenta of such highest weight states are given by 
\begin{equation}
 E_n = \frac{2\pi}{R}\left[\left(n\beta\right)^2 - \frac{1}{12} \right] \,,\qquad P_n = 0 \,.
\end{equation}
From each highest weight state, the descendants are obtained by applying the left and right modes of the field:
\begin{equation}
 \ket{n,a_L,a_R} = \underbrace{a_{-1}^{p_1}a_{-2}^{p_2}\dots}_{a_L}\; \underbrace{\bar a_{-1}^{q_1}\bar a_{-2}^{q_2}\dots}_{a_R} \ket{n} \,.
\end{equation}
Their energy and momentum are given by 
\begin{equation} \label{eq:w=0 E P}
 E = E_n + \frac{2\pi}{R}\sum_{j} j \left( p_j + q_j \right) \,, \qquad P = \frac{2\pi}{R} \sum_{j} j \left( p_j - q_j \right) \,.
\end{equation}
The restriction of the Hilbert space to energies below $E_{\text{tr}}$ leaves us thus with a finite dimensional Hilbert space.
Typically we will define the truncation of the Hilbert space by only keeping states whose energy is equal to or below 
\begin{equation}
E_{\text{tr}}=\frac{2\pi}{R}\left(2N_{\text{tr}}-\frac{1}{12}\right) \,,
\end{equation}
where $N_{\text{tr}}$ is an integer.

We will conclude this section by observing that one may think that for the LG theories it might be more convenient to 
use as a computational basis the eigenbasis of the free massive boson.  This is obviously true for the disordered phases of LG
theories where there is but a single vacuum.  
However such a computational basis is inappropriate for studying LG theories with multiple vacua -- thus we opt to
use a computational basis built around a massless boson where both broken and unbroken phases are treated on equal
footing.


\subsection{Normal Ordering}
\label{subsec:normal order}

As it is well known, for any scalar field theory in two dimensions with non-derivative interactions \cite{coleman1975}
the only ultraviolet (UV) divergences that occur at any order of perturbation theory come from contractions of two fields at 
the same vertex, that is, from any tadpole present as a subdiagram in the diagrammatic expansion.
Thus, all ultraviolet divergences can be removed by normal ordering the Hamiltonian.
When we do usual perturbation theory of $\Phi^4$ theory in the unbroken phase ($g_2=m_0^2>0$) around the free massive boson of 
mass $m_0$, 
there is an obvious choice for the normal ordering prescription of any operator.
In this case, we define ladder operators $a(k,m_0)$ and $a^\dag(k,m_0)$ via the usual mode expansions for the 
field $\phi$ and its conjugate momentum $\pi$:
\begin{equation} \label{eq:def a adag}
\begin{aligned}
 \phi(x) &= \int \frac{\ud k}{2\pi}\frac{1}{\sqrt{2E(k,m_0)}}\left[ a(k,m_0)\,e^{-i\,k\cdot x} + a^\dag(k,m_0)\,e^{i\,k\cdot x} \right] \,,\\
 \pi(x)  &= \int \frac{\ud k}{2\pi}\sqrt{\frac{E(k,m_0)}{2}} \left[ a(k,m_0)\,e^{-i\,k\cdot x} - a^\dag(k,m_0)\,e^{i\,k\cdot x} \right] \,,
\end{aligned}
\end{equation}
where here $k\cdot x = kx-E(k,m_0)t$ and $E(k,m_0) = \sqrt{k^2+m_0^2}$.
Any operator can be written as function of $\phi(x)$ and $\pi(x)$ and the corresponding normal ordered operator is obtained by
rearranging all the $a^\dag$'s so that they are found to the left of all $a$'s.
The normal ordered operator $:\phi^2:$ is then given by
\begin{equation} \label{eq:NO phi2}
 :\phi^2(x):\; = \lim_{x\rightarrow y}\Big( \phi(x)\phi(y) - \left[ \phi^+(x), \phi^-(x) \right] \Big) 
               = \lim_{x\rightarrow y}\Big( \phi(x)\phi(y) - D_{m_0}(x-y) \Big)\,,
\end{equation}
where $\phi^+$ contains all the $a$ operators and $\phi^-$ contains all the $a^\dag$'s, while $D_{m_0}(x-y)$ is the real space free-field propagator
\begin{equation}
 D_{m_0}(x-y) = \int \frac{\ud^2 k}{(2\pi)^2} \frac{4\pi e^{i\,k\cdot (x-y)}}{k^2+m_0^2} = 2 K_0(m_0(x-y))\,.
\end{equation}
From Eq.~\eqref{eq:NO phi2} we understand that to implement normal ordering we need to subtract the 
propagator computed at $x=y$, where it diverges logarithmically
\begin{equation}
 D_{m_0}(x) = -2 \ln(c m_0 x) + \mathcal{O}(x^2) \,,
\end{equation}
where $c$ is a constant.
We need to regularize the theory with a UV cutoff $\Lambda$, and we define the regularized propagator $D_{m_0,\Lambda}(x)$ by integrating only momenta up to the cutoff $\Lambda$.
The regularized propagator is well-defined at $x=0$
\begin{equation}
 D_{m_0,\Lambda}(0) = \ln\left(\frac{\Lambda^2+m_0^2}{m_0^2}\right) \,.
\end{equation}
If we normal order the kinetic and the $\phi^2$ term in the Hamiltonian, only (infinite) constant terms appear.
On the contrary, if we normal order the interaction Hamiltonian we obtain counterterms which  make the theory finite.
Applying Wick's theorem, the normal ordered $\phi^4$ term can be written as follows:
\begin{equation}
 :\phi^4(x):\; = \phi^4(x) - 6D_{m_0,\Lambda}(0)\,\phi^2(x) + 3D_{m_0,\Lambda}^2(0)\,.
\end{equation}
The first term is the usual quartic interaction, the third term is a simple constant, while the second term is a mass counterterm.
The normal ordered Hamiltonian is now
\begin{equation}
 :H:\; =\; :H_0: + \int\ud{x}\; \frac{g_4}{8\pi}:\phi^4:\; 
       =    H_0  + \frac{1}{8\pi}\int\ud{x}\left( g_4\phi^4 - 6g_4D_{m_0,\Lambda}(0)\,\phi^2 \right) + \text{const.}\,.
\end{equation}
The new term makes the theory UV finite at any order in perturbation theory.
At one loop, it will exactly cancel out the only diagram which contributes to the mass renormalization, the tadpole.
All the other divergencies in any other diagram come from tadpole subdiagrams and thus will
be canceled out as well by the counterterm.

All of this concerns the usual massive perturbation theory.
We now want to go a little further, but to do so let us return for the moment to the definitions in Eq.~\eqref{eq:def a adag}:
in principle, we could make the same definition using some mass $\mu$ in place of $m_0$.
Let us stress that $\mu$ is \emph{not} the mass parameter in the Hamiltonian.
We can then define a normal order prescription with respect to the new ladder operators $a(k,\mu)$ and $a^\dag(k,\mu)$.
Thanks to Wick's theorem, the two operators arising from normal ordering the same operator with the
two different prescriptions can be related one to the other \cite{coleman1975}.
For example, for the $\phi^2$ and $\phi^4$ operators we have
\begin{equation}
\begin{aligned}
 :\phi^2:_\mu\; &=\; :\phi^2:_{m_0} - \log\left(\frac{m_0^2}{\mu^2}\right) \,,\\
 :\phi^4:_\mu\; &=\; :\phi^4:_{m_0} - 6\log\left(\frac{m_0^2}{\mu^2}\right):\phi^2:_{m_0} 
 + 3\left(\log\left(\frac{m_0^2}{\mu^2}\right) \right)^2 \,,
\end{aligned}
\end{equation}
and the normal ordered Hamiltonian turns out to be (apart from constant terms)
\begin{equation}
 :H:_\mu = \frac{1}{8\pi}\int\ud{x}\Big( (\partial\phi(x))^2 + g_2\phi^2(x)
       - 6g_4D_{\mu,\Lambda}(0)\,\phi^2(x) + g_4\phi^4(x) \Big) \,.
\end{equation}
Here the subscript to the normal ordering symbol $:\cdot:$ indicates which mass scale has been used in the definitions 
of the ladder operators by which the normal ordering is defined.
When we do the usual perturbation theory with a normal ordering prescription with $\mu\neq m_0$, 
the tadpole diagram is no longer canceled out, but its contribution is still made finite by the counterterm.
The net effect of such a counterterm is to replace any tadpole subgraph in the diagrammatic expansion with the 
difference between the two tadpoles with different masses.
In practice, we should substitute any tadpole integral according to the following prescription:
\begin{equation}
 \int\frac{\ud^2 k}{k^2+m_0^2}\;\;\longrightarrow \;\;
 \int\ud^2 k\left( \frac{1}{k^2+m_0^2} - \frac{1}{k^2+\mu^2} \right) \,.
\end{equation}
These integrals are intended to be cutoff by $\Lambda$, but it is easy to see that the result is essentially independent of the cutoff
provided $\Lambda\gg \mu,\,m_0$.

As already discussed, we will use TCSA with the basis of the free massless boson as the computational basis.
We implicitly consider a normal ordering prescription corresponding to a massless boson.
We can use the scheme introduced so far but we need to account for the infrared
divergences of the massless theory.
We thus introduce an infrared (IR) cutoff, $\Lambda_{\text{IR}}$, to carry out the calculations.
When working on a finite geometry, the natural IR cutoff is $\pi/R$, with $R$ the system size.
When we normal order the $\phi^2$ operator as in Eq.~\eqref{eq:NO phi2}, we have to subtract the propagator computed with both the UV and the IR cutoff
\begin{equation} \label{eq:counter term}
 :\phi^2(x):\; = \phi^2(x) - D_{\Lambda_{\text{IR}},\Lambda}(0) = 
                 \phi^2(x) - \log\left( \frac{\Lambda^2}{\Lambda_{\text{IR}}^2} \right) \,.
\end{equation}
The same applies for the normal ordered $\phi^4$ operator and for the normal ordered interaction Hamiltonian.

To compare perturbation theory with the TCSA numerics we should also consider that being on a cylinder 
with periodic boundary conditions has the effect of quantizing momentum in the space direction.
When computing loop corrections we should substitute all integrals over 2-momentum with a sum over the quantized spatial momentum and an integral over the energy.
To reproduce the TCSA result, we must also impose the correct momentum cutoff which can take values between 
$-(2\pi/R)\,2N_{\text{tr}}$ and $(2\pi/R)\,2N_{\text{tr}}$.
This means all integrals over momenta must be replaced according to the following prescription:
\begin{equation} \label{eq:momentum integral}
 \int_0^{\Lambda}\frac{\mathrm{d}^2q}{(2\pi)^2}\; f(q_0,q_1) \;\longrightarrow\; \frac{1}{R}\sum_{n=-2N_{\text{tr}}}^{2N_{\text{tr}}}
 \int_{-\infty}^{\infty}\frac{\mathrm{d}q_0}{2\pi} \; f\left(q_0,\frac{2\pi}{R}n\right) \,.
\end{equation}
In the case of the massless tadpole, the above sum is further restricted by omitting the $n=0$ term.

\subsection{The perturbation}

Once the computational basis has been chosen, we need to compute the corresponding matrix elements of 
the perturbation $V$.
Perturbations of the LG theories are powers of the bosonic field $\phi(x)$.
We use a trick to write these perturbations as derivatives of a vertex operator (whose matrix elements are straightforward
to compute): 
\begin{equation}
 M^{(m)}_{ab} = \bra{n,a_L,a_R} :\phi(x)^m: \ket{n,b_L,b_R} 
 = (-i)^m \bra{n,a_L,a_R} \partial_\gamma^m\, :e^{i\gamma \phi}:\,\big|_{\gamma=0} \ket{n,b_L,b_R} \,.
\end{equation}
We thus need to compute the matrix element $\ave{:e^{i\gamma\phi}:}$ for arbitrary $\gamma$.
In order to compute the matrix element we can separate the left and right part of the vertex operator, as well as the zero mode
\begin{equation}
 \bra{n,a_L,a_R} :e^{i\gamma\phi}: \ket{n',b_L,b_R} = \bra{n}
 e^{i\gamma\phi_0}
 \left((a_L)^\dag\, :e^{i\gamma\phi_L}:\, b_L\right)
 \left((a_R)^\dag\, :e^{i\gamma\phi_R}:\, b_R\right) \ket{n'} \,.
\end{equation}
The commutation relations of $:e^{i\gamma\phi}:$ with the modes $a_n,\bar{a}_n$ are given by
\begin{equation} \label{eq:comm}
 \left[a_n,:e^{i\gamma\phi}:\right]=-\gamma\, :e^{i\gamma\phi}: \,.
\end{equation}
Exploiting these commutation relations, the matrix element can be rewritten in terms of two polynomials in $\gamma$, 
$P_L(\gamma)$ and $P_R(\gamma)$:
\begin{equation}
 \bra{n,a_L,a_R} :e^{i\gamma\phi}: \ket{n',b_L,b_R} = \bra{n} e^{i\gamma\phi_0} \ket{n'} P_L(\gamma)P_R(\gamma) \,,
\end{equation}
where $P_{L(R)}(\gamma) = \alpha_{0L(R)} + \alpha_{1L(R)}\gamma + \alpha_{2L(R)}\gamma^2 + \dots$.
These polynomials are related to associated Laguerre polynomials that arise in the use of coherent
states (see for example Ref. \cite{perelomov}).
The remaining matrix element involving the zero mode can be computed using the following representation:
\begin{equation} \label{eq:hwstate_me}
 \bra{n} f(\phi_0) \ket{n'} = \frac{\beta}{2\pi}\int_{-\pi/\beta}^{\pi/\beta} \ud[\phi_0]\, e^{i\beta\phi_0(n'-n)} f(\phi_0) \,.
\end{equation}
The full matrix element then takes the following form:
\begin{equation} \label{eq:Mab(gamma)}
 M^{(m)}_{ab} = (-i)^m\, \partial_\gamma^m \left\{
 \frac{\beta}{2\pi} \int_{-\pi/\beta}^{\pi/\beta}\ud[\phi_0]\,e^{i\left[\beta(n'-n)+\gamma\right]\phi_0}\;
 P_L(\gamma)\; P_R(\gamma) \right\}_{\gamma=0} \,.
\end{equation}
If we now perform the derivative with respect to $\gamma$ and the integration in $\phi_0$, we get
\begin{equation} \label{eq:Mab}
 \begin{split}
 M^{(m)}_{ab} =&
  (-1)^{\frac{m}{2}}m!\sum_{p=0}^{m}\left(\frac{\pi}{\beta}\right)^p
  \left\lbrace \delta_{n,n'} \frac{1+(-1)^p}{2} \frac{(-1)^{\frac{p}{2}}}{(p+1)!}+\right.\\
 &\left.(1-\delta_{n,n'})(1-\delta_{p,0})(-1)^{n-n'-1}
  \sum_{\substack{q=0\\p-q\; \text{odd}}}^{p-1}\frac{(-1)^{\frac{p+q+1}{2}}}{(p-q)!\left[\pi(n-n')\right]^{q+1}}
  \right\rbrace \sum_{k,j=0}^m\alpha_{kL}\alpha_{jR}\,\delta_{k+j,m-p}\,.
 \end{split}
\end{equation}
This last formula expresses all matrix elements of the operator $\phi(0)^m$ in terms of the coefficients $\{\alpha_{sL(R)}\}$.
We present the formula specialized to the case of even $m$, since in the following we will only address parity symmetric LG theories.
However the method straightforwardly applies to LG theories containing odd powers of $\phi$.
The matrix element for the simplest $\phi^2$ perturbation reduces to
\begin{equation} \label{eq:Mab_C1}
 \begin{split}
 M^{(2)}_{ab} =& 
 \left\{ \delta_{n,n'}\frac{\pi^2}{3\beta^2} + \left(1-\delta_{n,n'}\right)\frac{2(-1)^{n-n'}}{\beta^2(n-n')^2} \right\}
   \alpha_{0L}\alpha_{0R} - \\
 &\left(1-\delta_{n,n'}\right)\frac{2(-1)^{n-n'}}{\beta(n-n')} \left( \alpha_{0L}\alpha_{1R} + \alpha_{1L}\alpha_{0R} \right)
  - 2\delta_{n,n'}\left( \alpha_{0L}\alpha_{2R} + \alpha_{2L}\alpha_{0R} + \alpha_{1L}\alpha_{1R} \right) \,.
 \end{split}
\end{equation}

\begin{figure}
\centering
\subfigure[]{\includegraphics[width=0.48\textwidth]{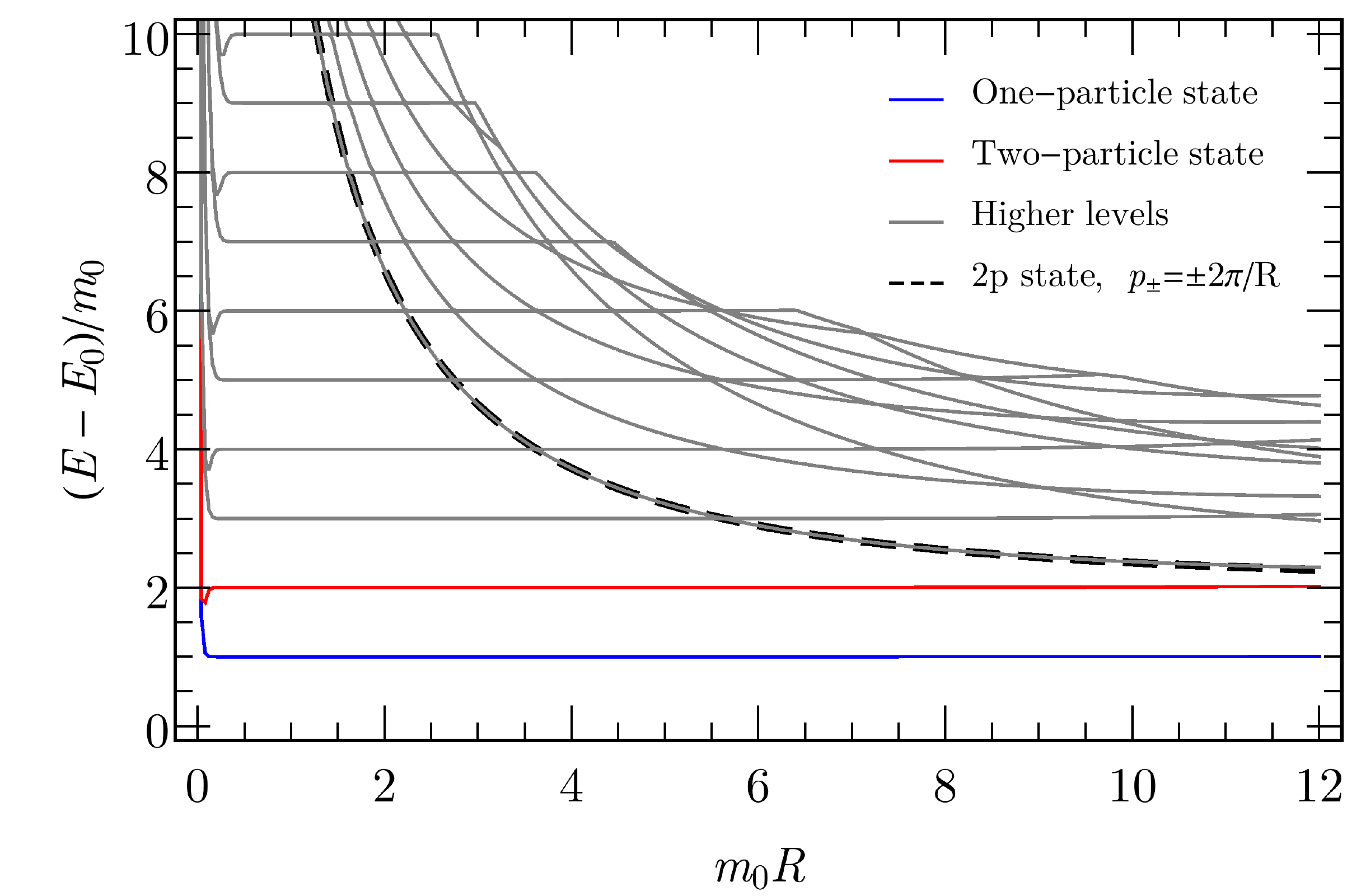}\label{subfig:phi2_sp}}\hspace{1mm}
\subfigure[]{\includegraphics[width=0.48\textwidth]{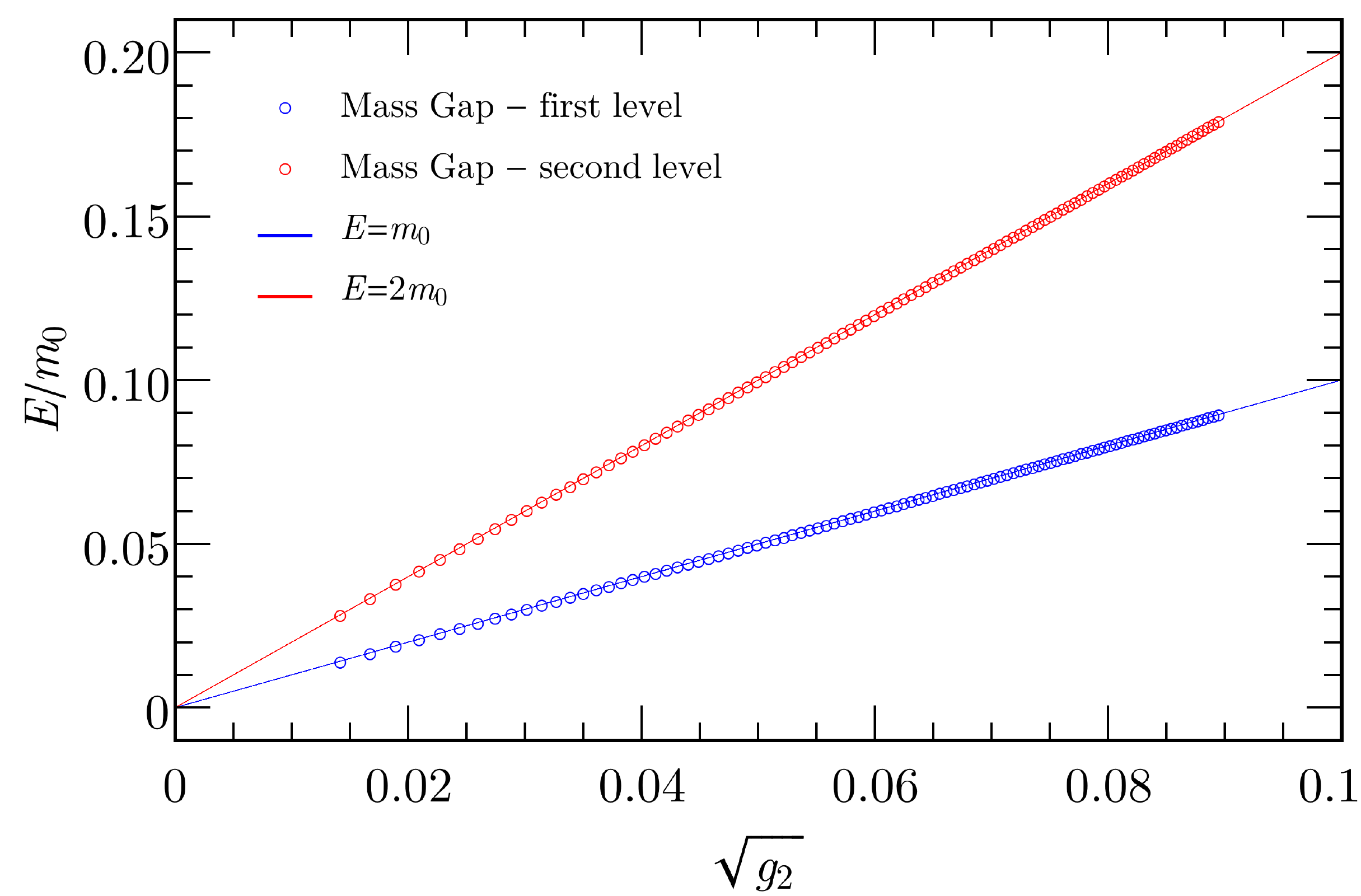}\label{subfig:phi2_mg}}\\
\caption{In panel \subref{subfig:phi2_sp} we show the energy levels (with the ground state energy subtracted) 
for the $\Phi^2$ perturbation.
Here $\beta=0.1$, $g_2=0.0016$, and $N_{\text{tr}}=6$.
The numerical mass coincides with the expected value of $m_0=\sqrt{g_2} = 0.04$.
The dashed black line represents the prediction for the energy of two particles moving with momentum $\pm 2\pi/R$, 
that is $E=2\sqrt{(2\pi/R)^2+m_0^2}$.  We see that it provides a good match to the numerical data.
In panel \subref{subfig:phi2_mg} we show the energy (points) 
of the first two levels as a function of $m_0=\sqrt{g_2}$ for a fixed $m_0 R=4$.
Their values are compared with the analytically expected values, $E=m_0$ and $E=2m_0$, plotted as solid lines.} 
\label{fig:phi2}
\end{figure}


\section{Reproducing the spectrum of a free massive boson}
\label{sec:phi2}

As a first check of our methodology, we study the free massive boson, i.e.\
\begin{equation}
\mathcal S=\frac{1}{8\pi}\int \ud^2x \left(:\partial_\mu\phi\partial^\mu\phi: + g_2:\phi^2:\right) \,.
\label{Sphi2}
\end{equation}
This is a non-trivial check as we use the states of a massless compactified boson as a computational basis.
In the zero momentum sector of this theory, the first excited state is a single particle state of mass $m_0=\sqrt{g_2}$.
The second excited state is the one with two particles at rest, and therefore with energy $2m_0$.
The higher energy excitations of this sector consist of some number of particles whose momenta sums to zero.

The spectrum that we obtain numerically in the zero momentum sector is shown in Fig.~\ref{subfig:phi2_sp}.
Here we plot the energy differences with respect to the ground state energy of the theory as a function of the cylinder radius $R$.
We have rescaled all energies as well as the cylinder radius by $m_0$, the natural mass scale of the theory.
The first level (blue) is the state with a single particle, while the second one (red) is the state with two particles, both at zero momentum.
This second state has energy of exactly $2m_0$, the threshold.
Indeed, since the theory is free the two particles do not interact and hence this state's energy has no dependence on
$R$ outside the small $R$ conformal region.
Above them, we see equally spaced horizontal lines corresponding to the $n$-particle states with all particles at rest, as well as the lines with some $1/R$ dependence, corresponding to the states where the single particles each have some momentum (which sums to zero).
In particular, the lowest momentum line corresponds to the state with a left moving and a right moving particle with momenta $\pm 2\pi/R$ and therefore with energy $2\sqrt{(2\pi/R)^2+m_0^2}$.
This prediction is shown in the plot as the black dashed curve.
In Fig.~\ref{subfig:phi2_mg} we also show the energy of the two first levels as function of the mass $m_0=\sqrt{g_2}$ at fixed cylinder radius $m_0 R$. The agreement with the expected values of $m_0$ and $2m_0$ again indicates the reliability of the method.

We also observe some additional features in the numerical spectrum:
\begin{itemize}
\item For small values of $m_0 R$, the Hamiltonian of the unperturbed CFT dominates and therefore the resulting spectrum is the conformal one with a $1/R$ dependence.
The degeneracies that we find are those expected for a free compactified boson:
there is a single ground state (not shown in the figure since it has been subtracted), while the low lying states are coupled in doublets each corresponding to the two primaries with $\pm n$ for increasing $n$.
In plotting more states, one would see the corresponding descendants with the correct degeneracies.

\item For larger values of $m_0 R$, we expect the perturbation to become more and more important until eventually 
truncation effects begin to dominate and the data are no longer reliable.  We see, for example, at $m_0R\sim 11$, the two-particle state in
red in Fig.~\ref{subfig:phi2_sp} begin to deviate from its expected value of $2m_0$.
Therefore to obtain physically sensible results from the numerics, 
one must look to an intermediate region in $R$, the so called {\em scaling region}.
\end{itemize}

\begin{figure}
\centering
\subfigure[]{\includegraphics[width=0.48\textwidth]{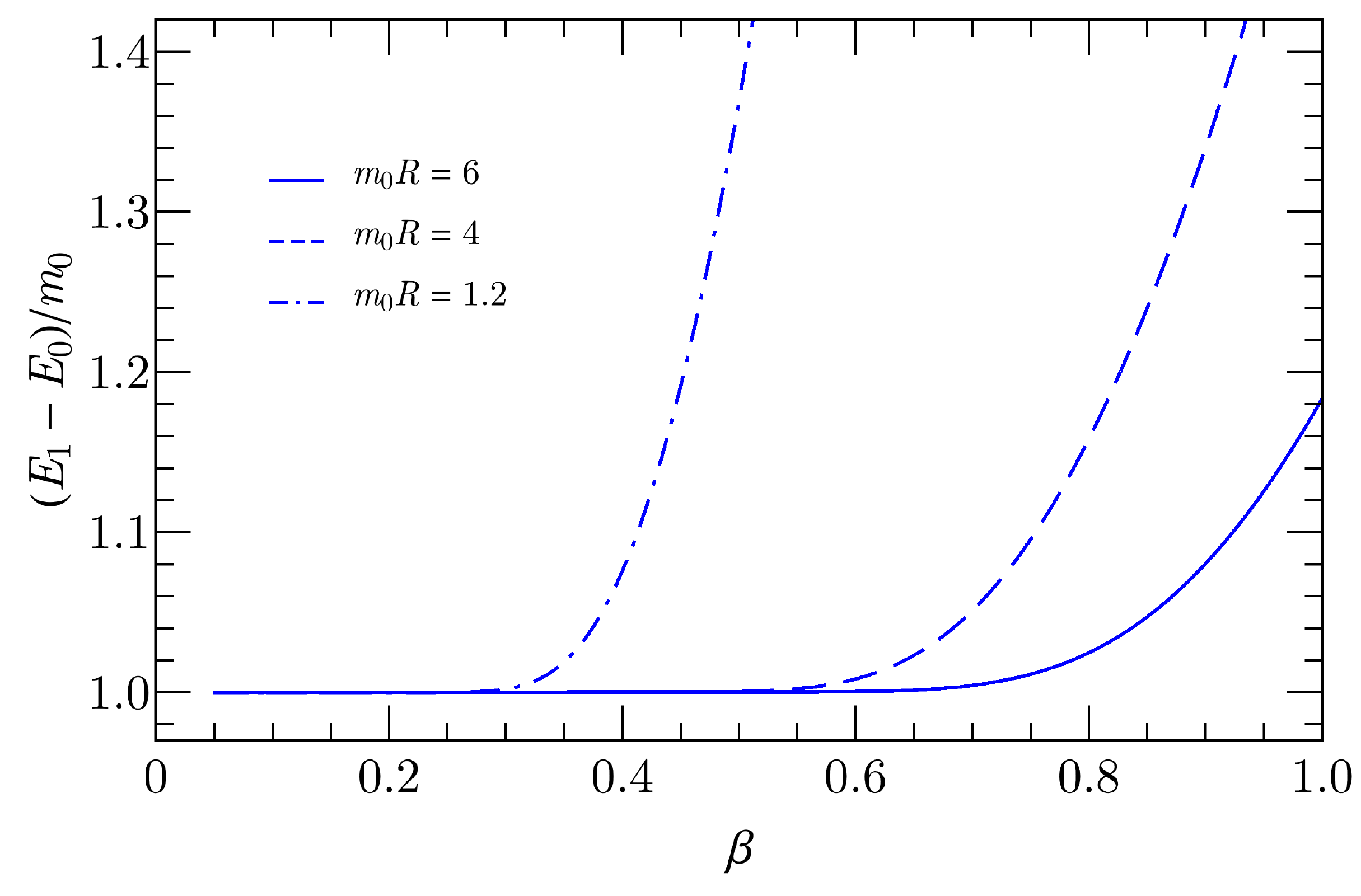}\label{subfig:phi2_beta_1}}\hspace{1mm}
\subfigure[]{\includegraphics[width=0.48\textwidth]{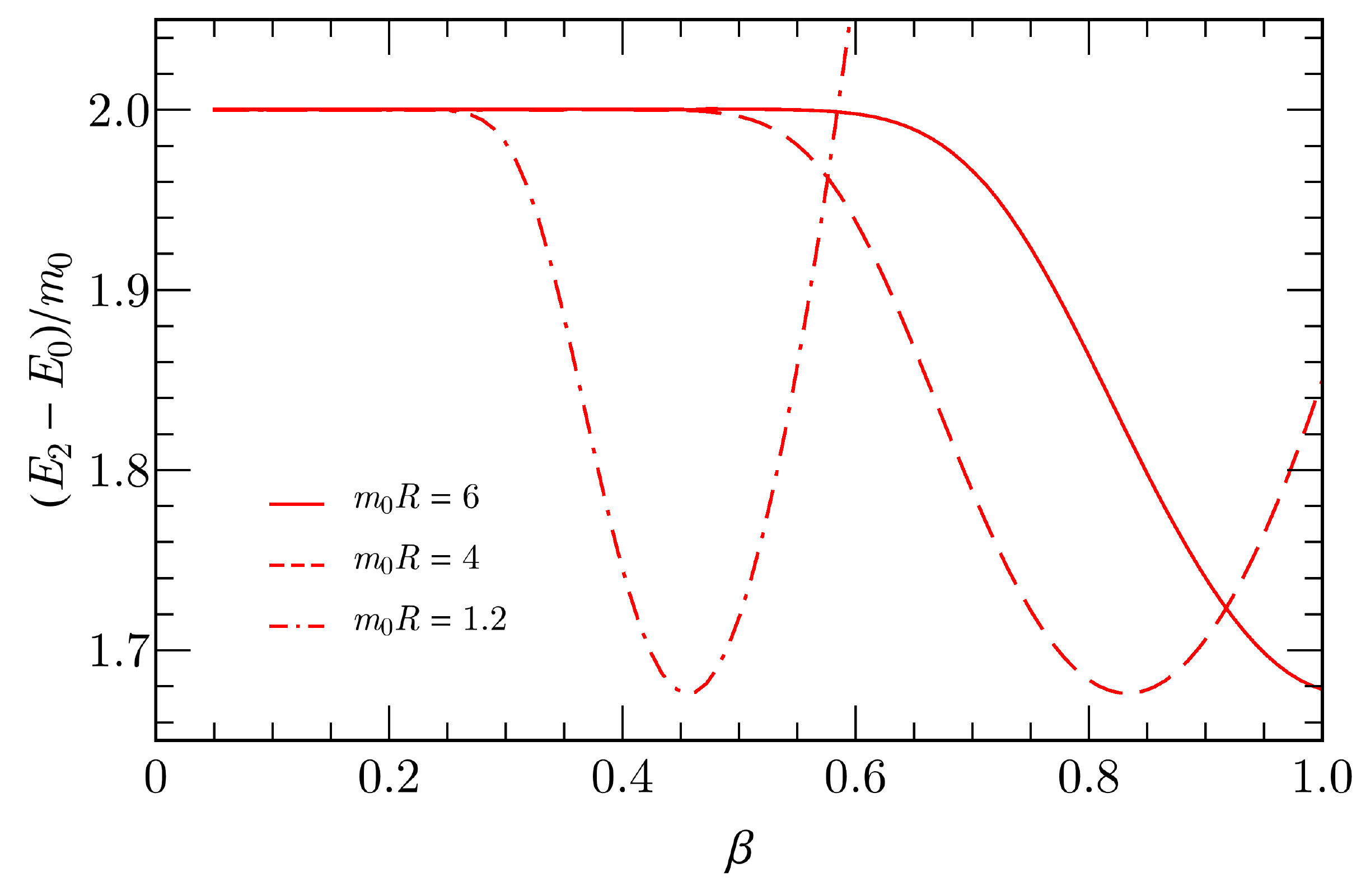}\label{subfig:phi2_beta_2}}\\
\caption{We show here for a $\Phi^2$ theory with $g_2=0.0016$, and
$N_{\text{tr}}=6$ how the first two excited states (
\subref{subfig:phi2_beta_1}: the single particle state of nominal mass $m_0$ and 
\subref{subfig:phi2_beta_2}: the two-particle state with nominal energy $2m_0$) in the zero momentum sector depend on the compactification
radius, $2\pi/\beta$.  We plot these energies as a function of $\beta$ at 
different fixed values of $m_0R$.  We see that the spectrum displays the
greatest sensitivity to non-zero values of $\beta$ when $m_0R$ is small.}
\label{fig:E_vs_beta}
\end{figure}

We conclude this section with a discussion of the effects of the boson's compactification radius upon the spectrum.
Clearly here we have chosen a value of the radius $1/\beta$ large enough that the low lying excitation spectrum is unaffected.
But we can ask the question, how large is large enough?  To answer this 
question, we plot in Fig.~\ref{fig:E_vs_beta} the first two excitations in the momentum
zero sector of a $\Phi^2$ theory as a function of $\beta$.  
The first feature we see in these plots is that the spectrum is most sensitive to $\beta$ being finite
for small values of $m_0R$.  For the smallest value of $m_0R$ plotted, $m_0R=1.2$, 
we see the energy levels do not obtain their infinite $\beta^{-1}$
values until $\beta \sim 0.3$, whereas for the largest value of $m_0R$ examined, $m_0R=6$, the $\beta^{-1}=\infty$ limits
are seen for $\beta \sim 0.6$. 

This behavior is relatively easy to understand, at least in the rough.  In order for a finite
compactification radius to not affect the low lying spectrum, we require the energy of field configurations that are sensitive
to compactification to be much greater than the energy of these lowest excitations.  One such field configuration would
be 
\begin{equation}
\phi(x) = \frac{\pi}{\beta},\quad 0\leq x < R.
\end{equation}
The (classical) energy of this field configuration is $m^2R\pi^2/(2\beta^2)$.  We thus want this energy to be much greater than $m$,
leading us to the condition, $mR \gg 2\beta^2/\pi^2$.  Thus for smaller values of $mR$ we see we need correspondingly smaller
values of $\beta$ for this condition to be met.

The second feature we observe is that once the energy of a low lying eigenstate begins to approach its $\beta^{-1}=\infty$ value,
it does so exponentially quickly.  Again this accords with the notion that only high energy field configurations ``know'' that
the boson is compactified.  From the viewpoint of any Euclidean path integral, such configurations should be suppressed
exponentially.

\section{The unbroken phase: comparison with massive perturbation theory}
\label{sec:posg2g4}

\tikzset{
particle/.style={thick,draw=black, postaction={decorate},
    decoration={markings,mark=at position #1 with {\arrow[black]{triangle 45}}}},
particle/.default=0.65,
particle_bend/.style={thick,draw=black, bend left=90,looseness=1.7, postaction={decorate},
    decoration={markings,mark=at position .58 with {\arrow[black,rotate=12,yshift=0.012cm]{triangle 45}}}},
cross/.style={path picture={ 
  \draw[black]
(path picture bounding box.south east) -- (path picture bounding box.north west) (path picture bounding box.south west) -- (path picture bounding box.north east);}},
label_ext/.style = {label=below:$ $},
}

\begin{figure}
\centering
\subfigure[]{\includegraphics[width=0.48\textwidth]{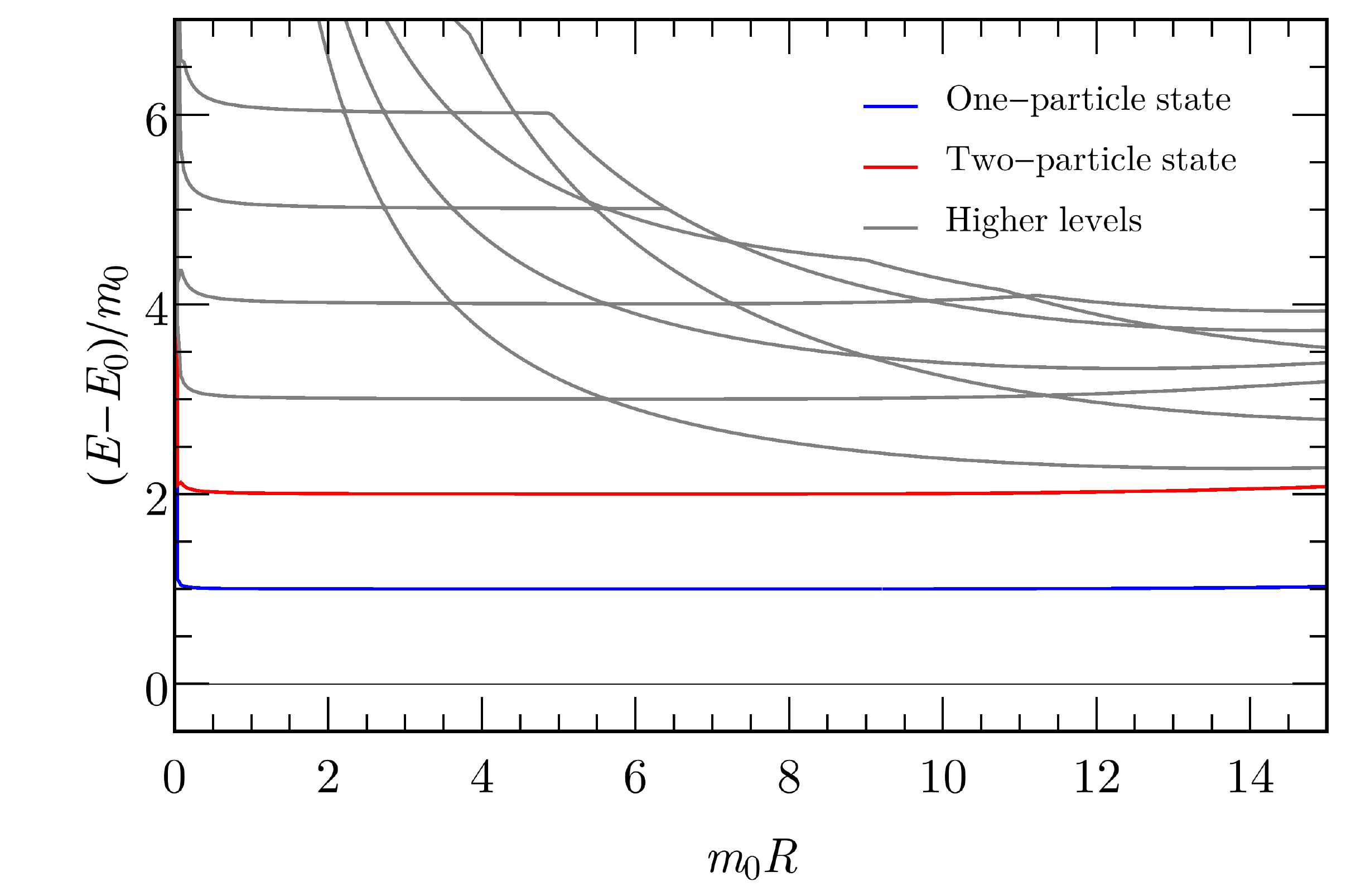}\label{subfig:phi4pos_sp_set1}}\hspace{1mm}
\subfigure[]{\includegraphics[width=0.48\textwidth]{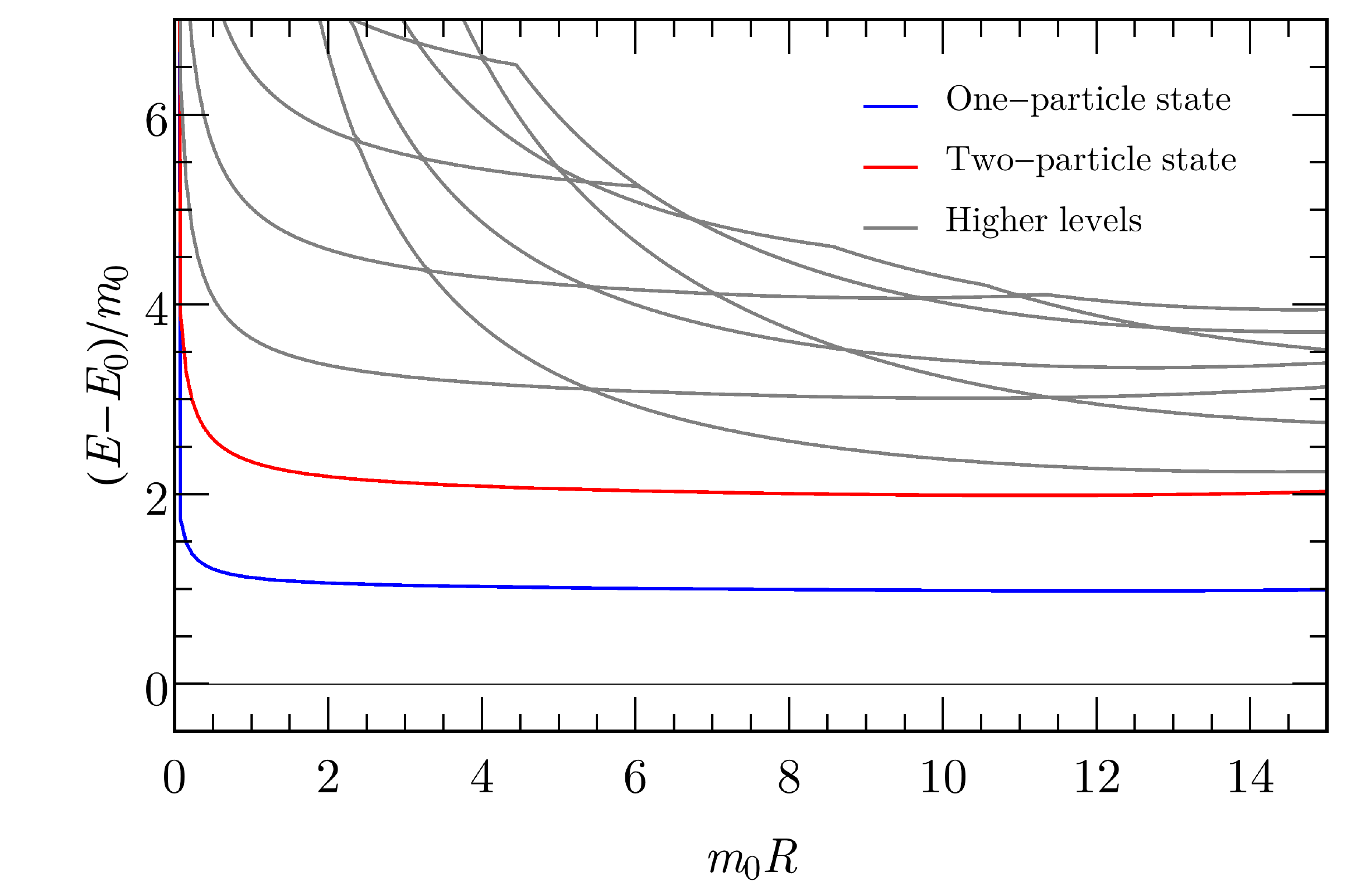}\label{subfig:phi4pos_sp_set9}}
\caption{Two spectra for $\Phi^4$ LG theories with \subref{subfig:phi4pos_sp_set1} $g_2=0.01$, $g_4=2\times 10^{-6}$ and \subref{subfig:phi4pos_sp_set9} $g_2=0.01$, $g_4=8\times 10^{-5}$.
In both cases $\beta=0.07$, $N_{\text{tr}}=6$ and the ground state energy has been subtracted.
All energies are normalized with respect to $m_0$ and are plotted as function of $m_0R$.
The first level (blue) is a one-particle state of a zero momentum particle, while the second level (red) is a state with two such particles.}
\label{fig:phi4pos_spectra}
\end{figure}

In this section we compare the TCSA numerics with the results of massive perturbation theory.
In particular, we consider the the unbroken phase of $\Phi^4$ LG theory.
This phase is described by the following Lagrangian
\begin{eqnarray}
{\cal L} = \frac{1}{8\pi}\Big(:(\partial\phi)^2: +\, g_2 :\phi^2: +\, g_4:\phi^4: \Big) \,,
\label{eq:phi4}
\end{eqnarray}
with $g_2,\,g_4>0$.
In this case, the spectrum has a single ground state and a single massive excitation.
At zeroth order in perturbation theory, this excitation has mass $m_0=\sqrt{g_2}$.
Multi-particle states are also present that are built out of this excitation.
The spectra for two different sets of coupling constants are shown in Fig.~\ref{subfig:phi4pos_sp_set1} and \ref{subfig:phi4pos_sp_set9}.

We can compare our result for the particle mass with the prediction given by standard perturbation theory around the free massive point.
As discussed in section~\ref{subsec:normal order}, this needs to be done with a little care.
First of all, we should keep in mind that we are working in a cylindrical geometry with periodic boundary conditions, hence the 
momenta need to be quantized.
Second, the correct cutoff must be imposed on the momentum sum.
This is summarized in Eq.~\eqref{eq:momentum integral}.

\begin{figure}
\centering
\subfigure[]{\includegraphics[width=0.48\textwidth]{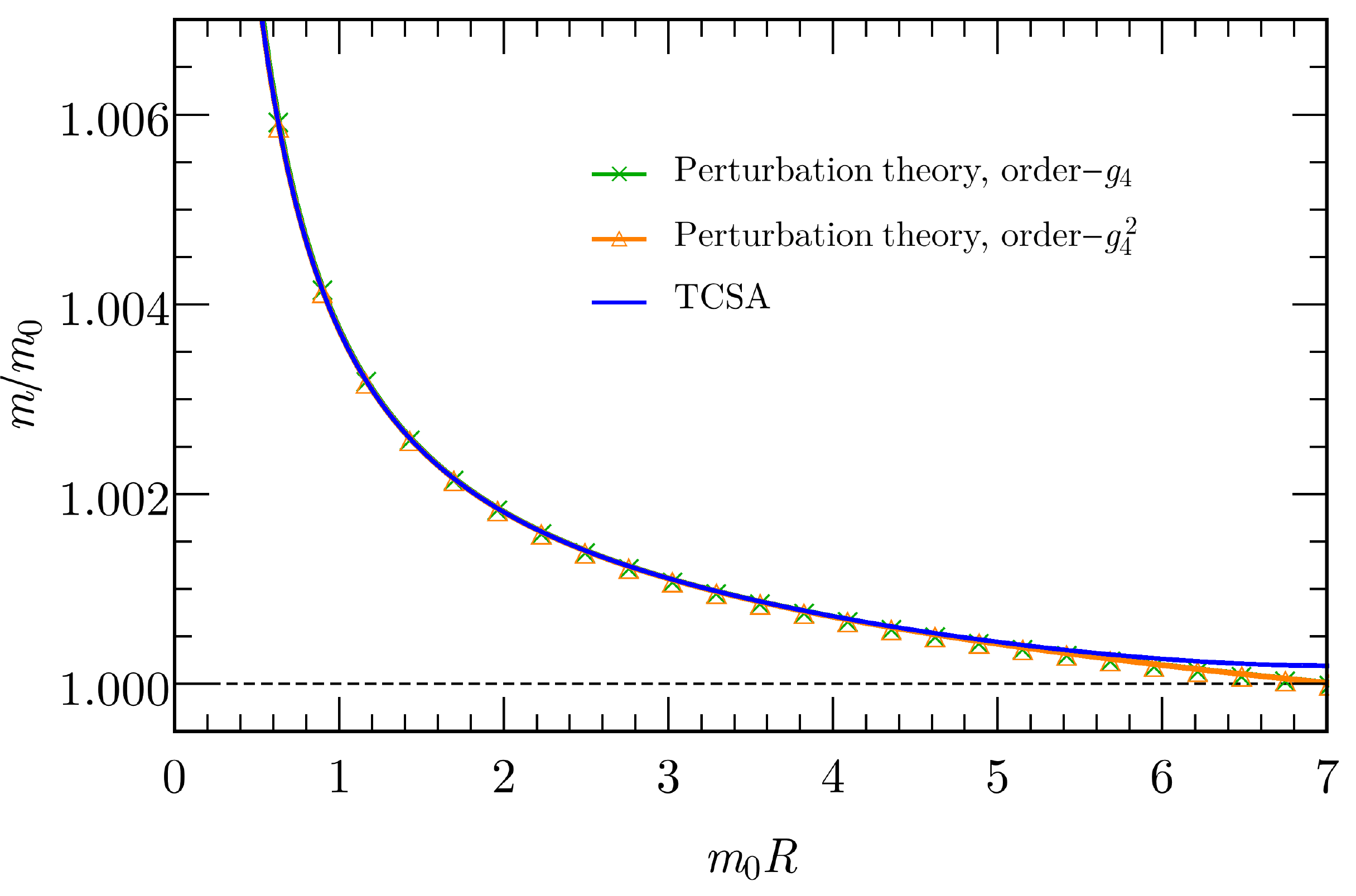}\label{subfig:phi4pos_IIo_set1}}\hspace{1mm}
\subfigure[]{\includegraphics[width=0.48\textwidth]{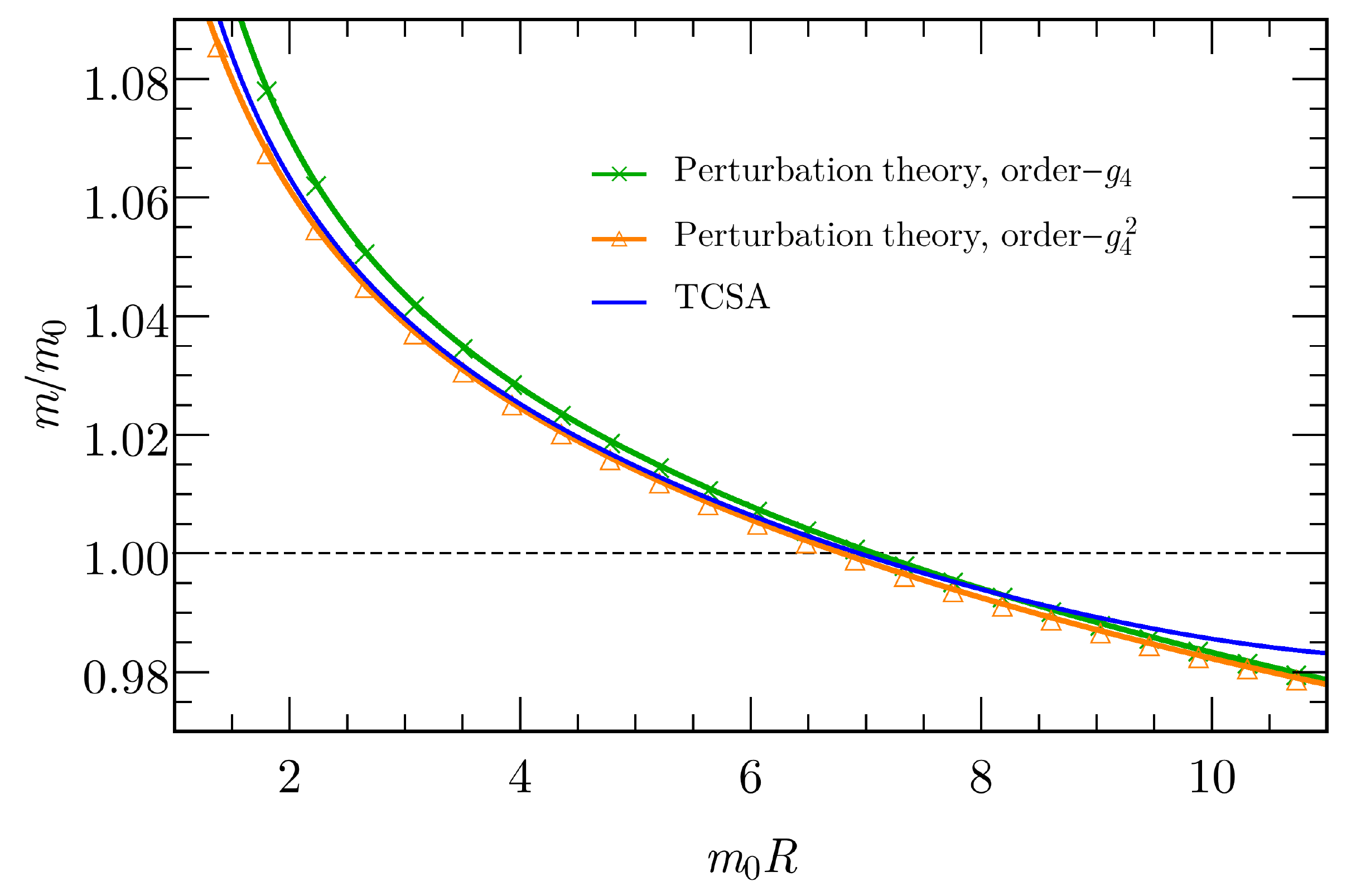}\label{subfig:phi4pos_IIo_set9}}\\[-2mm]
\subfigure[]{\includegraphics[width=0.48\textwidth]{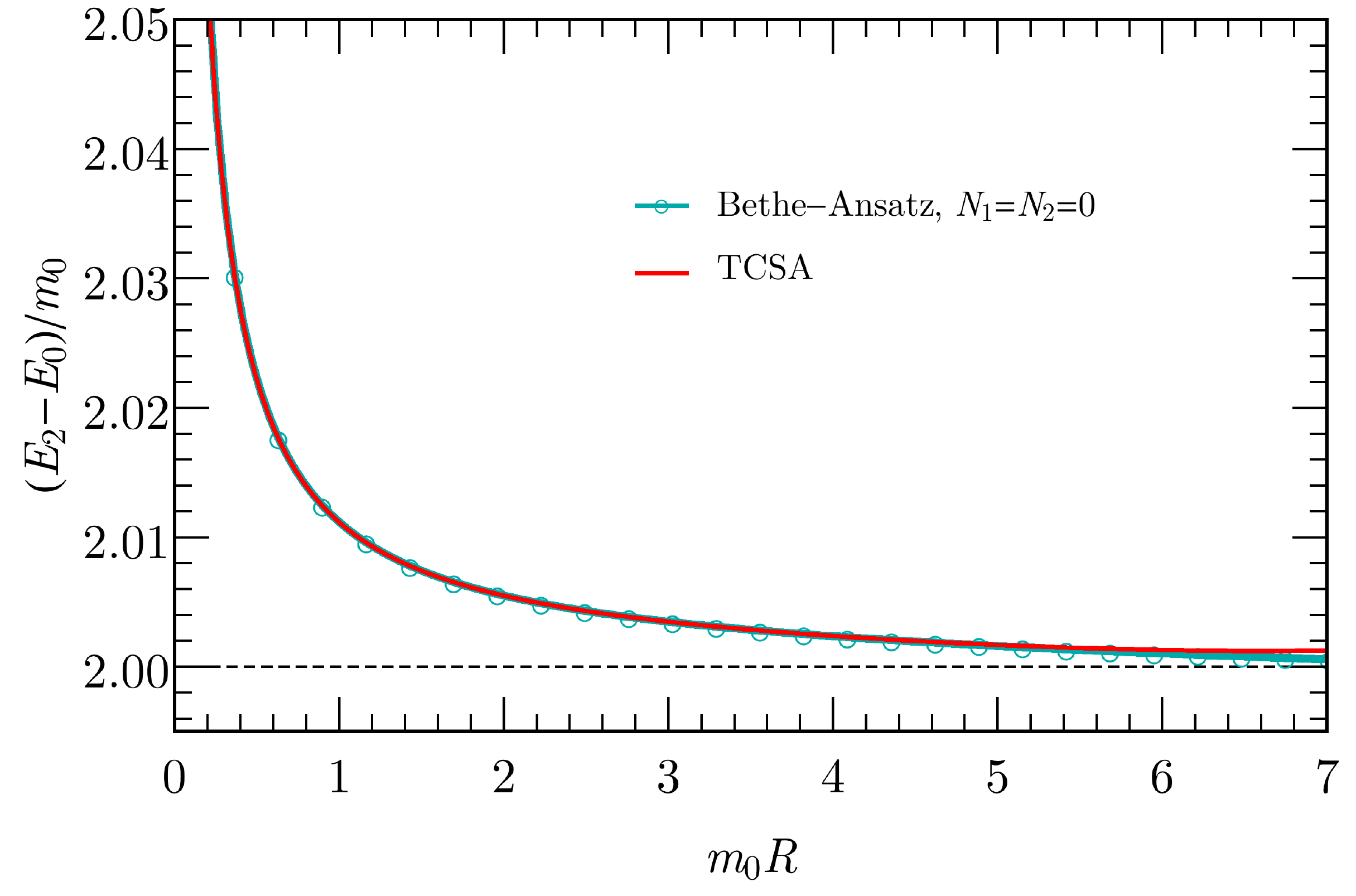}\label{subfig:phi4pos_2p_set1}}\hspace{1mm}
\subfigure[]{\includegraphics[width=0.48\textwidth]{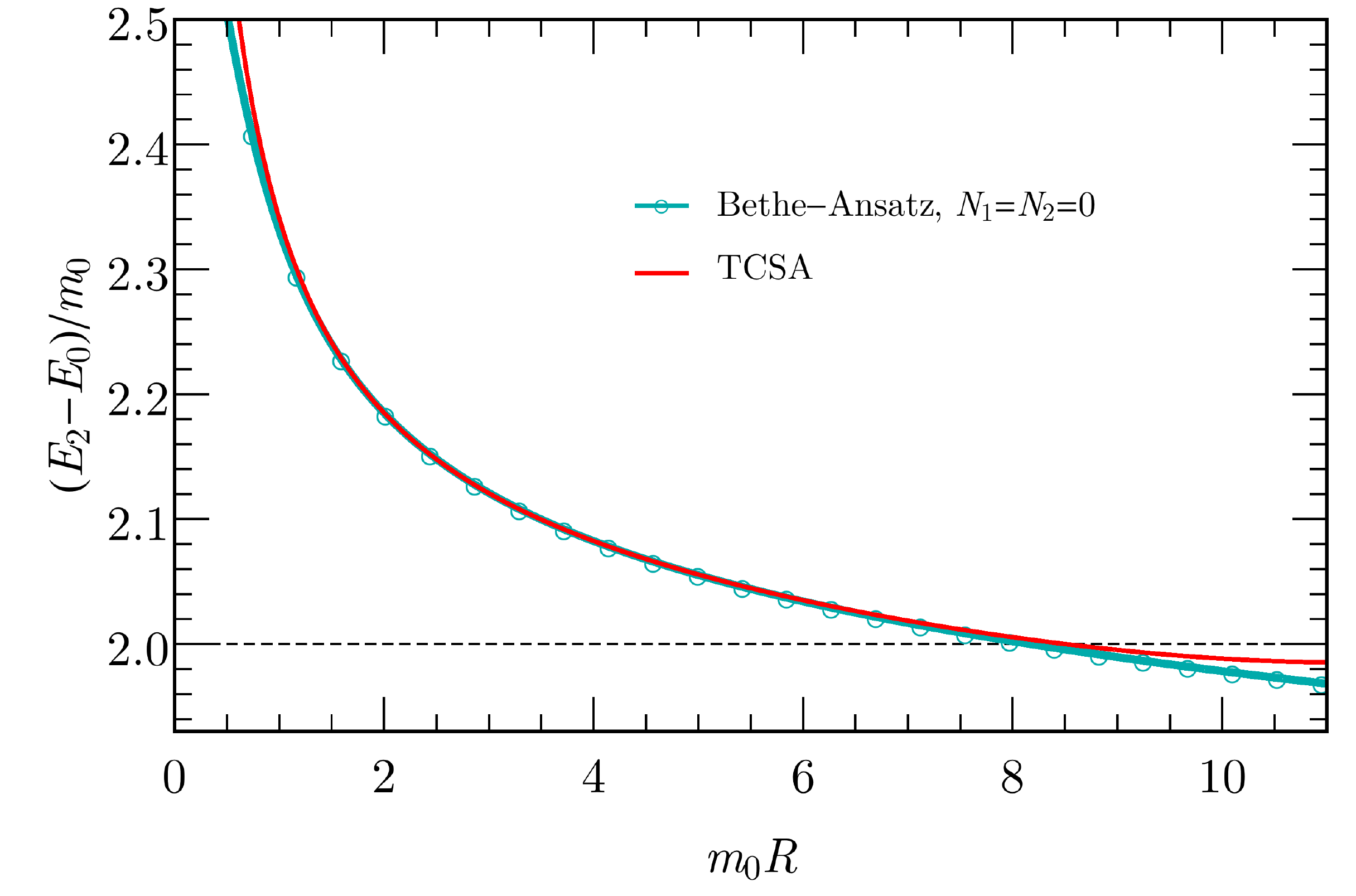}\label{subfig:phi4pos_2p_set9}}
\caption{The first two energy levels of $\Phi^4$ LG theories with the same sets of parameters as in Fig.~\ref{fig:phi4pos_spectra}, 
namely for \subref{subfig:phi4pos_IIo_set1} and \subref{subfig:phi4pos_2p_set1}: $g_2=0.01$, $g_4=2\times 10^{-6}$ while for
\subref{subfig:phi4pos_IIo_set9} and \subref{subfig:phi4pos_2p_set9}: $g_2=0.01$, $g_4=8\times 10^{-5}$.
In both cases $\beta=0.07$, $N_{\text{tr}}=6$ and the ground state energy has been subtracted.
All energies are normalized with respect to $m_0$ and are plotted as function of $m_0R$.
In \subref{subfig:phi4pos_IIo_set1} and \subref{subfig:phi4pos_IIo_set9} we show the comparison between the 
perturbative prediction for the mass (at first and second order in $g_4$) and the TCSA data.
In the regime of small $m_0R$, the agreement is found to be excellent.
For larger values of $m_0R$, the TCSA data begins to suffer from truncation effects and begins to deviate from the analytic predictions.
In \subref{subfig:phi4pos_2p_set1} and \subref{subfig:phi4pos_2p_set9} we show the comparison between the TCSA data for the second
excited state (consisting of two particles, each with momentum zero) and the analytic prediction obtained using the Bethe ansatz 
quantization of a two-particle state with quantum numbers, $N_1=N_2=0$ together with a purely elastic $S$-matrix (this computation 
is outlined in Appendix~\ref{app:2particle}).}
\label{fig:phi4pos_1p_2p}
\end{figure}

To obtain the renormalized mass, $m_{\text{pt},1}$, we compute the second derivative of the effective action $\Gamma^{(2)}(p^2)$, and solve the equation $\Gamma^{(2)}(p^2=-m_{\text{pt},1}^2)=0$.
In perturbation theory $\Gamma^{(2)}$ is obtained from the sum of all one-particle irreducible diagrams with two external legs to the desired order in the coupling constant $g_4$.
The first order contribution is independent of the external momentum $p$ and it is given by the tadpole diagram \cite{binney}:
\begin{equation} \label{eq:tadpole}
\begin{split}
\Gamma^{(2)}(p^2) &= 
p^2 + m_0^2
-
\begin{tikzpicture}[baseline=(i.base),outer sep=5pt]
\coordinate[] (i) at (0,0);
\coordinate[] (v) at (1.25,0);
\coordinate[] (o) at (2.5,0);
\coordinate[] (vc) at ($(v)+(0,0.625)$);
\coordinate[] (vup) at ($(v)+(0,1.25)$);
\draw[particle] (i) -- node[label_ext] {} (v);
\draw[particle] (v) -- node[label_ext] {} (o);
\draw[thick, postaction={decorate},decoration={markings,mark=at position 0.23 with {\arrowreversed[black,rotate=12,yshift=0.01cm]{triangle 45}}}] (vc) circle (0.625cm);
\node[] at ($(vup)+(0,0.3)$) {$q$};
\end{tikzpicture}
+\dots \\
  &= p^2 + m_0^2+ 4!\pi g_4\frac{1}{R}\sum_{n=-2N_{\text{tr}}}^{2N_{\text{tr}}}
 \int_{-\infty}^\infty\!\!\frac{\mathrm{d}q_0}{2\pi}\,\frac{1}{q^2_0+\left(\frac{2\pi}{R}n\right)^2+m_0^2}+\dots \,.
\end{split}
\end{equation}
The integral can be easily computed giving
\begin{equation}
 m_{\text{pt},1}^2 = m_0^2 + 6g_4\sum_{|n|<2N_{\text{tr}}} \frac{1}{\sqrt{n^2 + \left(\frac{m_0 R}{2\pi}\right)^2}}\,.
\end{equation}
In section~\ref{subsec:normal order}, we also argued that the tadpole is divergent when the cutoff is sent to infinity and that the TCSA method cures this divergence by normal ordering the interaction Hamiltonian with respect to the massless theory.
For this reason we have an extra Feynman rule which is a mass insertion proportional to $g_4$ (see Eq.~\eqref{eq:counter term}) and we represent this term with a crossed circle.
Such a counterterm is given by minus the massless tadpole, computed with an IR cutoff:
\begin{equation} \label{eq:tadpole counterterm}
\begin{tikzpicture}[baseline=-1mm,outer sep=5pt]
\coordinate[] (i) at (0,0);
\coordinate[] (v) at (1.25,0);
\coordinate[] (o) at (2.5,0);
\draw[particle] (i) -- node[label_ext] {} (v);
\draw[particle] (v) -- node[label_ext] {} (o);
\draw[cross,fill=white,cross] (v) circle (5pt);
\end{tikzpicture}
= 4!\pi g_4\frac{2}{R}\sum_{n=1}^{2N_{\text{tr}}}
 \int\frac{\mathrm{d}q_0}{2\pi}\,\frac{1}{q^2_0+\left(\frac{2\pi}{R}n\right)^2} =
 12g_4\sum_{n=1}^{2N_{\text{tr}}} \frac{1}{n} \,.
\end{equation}
Combining everything together, we get for the first order mass correction:
\begin{equation}  \label{eq:Iorder}
 \frac{m_{\text{pt},1}^2}{m_0^2} = 1+ 12\frac{g_4}{g_2}\left\{
 \sum_{n=1}^{2N_{\text{tr}}} 
 \left( \frac{1}{\sqrt{n^2+\left(\frac{m_0 R}{2\pi}\right)^2}} - \frac{1}{n} \right)
 + \frac{\pi}{m_0 R} \right\}\,.
\end{equation}
As discussed, the formula is finite when the cutoff $N_{\text{tr}}$ goes to infinity.
Moreover, it is only weakly dependent on the truncation level $N_{\text{tr}}$ and saturates very fast with increasing $N_{\text{tr}}$.
This is compatible with the TCSA data in the scaling region where we have found that the data has only a 
very weak dependence on the truncation (for sufficiently high $N_{\text{tr}}$).  We have also extended this computation
to second order.  The details may be found in Appendix \ref{app:2loop}.

In Fig.~\ref{fig:phi4pos_1p_2p} we compare the prediction of 
perturbation theory with the numerical data for the mass gap.
We can see that the comparison between the TCSA numerics and the one-loop corrections is remarkably good, 
at least in the scaling region and for the smaller of the two values of the coupling $g_4$ that we
study (i.e.\ Fig.~\ref{subfig:phi4pos_sp_set1}).
For the larger value of $g_4$ considered in Fig.~\ref{subfig:phi4pos_sp_set9}, 
the two-loop correction must be added in order to reproduce the numerical data.

It is also possible to estimate the first two-particle state in the theory 
(i.e.\ the second excited state in the theory).  
To estimate the effects of the interaction on this state, we quantize the particles' momenta using a purely elastic $S$-matrix.
In ignoring inelastic processes, we expect to reproduce the TCSA data below the particle production threshold, $E<E_{\text{th}}$,
or equivalently, $R > R_{\text{cr}}$, where $R_{\text{cr}}$ is the largest radius at which an avoided crossing takes place.
The details of this computation are given in Appendix \ref{app:2particle}.   In Figs.~\ref{subfig:phi4pos_2p_set1} and 
\ref{subfig:phi4pos_2p_set9} we show the results of this analysis.
We see that the agreement between the numerics and the analytics is remarkably good.
We have performed a similar analysis on higher excited states 
and again have found good agreement between the numerics and the analytics but for regions in the vicinity of an avoided crossing.

\subsection{Analysis of the finite momentum sector of the unbroken $\Phi^4$ LG theory}
\label{subsec:posg2g4_mom1}

The perturbative analysis tells us that the first levels (blue) of Figs.~\ref{subfig:phi4pos_sp_set1} and \ref{subfig:phi4pos_sp_set9} 
are single particle states with renormalized mass $m_{\text{pt},1}$.
We have also claimed that the other levels are multi-particle states formed from this single particle, where, for example,
the second levels (red) are two-particle states.
We can verify this by studying the spectra in the momentum-1 sector (i.e.\ states in the momentum-$n$ sector have total momentum $P=2\pi n/R$).

\begin{figure}
\centering
\subfigure[]{\includegraphics[width=0.48\textwidth]{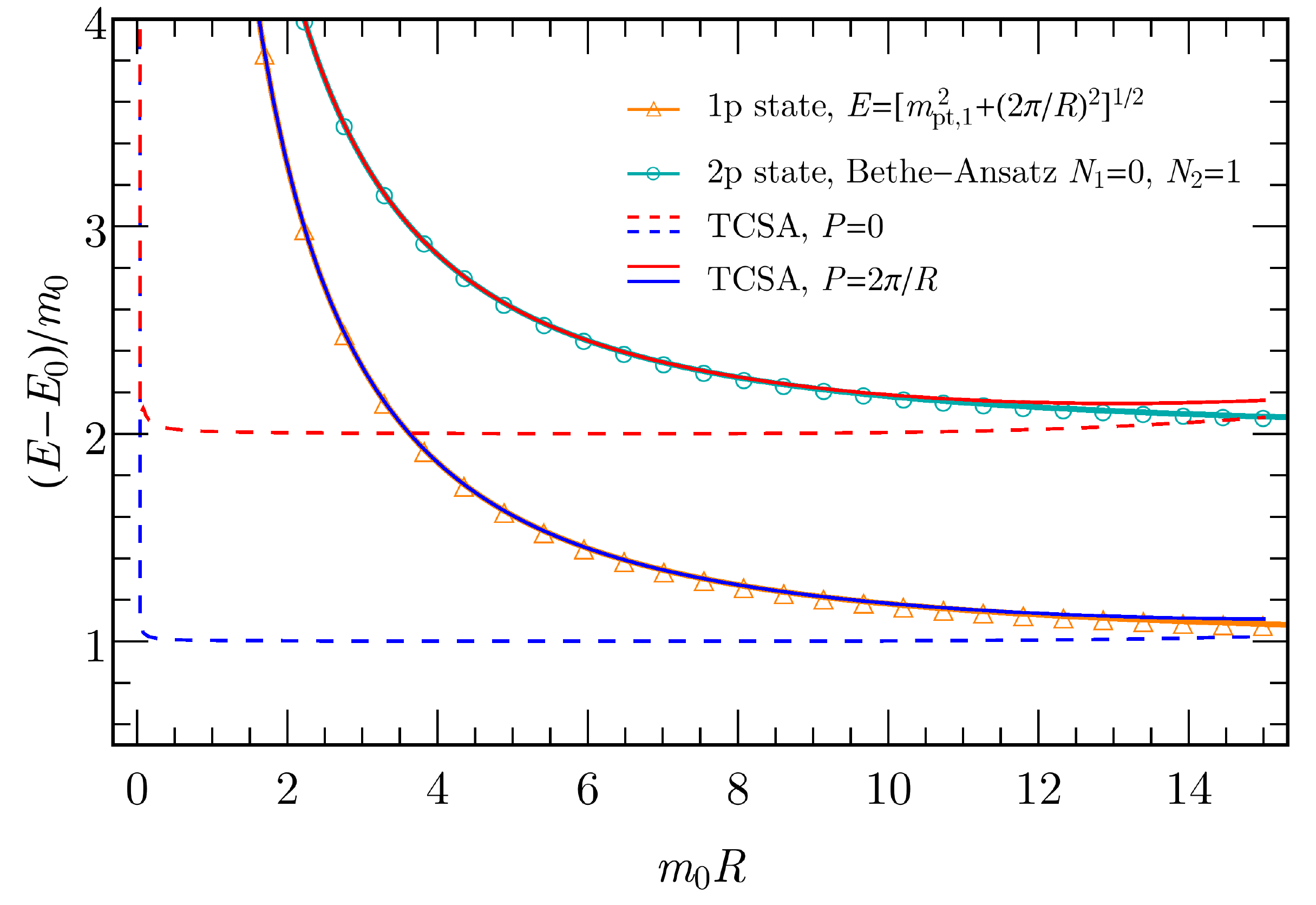}\label{subfig:phi4pos_mom1_set1}}\hspace{1mm}
\subfigure[]{\includegraphics[width=0.48\textwidth]{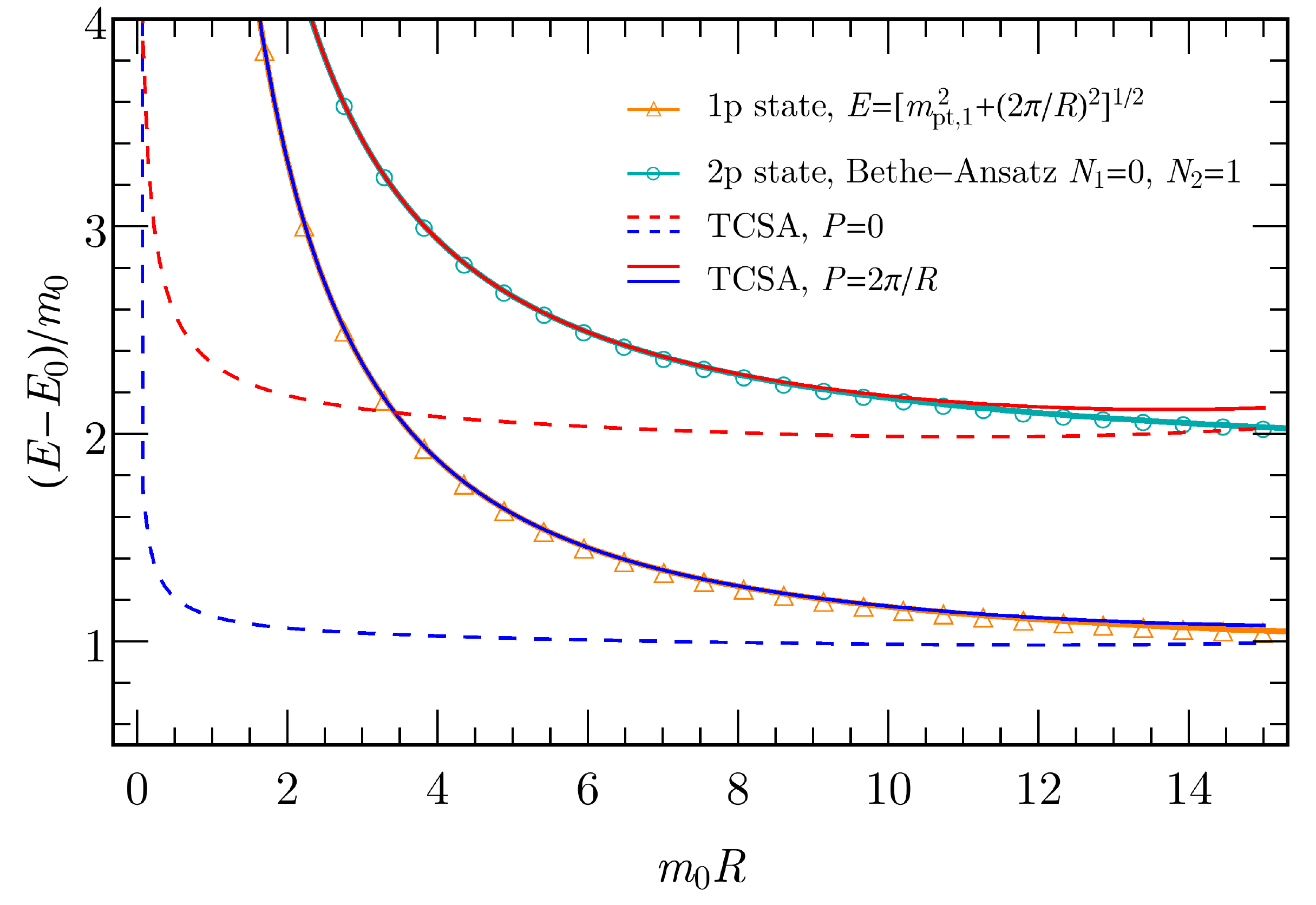}\label{subfig:phi4pos_mom1_set9}}
\caption{The two lowest momentum-1 energy levels (blue and red solid lines) for $\Phi^4$ LG theories with the same parameters 
of Fig.~\ref{fig:phi4pos_spectra} (\subref{subfig:phi4pos_mom1_set1}: $g_2=0.01$, $g_4=2\times 10^{-6}$ and 
\subref{subfig:phi4pos_mom1_set9}: $g_2=0.01$, $g_4=8\times 10^{-5}$).
In both cases $\beta=0.07$, $N_{\text{tr}}=6$ and the ground state energy belonging to the momentum-0 sector has been subtracted.
All energies are normalized with respect to $m_0$ and are plotted as a function of $m_0R$.
The first two excited levels of the momentum-0 sector are shown for reference as blue and red dashed lines.
The blue solid line is well described by the dispersion relation of a single particle of (renormalized) mass $m_{\text{pt},1}$ 
and momentum $2\pi/R$ (orange curve with triangles).
The red solid line is a two particle state, each particle having mass $m_{\text{pt},1}$.
The light blue curve with circles is computed using the quantization condition, Eq.~\eqref{eq:BetheAnsatzQuantization}, 
with $N_1=0$, $N_2=1$ and with the phase shift given by a purely elastic $S$-matrix 
(the details of this computation are given in Appendix~\ref{app:2particle}).
We see that the agreement for these finite momentum states
between the numerics and the analytics is very good.}
\label{fig:phi4pos_mom1}
\end{figure}

In Fig.~\ref{fig:phi4pos_mom1} we show the first two levels (solid blue and red lines) in the momentum-1 sector
of the unbroken phase of a $\Phi^4$ theory with the same choice of parameters as that considered in Fig.~\ref{fig:phi4pos_spectra}.
We also show the first two levels of the momentum-0 sector (blue and red dashed lines) for reference.

In both plots, the first momentum line (blue) is very well fitted by the dispersion relation of a single particle with 
the renormalized mass $m_{\text{pt},1}$ 
(which depends on the cylinder size $R$) and momentum $2\pi/R$, that is $E=\sqrt{m_{\text{pt},1}^2+(2\pi/R)^2}$ (orange curve with triangles).
In the infinite volume limit, it gets close to the momentum-0 line corresponding to a single static particle of mass 
$m_{\text{pt},1}$, their difference going as $1/R$.
The second level (red solid line) is a two-particle state.
To estimate its energy we can once again need to solve the quantization conditions for the momenta of the particles,
i.e.\ Eq.~\eqref{eq:BetheAnsatzQuantization}.
Using $N_1=1$, $N_2=0$ as the quantum numbers in these quantization conditions, we find as
a result the light blue curves with circles (see Fig.~\ref{fig:phi4pos_mom1}).
We see that for larger values of $R$ where our use of an elastic $S$-matrix is expected to be valid, 
the analytics and the numerics are in good agreement.

\section{Double well potential and stable neutral bound states}
\label{sec:double_well} 

In this section we present a series of results which support the conjecture on the number of stable neutral bound states discussed in Section~\ref{sec:semi} for double well potentials.
We will mainly focus on $\Phi^4$ theory while only briefly commenting on the two-well phase of 
$\Phi^6$ theory in Section~\ref{subsec:phi6}.

Let us start our analysis with the Lagrangian of Eq.~\eqref{eq:phi4}, but this time we are interested in the regime where
two wells are present, i.e.\ $g_2<0$, and $g_4>0$.
The perturbative mass is given by the curvature of the potential in the two minima:
\begin{equation}
U(\phi) = \frac{1}{2} \left( g_2 :\phi^2: +\, g_4 :\phi^4: \right) \,,\qquad m_0 = \partial_\phi U(\phi)\big|_{\phi^{(0)}} = \sqrt{-2g_2}\,.
\end{equation}
It would be very interesting to study a large area in the space of parameters $g_2$ and $g_4$ in order to see how the energy spectrum changes in different regimes.
Unfortunately, for large values of $g_4/g_2$, severe truncation effects are present and the scaling physical region rapidly shrinks and finally disappears.
In order to obtain meaningful data, we are forced to keep the values of the coupling constants small (and
so place ourselves in the regime where $\xi \ll 1$ (see Eq.~\eqref{definitiong} for the definition of $\xi$)).
In this regime two stable neutral bound states are predicted to exist for each well.
It turns out to be difficult to study the other regimes where less than two bound states are present.

\begin{figure}
\centering
\subfigure[]{\includegraphics[width=0.48\textwidth]{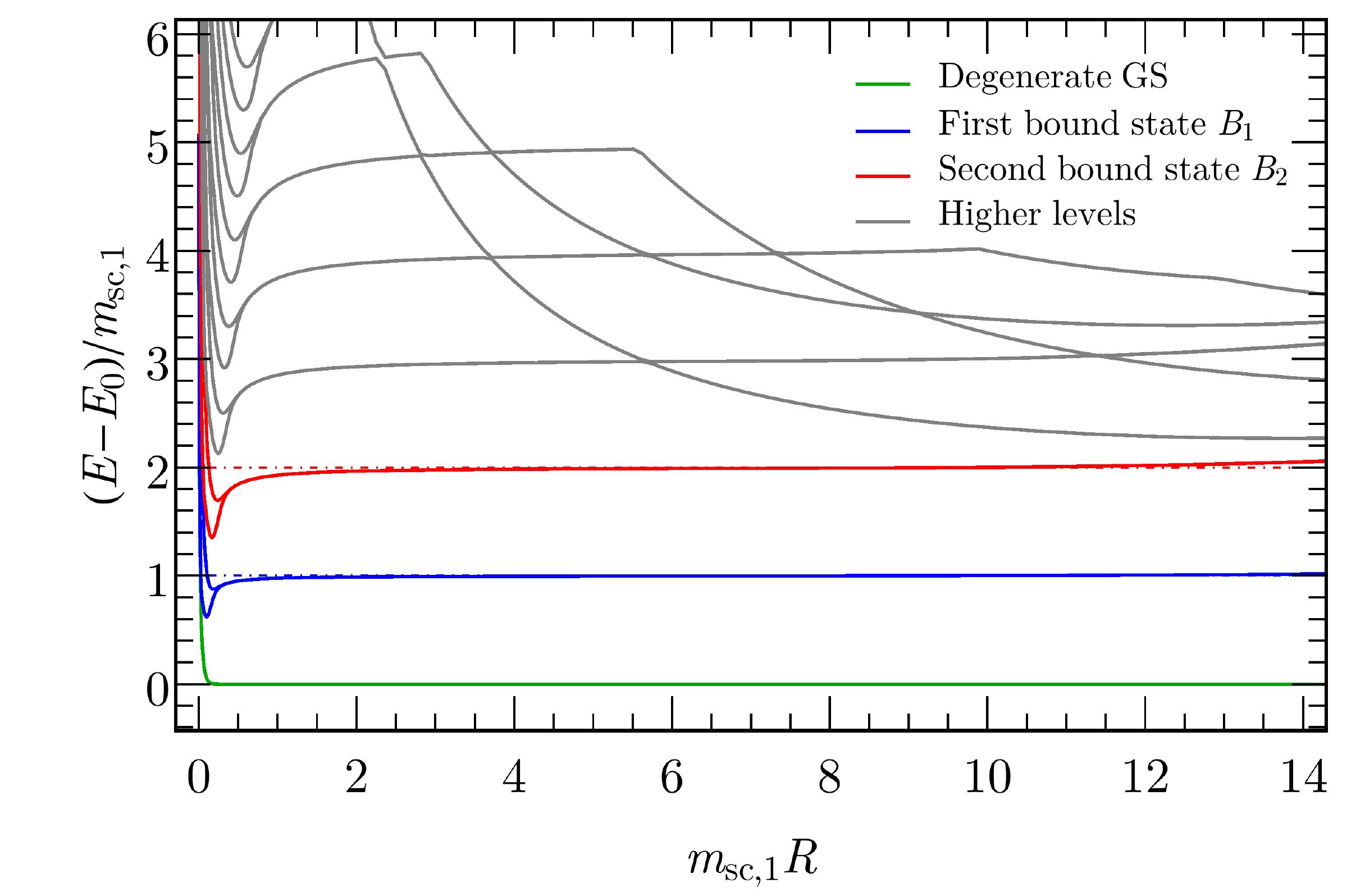}\label{subfig:phi4_sp_set11}}\hspace{1mm}
\subfigure[]{\includegraphics[width=0.48\textwidth]{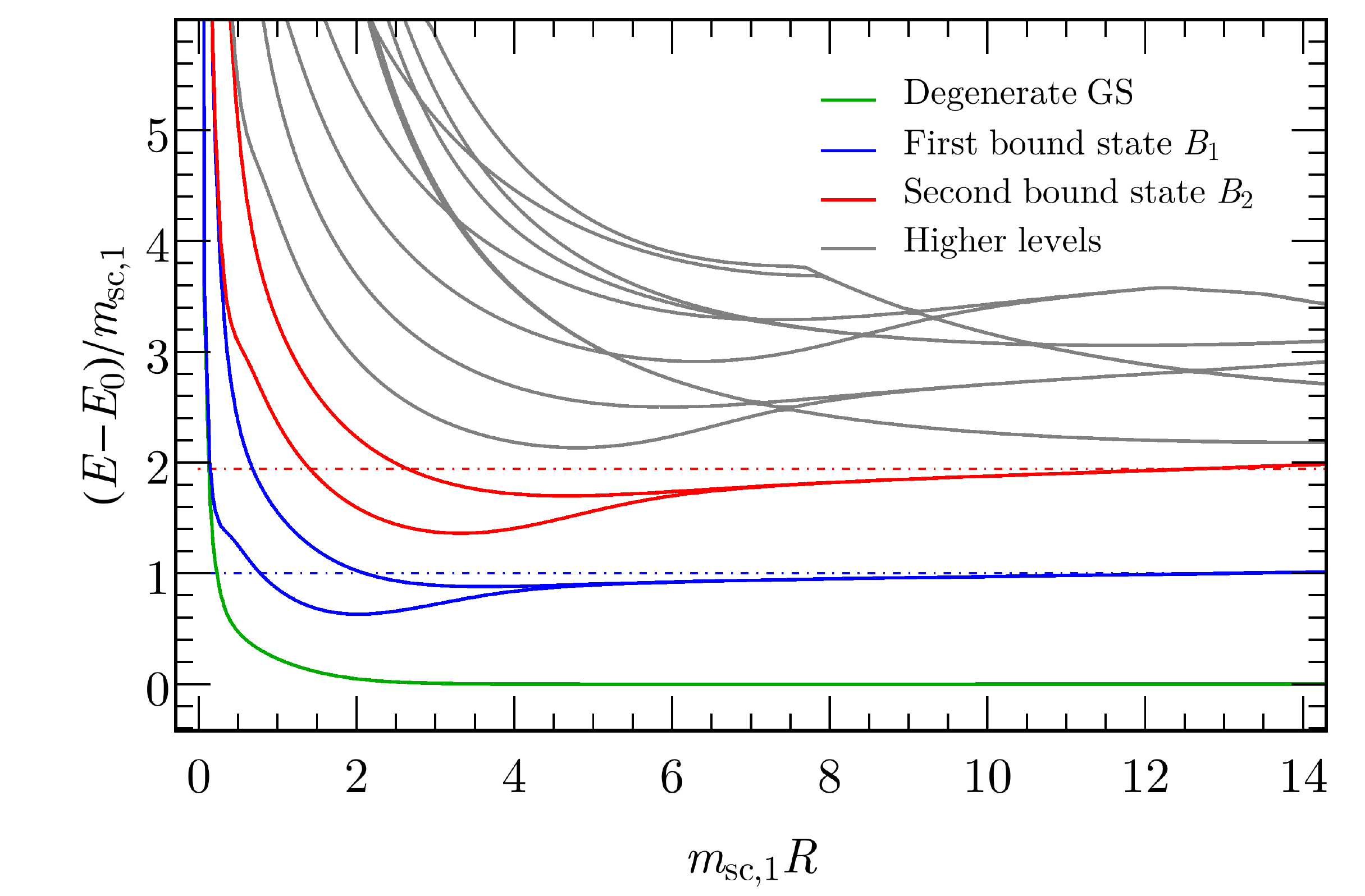} \label{subfig:phi4_sp_set6}}\\[-2mm]
\subfigure[]{\includegraphics[width=0.48\textwidth]{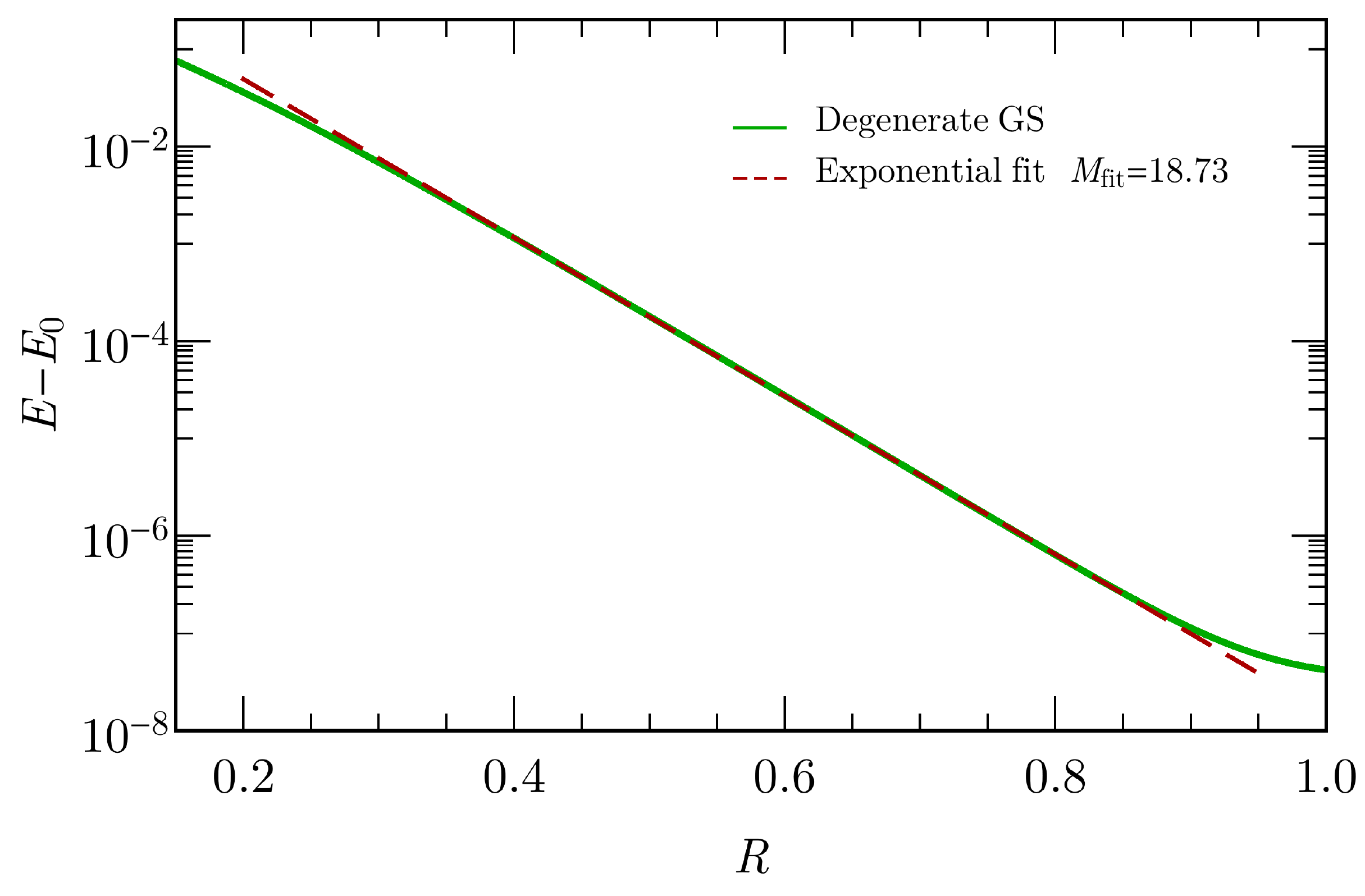}\label{subfig:phi4_GS_set11}}\hspace{1mm}
\subfigure[]{\includegraphics[width=0.48\textwidth]{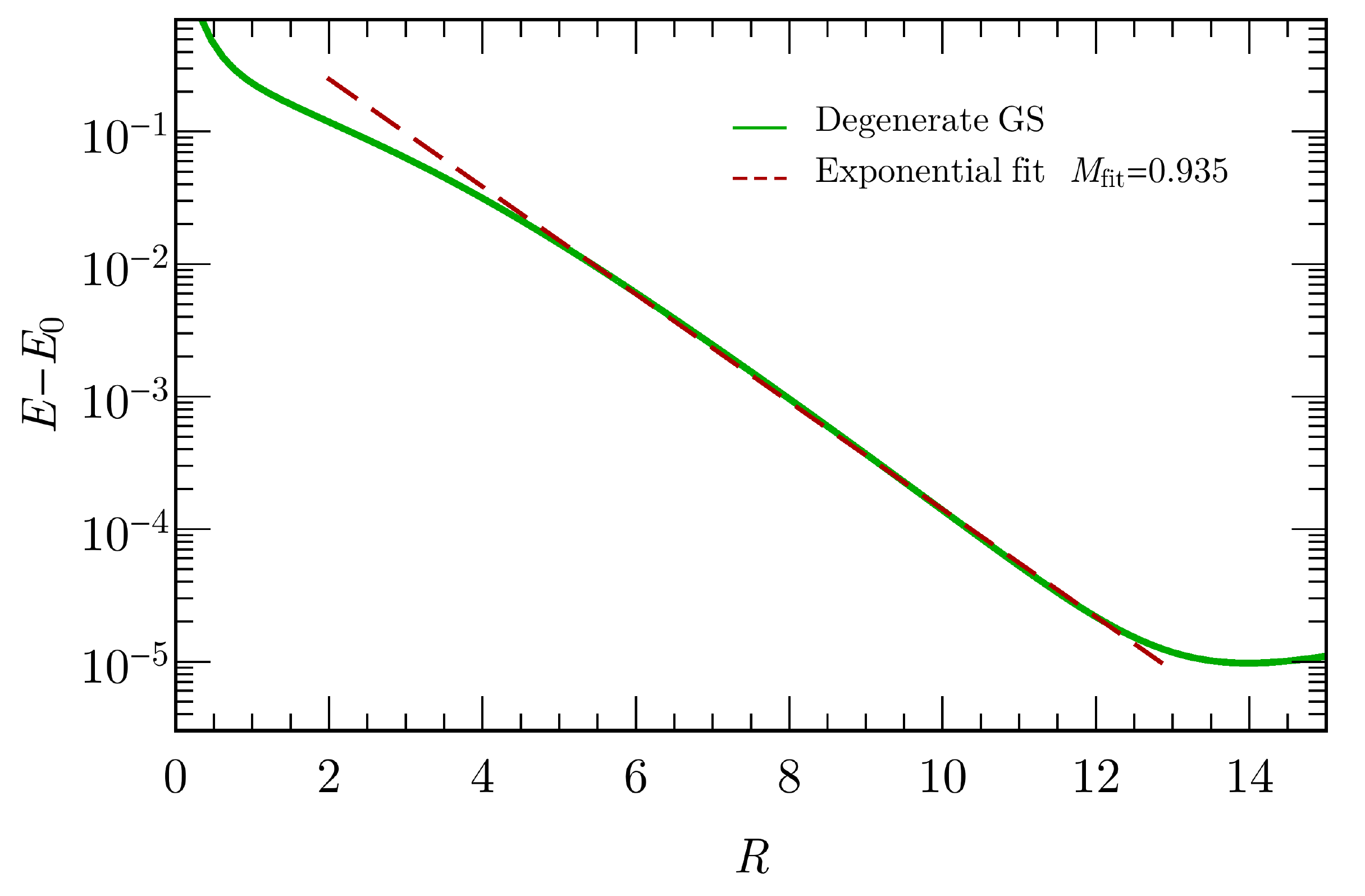} \label{subfig:phi4_GS_set6}}
\caption{The spectra for two $\Phi^4$ LG theories 
with \subref{subfig:phi4_sp_set11} $g_2=-0.1$, $g_4=6\times 10^{-5}$, $\beta=0.06$ 
and $N_{\text{tr}}=6$ and \subref{subfig:phi4_sp_set6} $g_2=-0.1$, $g_4=1.2\times 10^{-3}$, $\beta=0.2$ and $N_{\text{tr}}=7$.
As before, we show the energy levels with the ground state energy subtracted.
In both cases the zeroth order mass is $m_0=0.4472$.
The first energy level above the ground state (green line) becomes exponentially degenerate with the ground state.
Their difference is shown in logarithmic scale in panels \subref{subfig:phi4_GS_set11} and \subref{subfig:phi4_GS_set6} respectively.
By both performing an exponential fit and 
using the semiclassical formula in Eq.~\eqref{mass1phi4}, we can obtain two estimates for the kink mass:
in \subref{subfig:phi4_GS_set11} the semiclassical value for this mass is $M=19.62$, 
while numerically we find $M_{\text{fit}}=18.7 \pm 0.2$.
And in \subref{subfig:phi4_GS_set6}, the semiclassics give $M=0.840$, while numerically we have $M_{\text{fit}}=0.94 \pm 0.09$.
From these values of the kink masses, the bound state masses 
for \subref{subfig:phi4_GS_set11} are given by $m_{\text{sc},1}=0.4472$, $m_{\text{sc},2}=0.8943$,
while for \subref{subfig:phi4_GS_set6}, $m_{\text{sc},1}=0.4430$, $m_{\text{sc},2}=0.8609$.
All energies are normalized by the mass scale $m_{\text{sc},1}$ and are plotted as a function of $m_{\text{sc},1}R$.
The dot-dashed blue and red lines are set at 1 and $m_{\text{sc},2}/m_{\text{sc},1}$ respectively.
The first doubly-degenerate level (solid blue) is believed to be the neutral bound state $B_1$, 
while the second doubly-degenerate level (red) is the second bound state $B_2$.}
\label{fig:phi4_spectra}
\end{figure}

\begin{figure}
\centering
\subfigure[]{\includegraphics[width=0.48\textwidth]{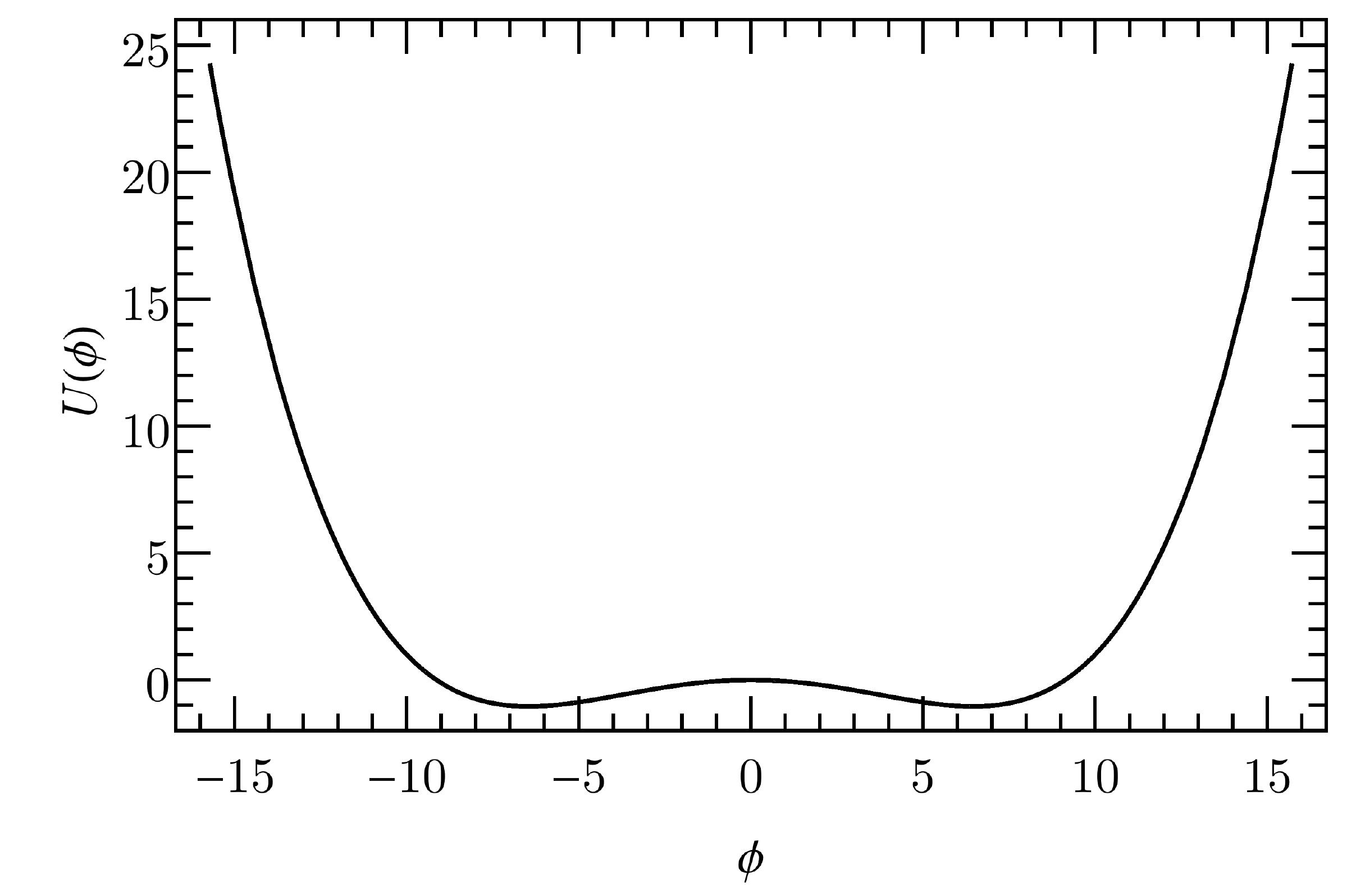}\label{subfig:phi4_pot_set6}}\hspace{1mm}
\subfigure[]{\includegraphics[width=0.48\textwidth]{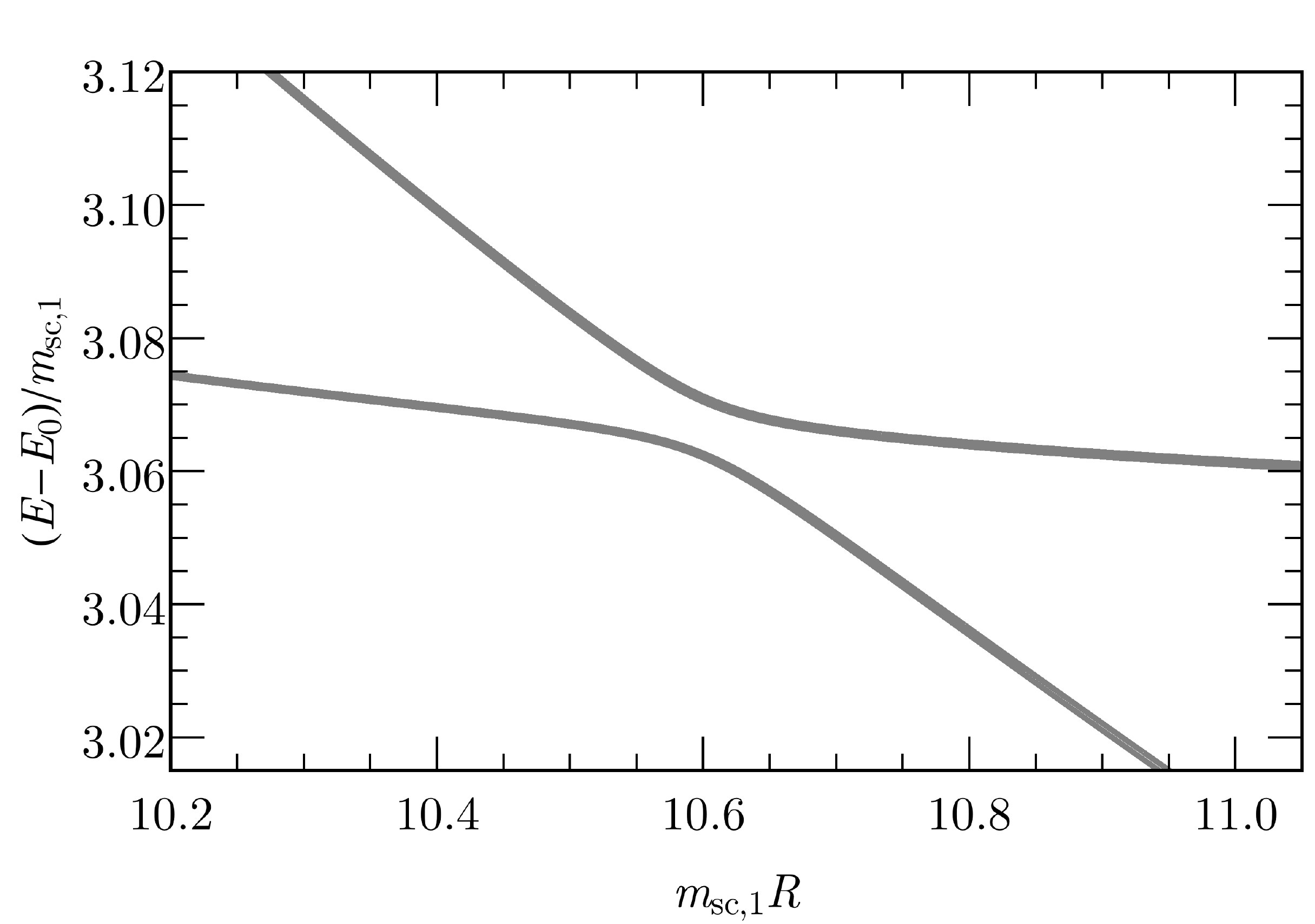} \label{subfig:phi4_cross_set6}}
\caption{\subref{subfig:phi4_pot_set6} The potential for the configuration of Fig.~\ref{subfig:phi4_sp_set6} where  
$g_2=-0.1$, $g_4=1.2\times 10^{-3}$ and $\beta=0.2$.
Since $\phi$ is compactified, the potential is periodic with period $[-\pi/\beta,\pi/\beta]$ (in the plot a single period is shown).
\subref{subfig:phi4_cross_set6} An expanded snapshot of the spectrum in Fig. \ref{subfig:phi4_sp_set6} showing the presence of 
an avoided crossing due to interactions.}
\label{fig:phi4_pot_cross}
\end{figure}

We have studied the energy spectrum for many values of the constants $g_2$ and $g_4$ in this regime and we always find
the same main characteristics.
Two representative examples are shown in Figs.~\ref{subfig:phi4_sp_set11} and \ref{subfig:phi4_sp_set6}.
In Fig.~\ref{subfig:phi4_pot_set6} we also show the typical shape of a studied $\Phi^4$ potential
(this potential gives rise to the spectrum of Fig.~\ref{subfig:phi4_sp_set6}).
As with the previous spectra, we subtract the ground state energy $E_0(m_{\text{sc},1}R)$ from 
all the other energy levels and we rescale these differences with the mass of the first bound state $m_{\text{sc},1}$.
Before discussing in detail how $m_{\text{sc},1}$ was computed, let us analyze the main features of these plots.
As usual, in the small $m_{\text{sc},1}R$ region the conformal part of the Hamiltonian dominates and the low energy levels have
the correct degeneracy of the unperturbed CFT:
we have a single ground state corresponding to the identity operator and double degenerate states corresponding 
to the other primary fields.
On the other hand, for large values of $m_{\text{sc},1}R$ (in this case $m_{\text{sc},1} R \gtrsim 12-14$), truncation effects become important 
and the data can no longer simply be related to that of a $\Phi^4$ LG theory.

The beginning of the scaling region of the theory at intermediate values of $R$ can be identified 
by observing where the first excited state becomes degenerate with exponential accuracy with the ground state (green). 
At finite $m_{\text{sc},1}R$ the exponential difference between the two states is a result of a finite energy barrier between the two vacua
given by $MR$ with $M$ the kink mass. 
In Fig.~\ref{subfig:phi4_GS_set11} and Fig.~\ref{subfig:phi4_GS_set6} we show this energy difference on a logarithmic scale 
between the two nearly degenerate ground states of the two spectra found in
Figs.~\ref{subfig:phi4_sp_set11} and \ref{subfig:phi4_sp_set6} respectively.
From fitting the numerical data, we find values $M_{\text{fit}}$ that are compatible with the ones expected 
from the semiclassical analysis of the $\Phi^4$ LG theory, Eq.~\eqref{mass1phi4}.
This is a strong indication that the TCSA is indeed correctly seeing the kink excitations connecting the two ground states.

Using the values of the kink mass obtained from the fit, we can compute the bound state masses $m_{\text{sc},n}$ from Eq.~\eqref{massphi4}.
In the region of coupling constants that we can simulate, $m_0/M$ is very small 
and therefore the formulae for the bound state masses essentially reduces to $m_{\text{sc},n} \sim n m_0$. 
The results obtained for $m_{\text{sc},1}$ are indeed very close to the perturbative mass $m_0$ and we see deviations on 
the order of at most of $\sim 1\%$.  In this regime $m_{\text{sc},2}$ is also very close to $2 m_{\text{sc},1}$ (with deviations on the order of $\sim 3\%$).

In Figs.~\ref{subfig:phi4_sp_set11} and~\ref{subfig:phi4_sp_set6} we see that directly above the doubly degenerate ground states 
there are two nearly degenerate excited states (blue lines) and above these, two more nearly degenerate levels (red). 
These red and blue lines correspond to the neutral bound states $B_1$ and $B_2$ predicted by the semiclassical analysis. 
The double degeneracy of all levels comes from the $Z_2$ symmetry of the theory which, in the broken phase, gives rise to 
a doubly-degenerate spectrum of excitations which are exponentially split in finite volume.   Unfortunately
because the binding energy of $B_2$ is small, it is difficult to distinguish this state from a state consisting
of two $B_1$ particles, each with zero momentum.  To demonstrate that the $B_2$ state is really a bound state and
not a two-particle state we will need to turn to an analysis of the data in a finite momentum sector.  We will
do this in Section \ref{fma}.

Above the first two levels, we can see constant in $R$ energy levels 
roughly spaced by an energy $\Delta E\approx m_{\text{sc},1}$.  These can be interpreted as the states 
predicted by Eq.~(\ref{massphi4}) with $n\geq 3$ (although we stress these are only resonances -- they appear
as single particle states because we are working in finite volume).
We also see that there are energy levels that are falling off as $1/R$ that are 
composed of multiple particles with some carrying finite momentum.

We notice that in Fig.~\ref{fig:phi4_spectra} we have what appear to be level crossings, but this is only
because we are looking at $\Phi^4$ theories with small values of $g_4$:  the avoiding crossing is 
proportional to the magnitude of $g_4$.
We see this in Fig.~\ref{subfig:phi4_cross_set6} where we present the spectrum of Fig.~\ref{subfig:phi4_sp_set6}
on a finer scale.

\subsection{Perturbation theory in the broken phase}
\label{subsec:Iorder broken}

\begin{figure}
\centering
\subfigure[]{\includegraphics[width=0.48\textwidth]{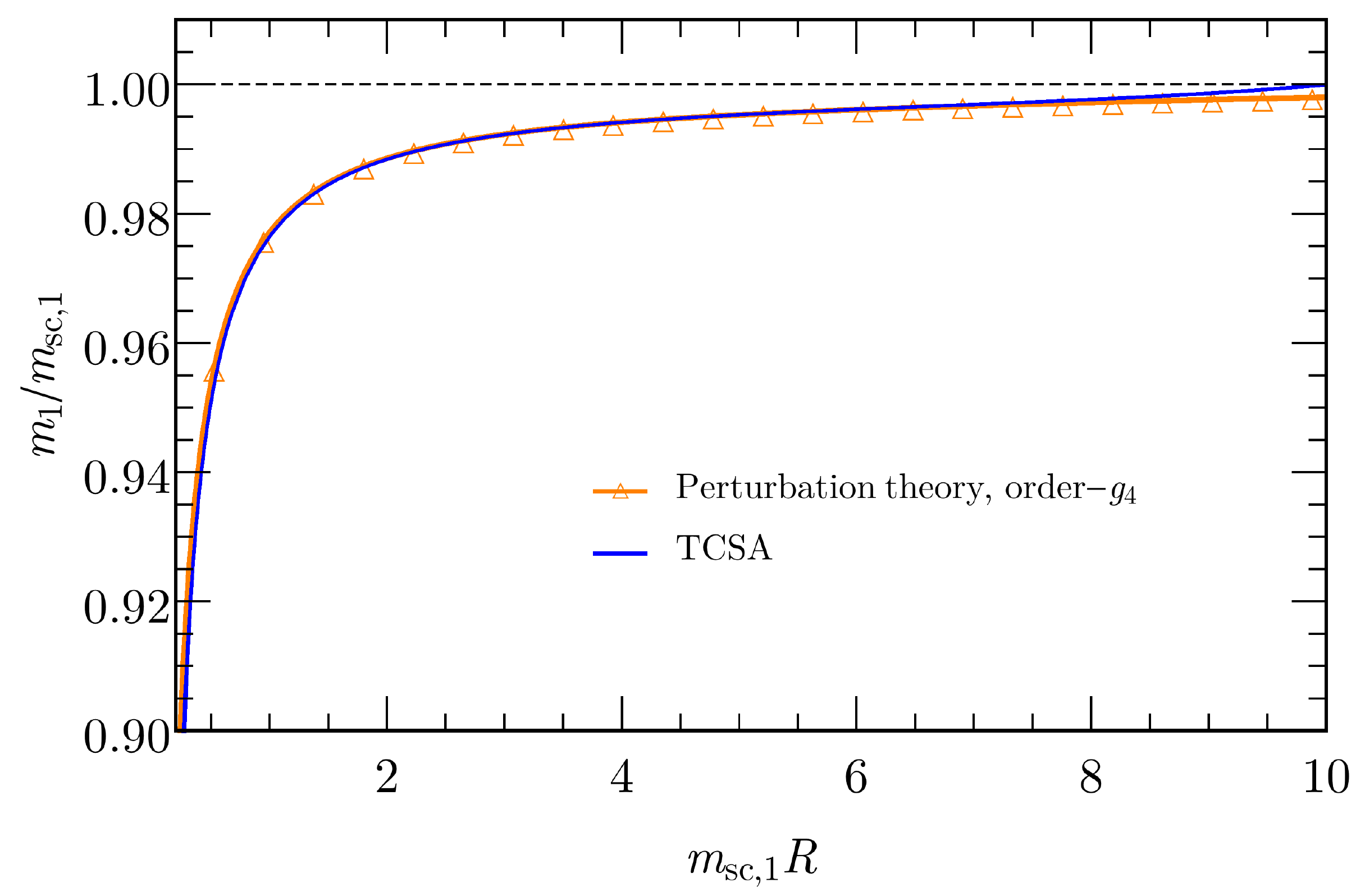}\label{subfig:phi4_Io_set11}}\hspace{1mm}
\subfigure[]{\includegraphics[width=0.48\textwidth]{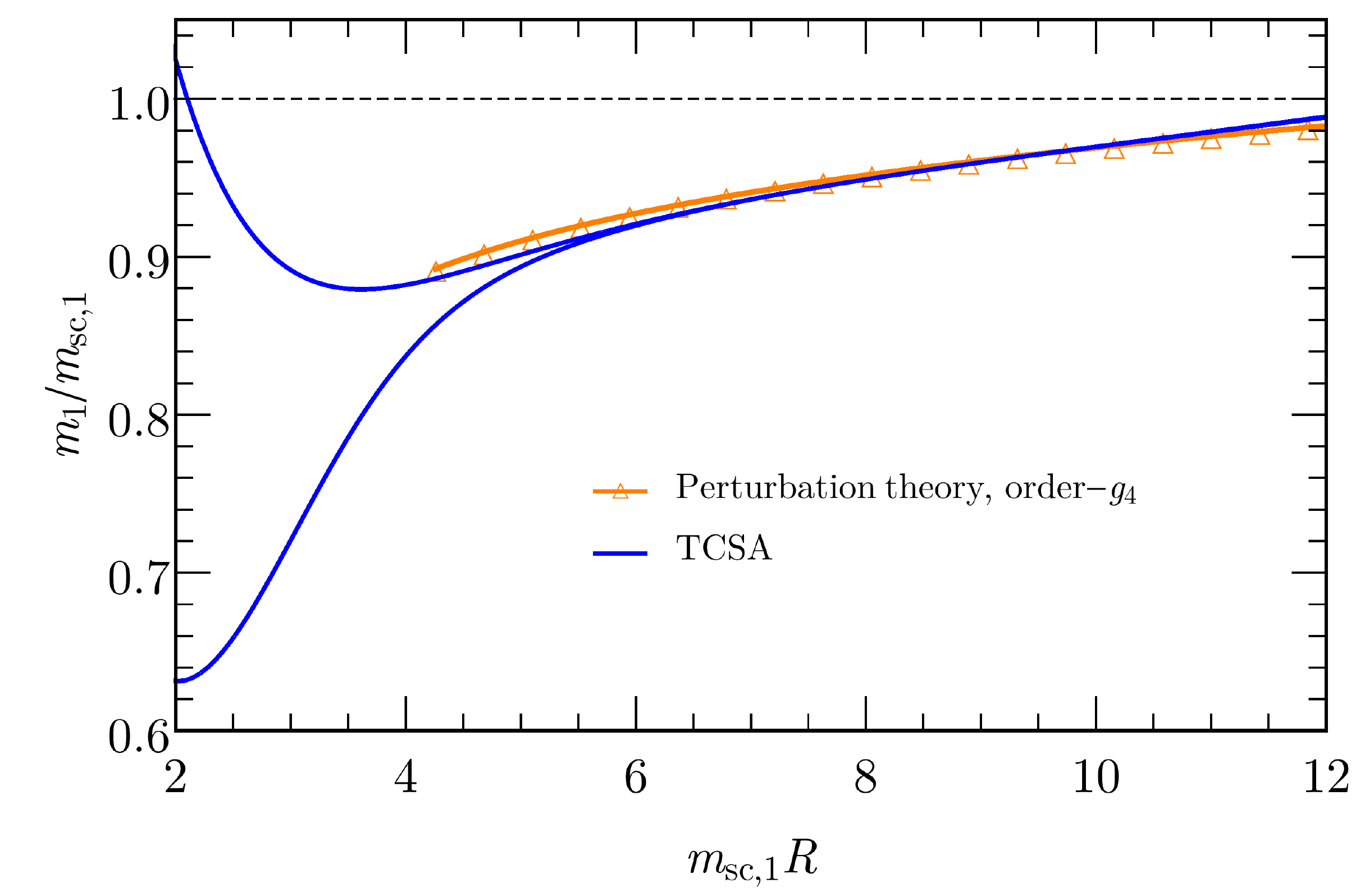}\label{subfig:phi4_Io_set6}}\\[-2mm]
\subfigure[]{\includegraphics[width=0.48\textwidth]{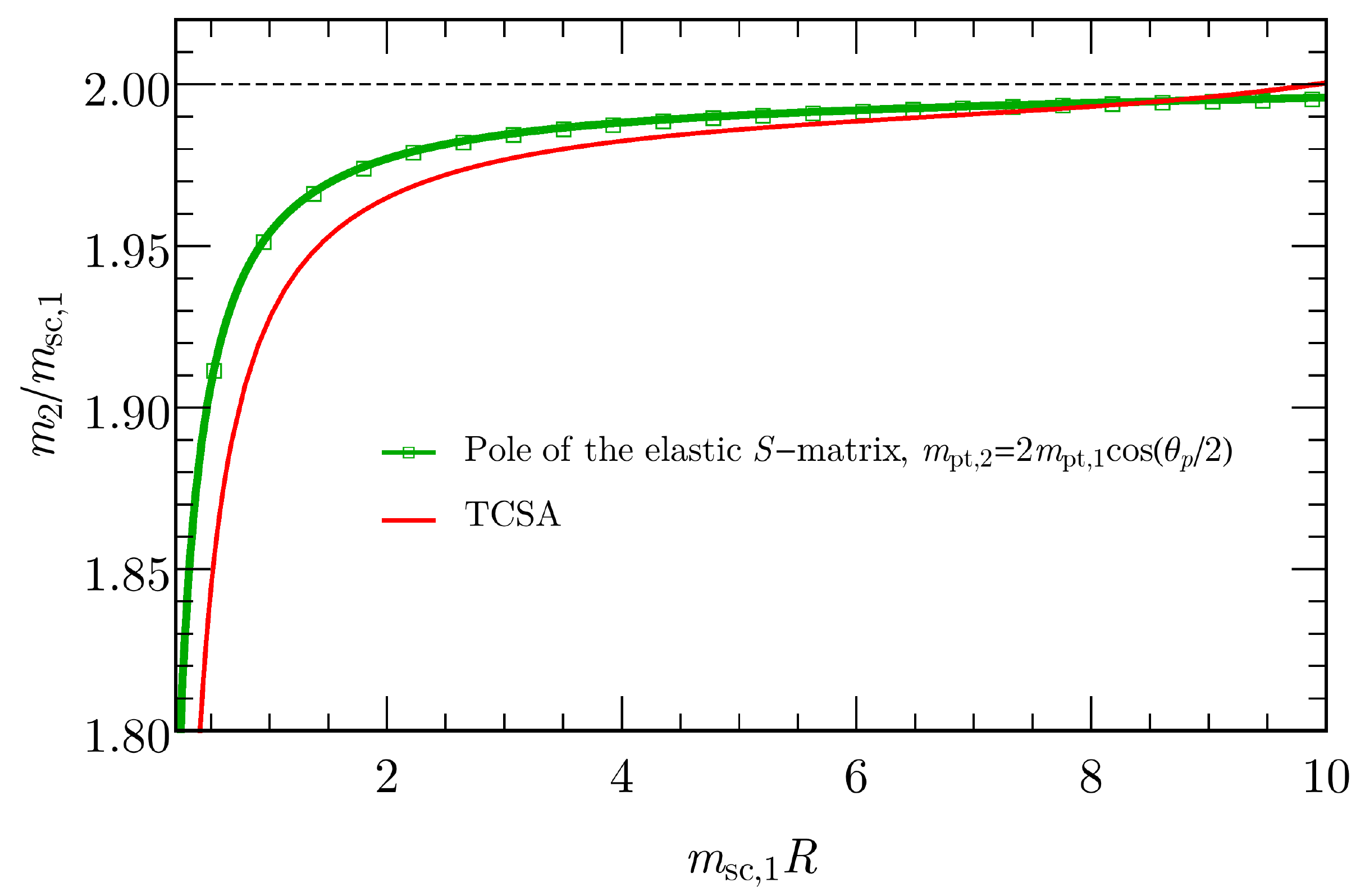}\label{subfig:phi4_2p_set11}}\hspace{1mm}
\subfigure[]{\includegraphics[width=0.48\textwidth]{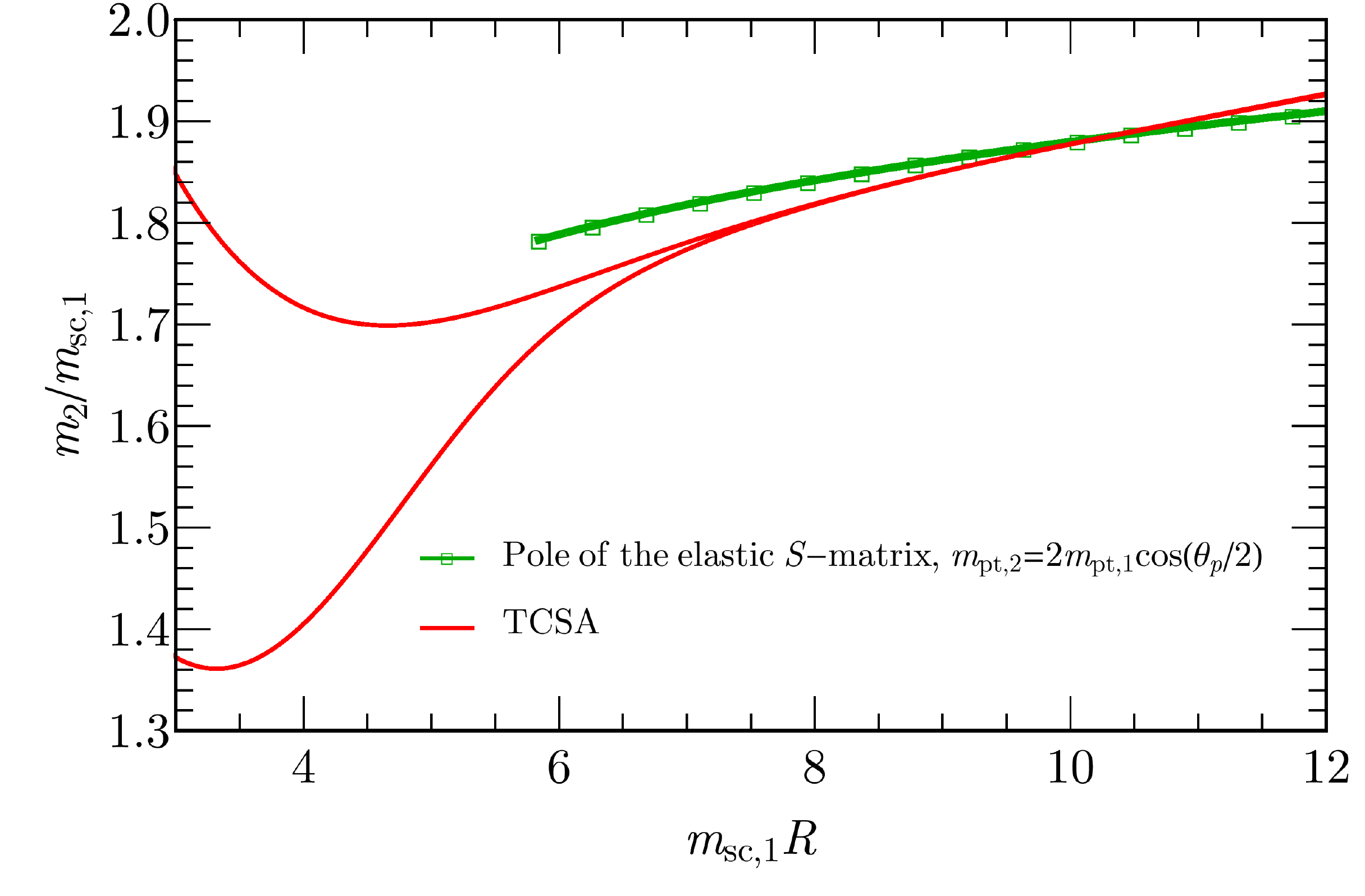}\label{subfig:phi4_2p_set6}}
\caption{Here we study the first two degenerate excited states for the same two sets of 
parameters as in Fig.~\ref{fig:phi4_spectra}, comparing our numerical results to perturbation theory.
In panel \subref{subfig:phi4_Io_set11} we show that our numerical data matches well with the perturbative prediction of
$m_{\text{pt},1}$ for the set of parameters first used in Fig. \ref{subfig:phi4_sp_set11}.  Similarly in
panel (b) we find good agreement between numerics and perturbation theory for the
set of parameters given in Fig. \ref{subfig:phi4_sp_set6}.
As expected, perturbation theory works very well for small values of the interaction, while it is less precise for larger values of 
$g_4$, where higher order contributions would be expected to contribute.
In \subref{subfig:phi4_2p_set11} and \subref{subfig:phi4_2p_set6} we show the comparison between the TCSA data 
for the second degenerate excited state, and the mass extracted from the pole of the purely elastic $S$-matrix, 
Eq.~\eqref{eq:Smatrix broken}, which agrees, at lowest order in perturbation theory, with the semiclassical prediction.
The mass, $m_{\text{pt},1}$, used in Eq.~\eqref{eq:bound state} is the one computed perturbatively at first order in $g_4$.}
\label{fig:phi4_1p_2p}
\end{figure}

We can use perturbation theory to give an estimate of the energy of the first massive excitation (blue line in Fig.~\ref{fig:phi4_spectra}) as an alternative to the semiclassical picture.
The advantage of perturbation theory is that we can compute it at finite $R$ thus obtaining an $R$ dependence for the 
estimated mass $m_{\text{pt},1}$.
However, since perturbation theory is valid locally around each of the
two classical vacua, it cannot reproduce
features related to the topological nature of the excitations.  In
particular 
it cannot reproduce the splitting of the levels that occurs for small values of $R$.

We implement the perturbative corrections to the mass only up to order $g_4$.
To do so, we need to rewrite the Hamiltonian around one of the two vacua.
The minimum of the classical potential is in $\phi_{\pm}^{(0)} =\pm \sqrt{-g_2/(2g_4)}$.
Expanding about one of them, $\phi =\phi_{\pm}^{(0)} + \eta$, gives
\begin{equation} \label{eq:potentialphi4 pert}
U(\phi_{\pm}^{(0)} + \eta) = - g_2 \,\eta^2 \pm \sqrt{- 2 g_2 g_4}\,\eta^3 + \frac{g_4}{2}\, \eta^4 \,.
\end{equation} 
As already discussed, with respect to the symmetric phase, at
zeroth-order its basic excitation has mass $m_0=\sqrt{-2g_2}$.  We
see from the above that a new term proportional to $\eta^3$ is present breaking the $Z_2$ symmetry.
This leads to a new 3-leg vertex, proportional to $\sqrt{g_4}$.
At first order in $g_4$, we should therefore consider all diagrams with one 4-leg vertex or with two 3-leg vertices.

To obtain the renormalized mass, $m_{\text{pt},1}$, we again compute 
the second derivative of the effective action $\Gamma^{(2)}(p^2)$, and solve the equation $\Gamma^{(2)}(p^2=-m_{\text{pt},1}^2)=0$.
The following diagrams contribute to $\Gamma^{(2)}$.
With the 4-leg vertex we can build the same divergent tadpole as in the symmetric phase, but with a different bare mass $m_0^2=-2g_2$.
In addition, with two 3-leg vertices we can build another divergent tadpole.
To take care of these divergencies and make a comparison with the TCSA data, we need to 
consider the correct normal ordering prescription.
As extensively discussed in Section~\ref{subsec:normal order}, our Hamiltonian is normal ordered with respect to the underlying CFT.
In the symmetric phase, the normal ordering of the $\phi^4$ term produces a counterterm proportional to $\phi^2$, which contributes to the Feynman diagram in Eq.~\eqref{eq:tadpole counterterm}.
In the broken phase we have the same counterterm coming from the normal ordering of the $\eta^4$ term, while
we also have a second counterterm, proportional to $\eta$, coming from the normal ordering of $\eta^3$:
\begin{equation}
 :\eta^3: \,= \eta^3 - 3D_{\Lambda_{\text{IR}},\Lambda}(0)\,\eta \,.
\end{equation}
We represent the new counterterm as a crossed circle with a single incoming particle line.
It is proportional to $\sqrt{g_4}$ and as with the previous case, is divergent for $\Lambda\rightarrow\infty$.
When contracted with a 3-leg vertex, it precisely cancels out the
divergence coming from the tadpole diagram 
built with two 3-leg vertices:
\begin{equation} \label{eq:tadpole 3leg}
\begin{tikzpicture}[baseline=(i.base),outer sep=5pt]
\coordinate[] (i)  at (0,0);
\coordinate[] (v1) at (1,0);
\coordinate[] (v2) at (1,0.75);
\coordinate[] (o)  at (2,0);
\coordinate[] (vc) at ($(v2)+(0,0.5)$);
\coordinate[] (vup) at ($(v2)+(0,1)$);
\draw[particle] (i)  -- node[label_ext] {} (v1);
\draw[particle] (v1) -- node[label_ext] {} (o);
\draw[particle] (v1) -- (v2);
\draw[thick, postaction={decorate},decoration={markings,mark=at position 0.23 with {\arrowreversed[black,rotate=12,yshift=0.01cm]{triangle 45}}}] (vc) circle (0.5cm);
\node[] at ($(vup)+(0,0.3)$) {$q$};
\end{tikzpicture}
+
\begin{tikzpicture}[baseline=(i.base),outer sep=5pt]
\coordinate[] (i)  at (0,0);
\coordinate[] (v1) at (1,0);
\coordinate[] (v2) at (1,1);
\coordinate[] (o)  at (2,0);
\draw[particle] (i)  -- node[label_ext] {} (v1);
\draw[particle] (v1) -- node[label_ext] {} (o);
\draw[particle] (v1) -- (v2);
\draw[cross,fill=white,cross] (v2) circle (5pt);
\end{tikzpicture}
=
 36 g_4 \left\{ \sum_{n=1}^{2N_{\text{tr}}} 
 \left( \frac{1}{\sqrt{n^2+\left(\frac{m_0 R}{2\pi}\right)^2}} - \frac{1}{n} \right)
 + \frac{\pi}{m_0 R} \right\}\,.
\end{equation}
The other diagram that appears is a bubble diagram, which is convergent. 
This diagram depends on the external momentum and should be computed for $p^2=-m_{\text{pt},1}^2$.
Since we are interested only in first order corrections, we can evaluate it at $p^2=-m_0^2$.
The result is given by
\begin{equation} \label{eq:bubble}
\begin{tikzpicture}[baseline=-1mm,outer sep=5pt]
\coordinate[] (i)  at (0,0);
\coordinate[] (v1) at (0.875,0);
\coordinate[] (v2) at (2.125,0);
\coordinate[] (o)  at (3,0);
\coordinate[] (vc) at (1.5,0);
\draw[particle] (i)  -- node[label_ext] {} (v1);
\draw[particle] (v2) -- node[label_ext] {} (o);
\draw[thick, postaction={decorate},%
decoration={markings,mark=at position 0.24 with {\arrowreversed[black,rotate=4,yshift=0.00cm]{triangle 45}},
                     mark=at position 0.79 with {\arrow[black,rotate=-14,yshift=0.00cm]{triangle 45}}}] (vc) circle (0.625cm);
\node[] at ($(i)+(0.4,0.3)$) {$p$};
\node[] at ($(o)+(-0.5,0.3)$) {$p$};
\node[] at ($(vc)+(0,0.9)$) {$q_1$};
\node[] at ($(vc)-(0,0.9)$) {$p-q_1$};
\end{tikzpicture}
= 9 g_4 \left(\frac{m_0R}{2\pi}\right)^2 \sum_{|n|<2N_{\text{tr}}}\frac{1}{\sqrt{n^2+\left(\frac{m_0R}{2\pi}\right)^2}\left[n^2+\frac{3}{4}\left(\frac{m_0R}{2\pi}\right)^2\right]} \,.
\end{equation}
Collecting all contributions, we find the following expression for the mass of the elementary excitation:
\begin{equation} \label{eq:mass broken}
\begin{split}
\frac{m_{\text{pt},1}^2}{m_0^2} = 1 &- 24 \frac{g_4}{m_0} \left\{ \sum_{n=1}^{2N_{\text{tr}}} 
 \left( \frac{1}{\sqrt{n^2+\left(\frac{m_0 R}{2\pi}\right)^2}} - \frac{1}{n} \right)
 + \frac{\pi}{m_0 R} \right\} \\
 &- 9 \frac{g_4}{m_0} \left(\frac{m_0R}{2\pi}\right)^2 \sum_{|n|<2N_{\text{tr}}}\frac{1}{\sqrt{n^2+\left(\frac{m_0R}{2\pi}\right)^2}\left[n^2+\frac{3}{4}\left(\frac{m_0R}{2\pi}\right)^2\right]}\,.
\end{split}
\end{equation}
Notice the change of sign in the mass correction in comparison to the symmetric phase.

The comparison of the perturbative formula Eq.~\eqref{eq:mass broken}
with the numerical TCSA data is 
shown in Figs.~\ref{subfig:phi4_Io_set11} and
\ref{subfig:phi4_Io_set6} for the same set of parameters as in
Figs.~\ref{subfig:phi4_sp_set11} and \ref{subfig:phi4_sp_set6}.
The first two degenerate levels above the ground states computed through TCSA are plotted as the blue solid line.
The orange curve with triangles is the renormalized mass at first order in $g_4$.
As expected from perturbation theory, the results are very good for
small values of $g_4$, while they gets worse for 
larger values.

In Figs.~\ref{subfig:phi4_2p_set11} and \ref{subfig:phi4_2p_set6}
we consider the second excited state from the spectra of 
Figs.~\ref{subfig:phi4_sp_set11} and \ref{subfig:phi4_sp_set6}.
We recall that in the symmetric phase the energy of these states turned out to be very well described by the 
solution of the Bethe equation in the center of mass, Eq.~\eqref{eq:Bethe Eq in CM}
quantized according to the quantum numbers $N_1=N_2=0$.
With $g_2<0$, on the contrary, these states are bound states.  (We can
see that everything is consistent because for the given S-matrix in the
broken phase, the Bethe quantization condition does not have solutions corresponding to $N_1=N_2=0$.)
Instead we give a rough estimate of the energy of this level in finite volume, $R$, by considering the pole of 
the elastic two-particle $S$-matrix computed in Appendix~\ref{app:2particle} for the broken phase, 
Eq.~\eqref{eq:Smatrix broken}.
The mass of the bound state corresponding to a pole in the two-particle $S$-matrix is given by 
\begin{equation} \label{eq:bound state}
 m_{\text{pt},2} = 2m_{\text{pt},1}\,\cos\left(\theta_p/2\right)\,, \qquad\text{where}\qquad \theta_p = 24\pi \frac{g_4}{m_{\text{pt},1}^2} \,.
\end{equation}
From the plots we see that our estimate of the bound state energy matches the general behavior of the TCSA data, 
but fails to reproduce the exact numerical values.
This is not surprising given the level of approximation in this estimate.

\subsection{Analysis of the finite momentum sector of the broken $\Phi^4$ LG theory}
\label{fma}

It is very difficult to argue for the presence of the bound states predicted by the semiclassical analysis 
from the momentum-0 spectrum alone.
Because the binding energy of $B_2$ is relatively small for the weak values of the coupling $g_4$ studied,
it is difficult to distinguish a putative $B_2$ state from a state involving two $B_1$ particles, both with zero
momentum.
However in the momentum-1 sector, a state involving $B_2$ has an energy equal to $\sqrt{m_{2}^2+(2\pi/R)^2}$ while
a state involving two $B_1$ particles has energy $m_1+\sqrt{m_1^2+(2\pi/R)^2}$.  These energies are
different and so give us a way to discriminate the states from one another.

\begin{figure}
\centering
\subfigure[]{\includegraphics[width=0.48\textwidth]{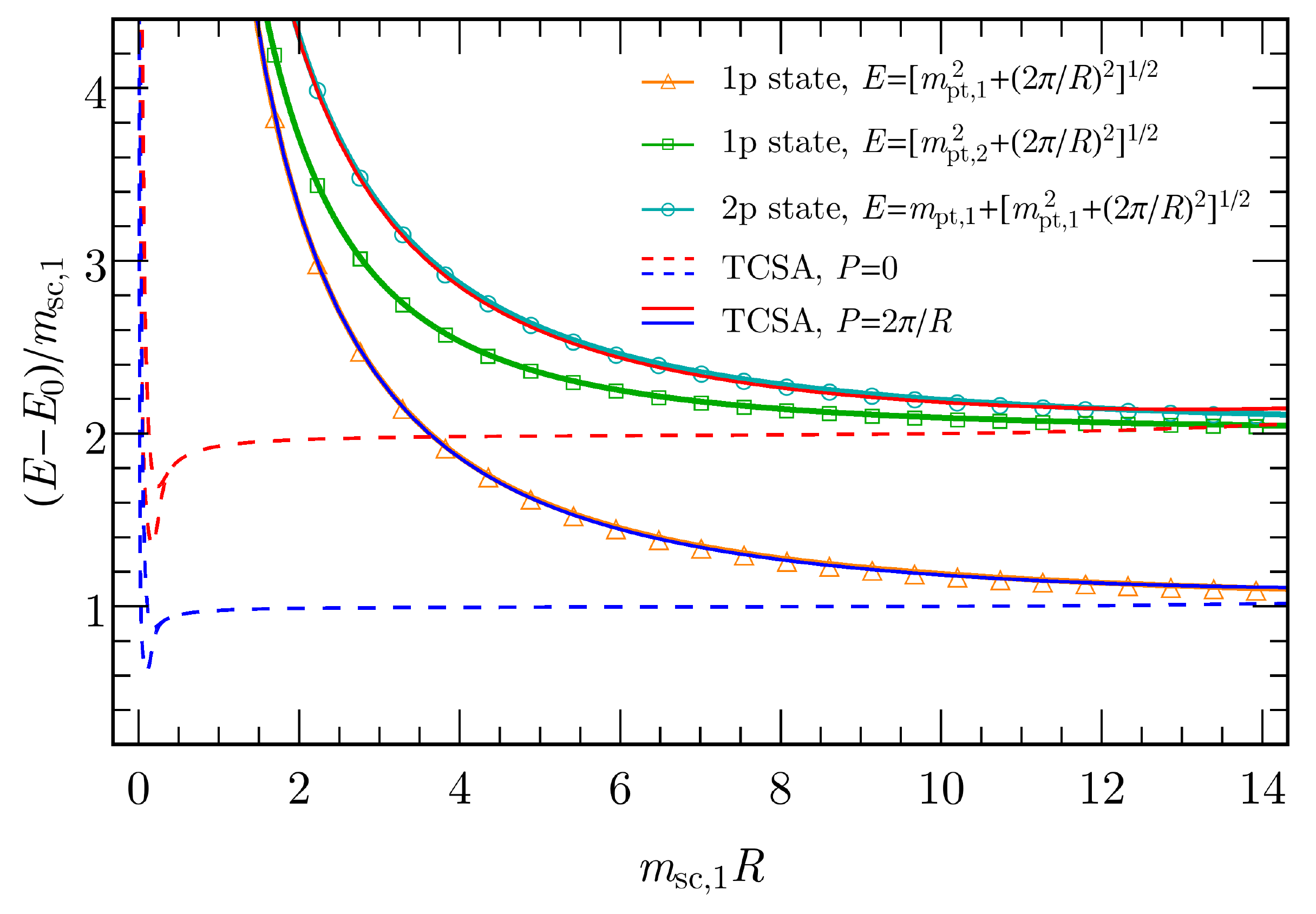}\label{subfig:phi4_mom1_set11}}\hspace{1mm}
\subfigure[]{\includegraphics[width=0.48\textwidth]{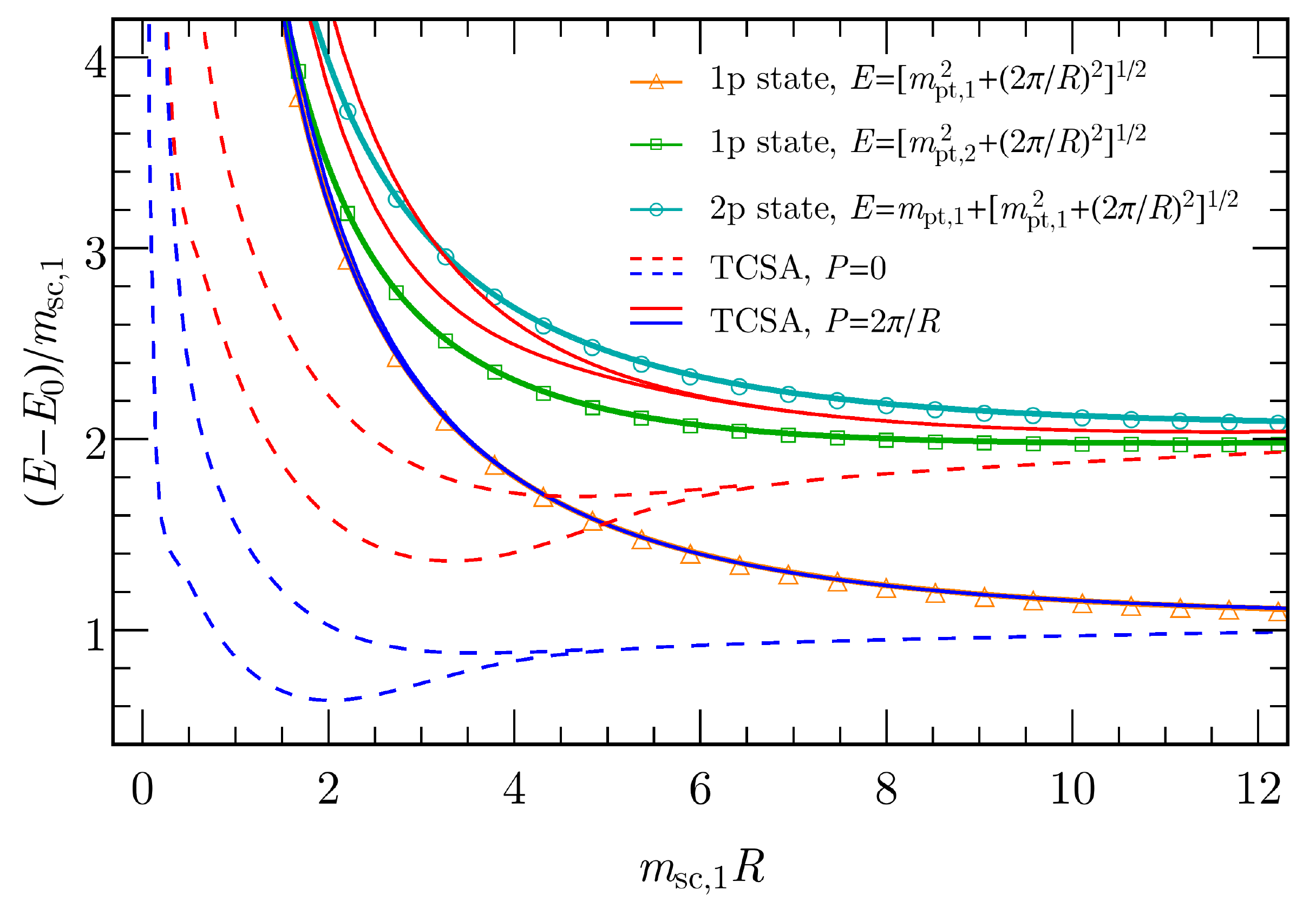}\label{subfig:phi4_mom1_set6}}
\caption{The low lying energy levels in the momentum-1 sector (blue and red solid lines) for $\Phi^4$ LG theories with the same 
parameters as Fig.~\ref{fig:phi4_spectra}, namely for Fig. \subref{subfig:phi4_mom1_set11} $g_2=-0.1$, $g_4=6\times 10^{-5}$, $\beta=0.06$, and $N_{\text{tr}}=6$, while for Fig. 
\subref{subfig:phi4_mom1_set6} $g_2=-0.1$, $g_4=1.2\times 10^{-3}$, $\beta=0.2$, and $N_{\text{tr}}=7$.
The ground state energy belonging to the momentum-0 sector has been subtracted.
All energies are normalized with respect to $m_{\text{sc},1}$ and are plotted as function of $m_{\text{sc},1}R$.
The first two levels of the momentum-0 sector are shown for reference as the blue and red dashed line.
In both plots, the blue solid line is well described by the dispersion relation of a single particle of mass, $m_{\text{pt},1}$, and
momentum $2\pi/R$ (orange curve with triangles).
(Here $m_{\text{pt},1}$ is the renormalized mass at first order in perturbation theory (see Section~\ref{subsec:Iorder broken})).
For the smaller value of the interaction parameter $g_4$ (panel \subref{subfig:phi4_mom1_set11}), 
the second momentum-1 state (red solid line) is well described by the non-interacting energy of a two-particle $B_1$
state, one particle with momentum $0$ and one with momentum $2\pi/R$ (light blue curve with circles).
For the larger value of $g_4$ (panel \subref{subfig:phi4_mom1_set6}), the red solid line moves 
towards the dispersion relation of a $B_2$ particle of mass $m_{\text{pt},2}$ with momentum $2\pi/R$ (green curve with squares) at
large values of $R$.  Here the mass $m_{\text{pt},2}$ is estimated through Eq.~\eqref{eq:bound state}.}
\label{fig:phi4_mom1}
\end{figure}

In Fig.~\ref{fig:phi4_mom1} we show the first two pairs of doubly degenerate levels (blue and red solid lines) in the 
momentum-1 sector for the set of parameters found in Fig.~\ref{fig:phi4_spectra}.
We also show the lowest energy levels of the momentum-0 spectra as red and blue dashed lines.
In Fig.~\ref{subfig:phi4_mom1_set11}, where the coupling $g_4$ is very small, the first doubly degenerate momentum-1 energy level 
is well described by the dispersion relation of a single particle of mass $m_{\text{pt},1}$ moving with momentum 
$2\pi/R$ (orange curve with triangles), that is $E=\sqrt{m_{\text{pt},1}^2+(2\pi/R)^2}$.
Here $m_{\text{pt},1}$ is the renormalized mass at first order in perturbation theory, as computed in Section~\ref{subsec:Iorder broken}.
The second doubly degenerate momentum-1 energy level is fitted by the energy of two free particles of mass $m_{\text{pt},1}$, one static and 
the other moving with momentum $2\pi/R$ (light blue curve with circles), namely $E=m_{\text{pt},1}+\sqrt{m_{\text{pt},1}^2+(2\pi/R)^2}$.
The green curve with squares is the dispersion relation of a single particle of type $B_2$ with mass $m_{\text{pt},2}$, 
computed by Eq.~\eqref{eq:bound state}.  From our numerical data,
it is clear that this second excited state is behaving more like
a state involving two $B_1$ particles rather than a single $B_2$ particle.

However the semiclassical prediction for the bound states is strictly valid only in the infinite volume limit
whereas we are working in finite volume.  There may however be a crossover at large $R$ where the state begins
to behave more like a one-particle state.  For the parameters in Fig.~\ref{subfig:phi4_mom1_set11}, the value of $g_4$
is small enough that this putative value of $R$ is beyond our ability to simulate.  However this crossover can be
seen for larger values of $g_4$ as in Fig.~\ref{subfig:phi4_mom1_set6}.  There the red solid line marks the
second pair of excited states in the momentum-1 sector.  For smaller values of $R$ it hues closer to the line with
energy $E=m_{\text{pt},1}+\sqrt{m_{\text{pt},1}^2+(2\pi/R)^2}$, i.e.\ the energy of two $B_1$ particles.  For larger values of $R$
it moves towards a line with energy $E=\sqrt{m_{\text{pt},2}^2+(2\pi/R)^2}$, i.e.\ the energy of a particle $B_2$ with
momentum $p=2\pi/R$.  Thus we conclude that indeed the semiclassical prediction of two stable bound states
in the broken phase of $\Phi^4$ is correct.  However to see it in the TCSA data one needs to go to large enough
$g_4$ and $R$.

\subsection{The $\Phi^6$ LG theory in the two well potential}
\label{subsec:phi6}

\begin{figure}
\centering
\subfigure[]{\includegraphics[width=0.48\textwidth]{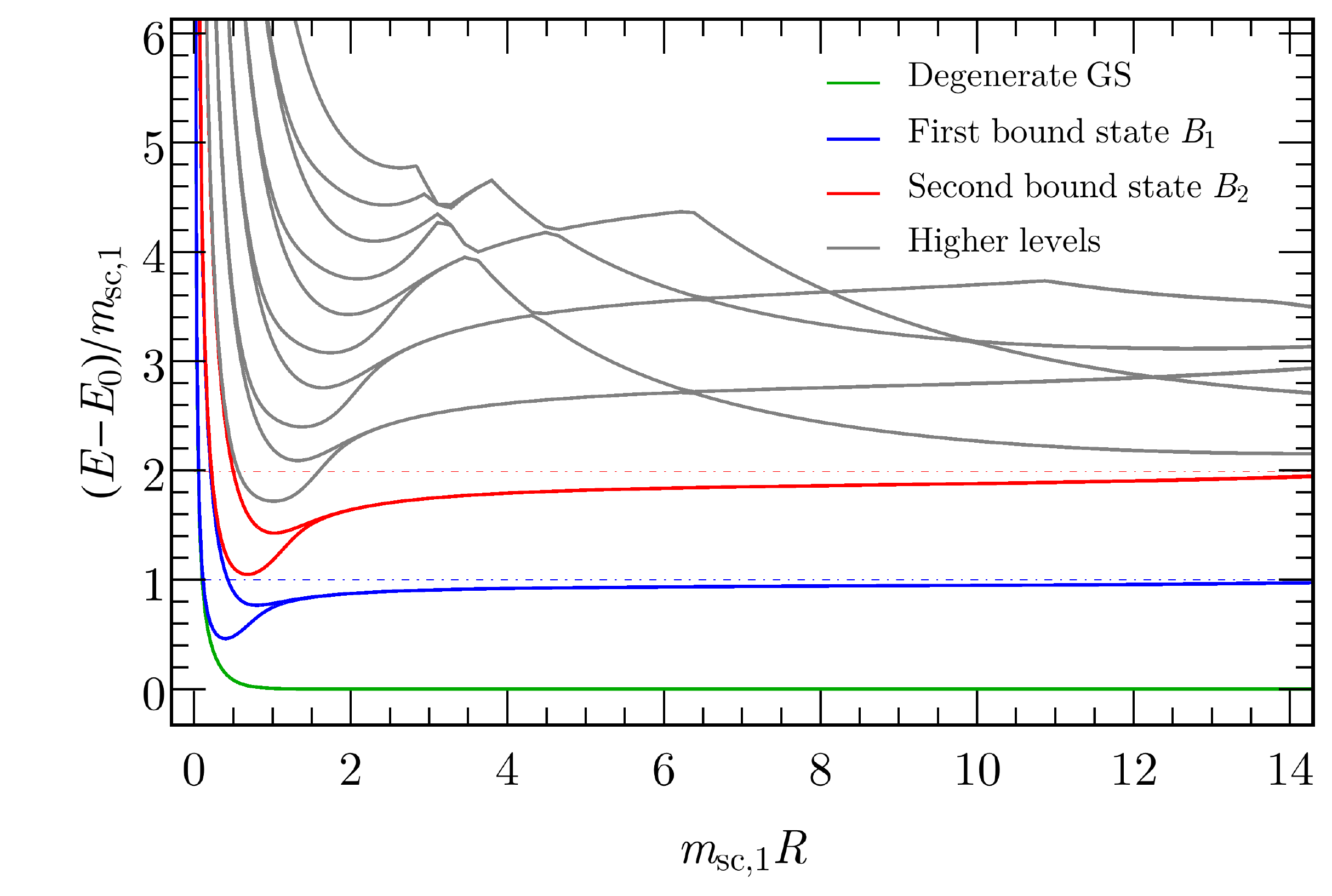}\label{subfig:phi6_sp}}\hspace{1mm}
\subfigure[]{\includegraphics[width=0.48\textwidth]{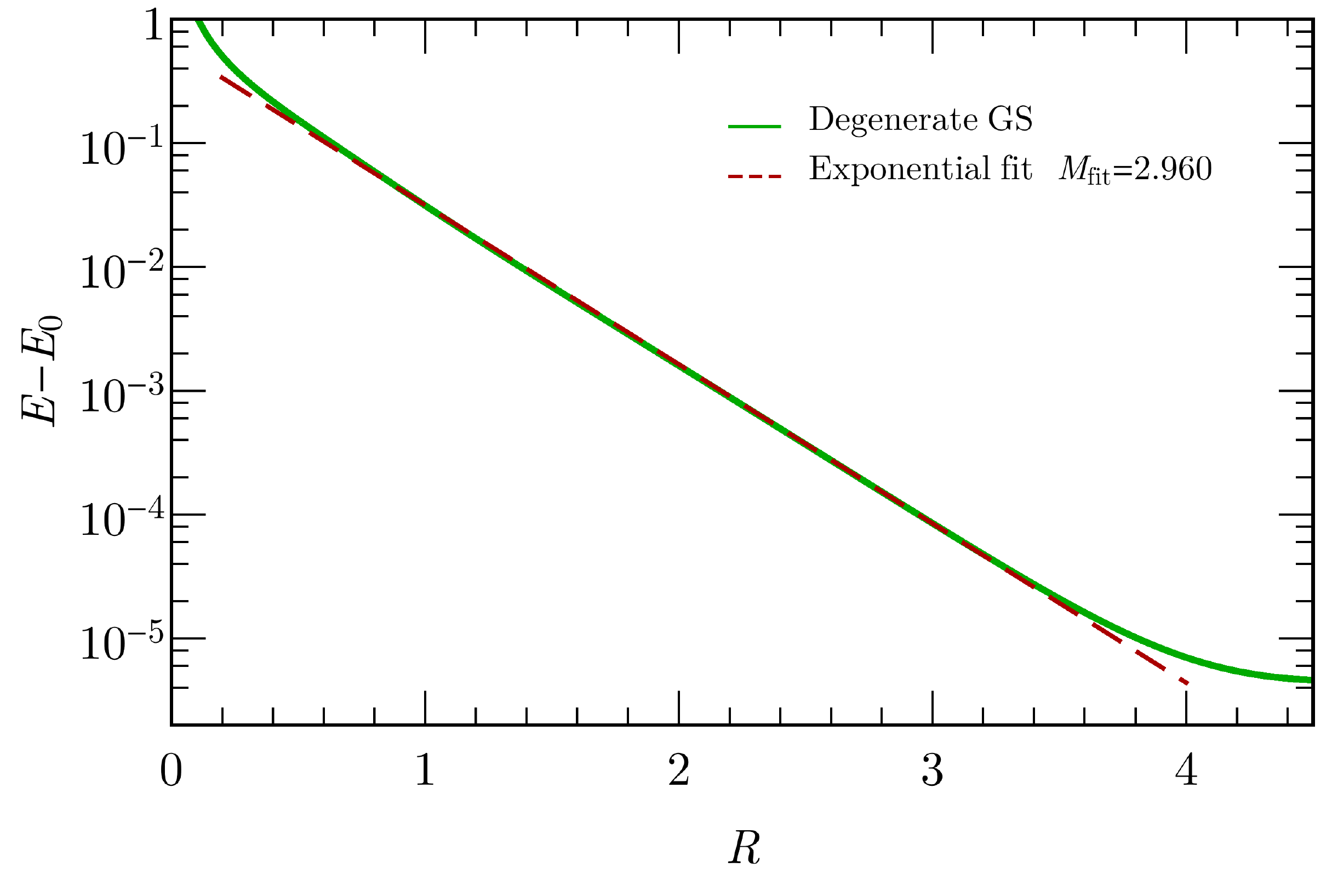}\label{subfig:phi6_GS}}\\
\subfigure[]{\includegraphics[width=0.48\textwidth]{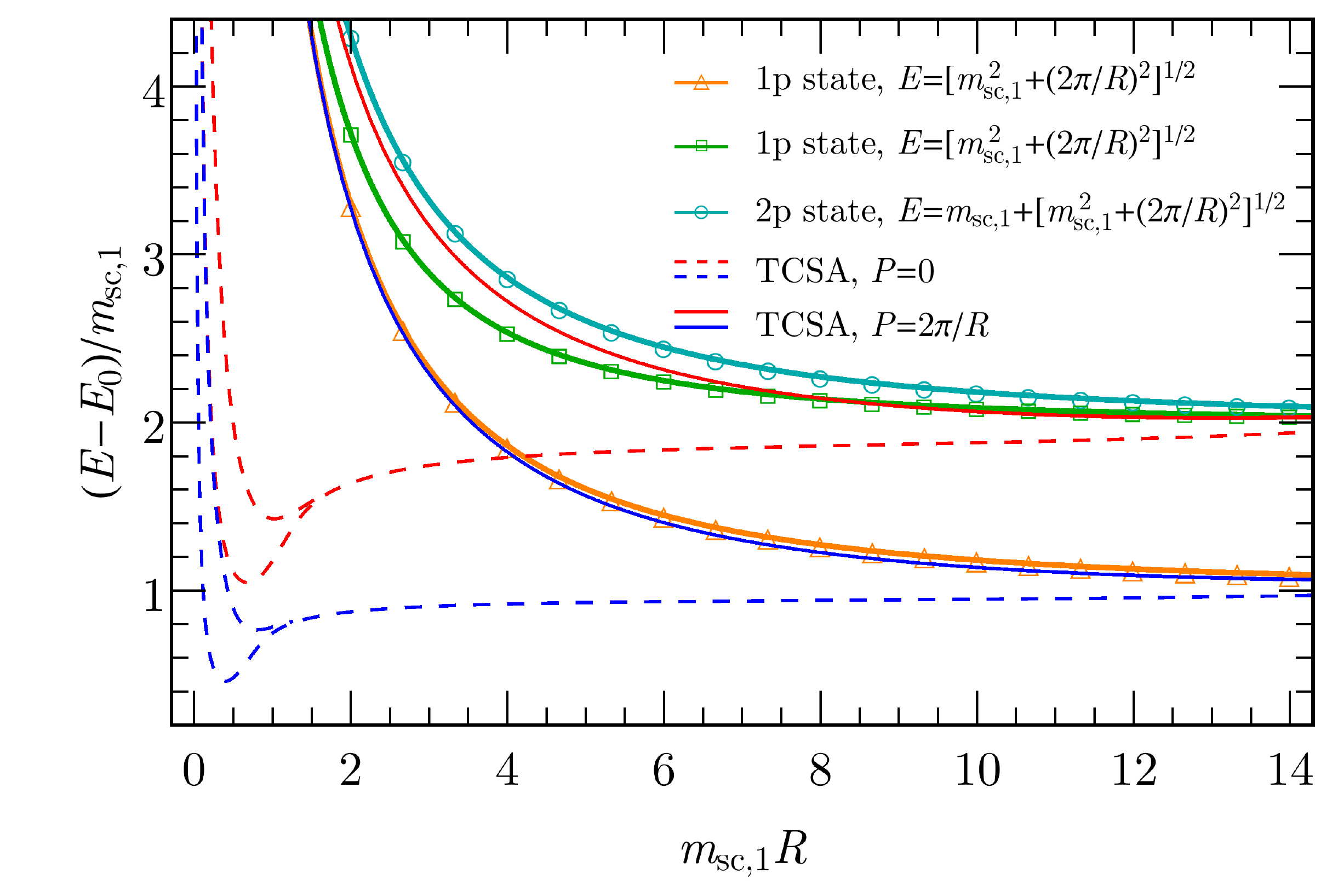}\label{subfig:phi6_mom1}}\hspace{1mm}
\subfigure[]{\includegraphics[width=0.48\textwidth]{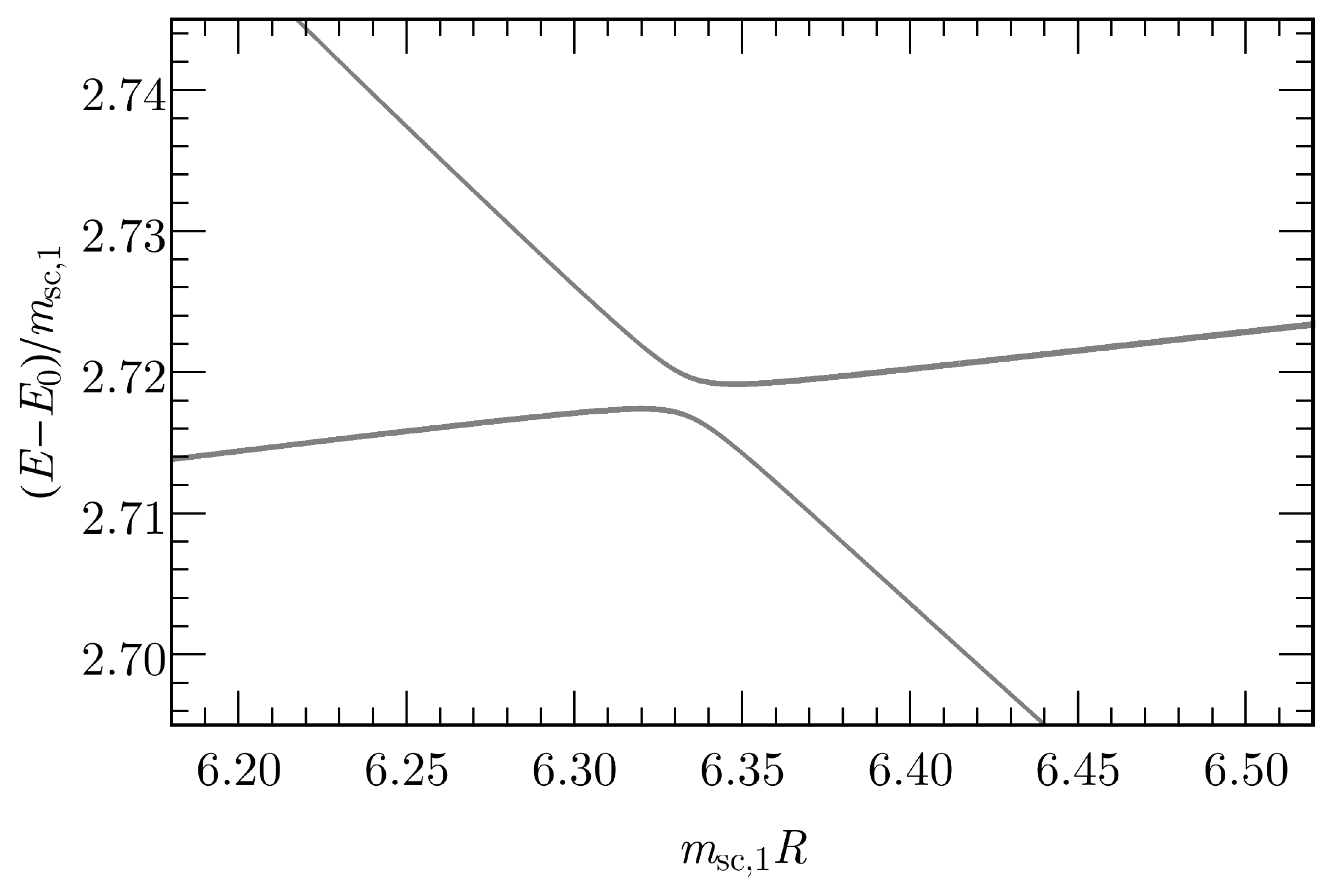}\label{subfig:phi6_cross}}
\caption{\subref{subfig:phi6_sp}: The spectrum of a double-welled $\Phi^6$ LG theory with
$g_2=-0.06$, $g_4=-6\times 10^{-5}$, and $g_6=4\times 10^{-7}$.
The mass for this theory is $m_0=0.5727$.
The first excited state (green line) becomes exponentially degenerate with the ground state.
All energies are normalized with the mass scale $m_{\text{sc},1}$ and are plotted as function of $m_{\text{sc},1}R$.
The dot-dashed blue and red lines are given by 1 and $m_{\text{sc},2}/m_{\text{sc},1}$.
The first double-degenerate level (blue) is believed to be the neutral bound state $B_1$, 
while the second double-degenerate level (red) is the second bound state $B_2$.
\subref{subfig:phi6_GS}: By parameterizing this energy difference with an exponential fit,
we can estimate the mass of the kink to be $M=2.960$, and from this the masses of the first two 
bound states $m_{\text{sc},1}=0.5718$ and $m_{\text{sc},2}=1.138$.
\subref{subfig:phi6_mom1}: The low-lying spectrum in the momentum-1 sector.
The first two doubly degenerate levels of the momentum-0 sector are shown as the blue and red dashed lines.
The blue solid line corresponds to the numerical data describing a single particle of mass $m_{\text{sc},1}$ moving with momentum $2\pi/R$.
It is well fitted by the corresponding dispersion relation (orange curve with triangles).
The red solid line is believed to correspond to a second bound state $B_2$ in the infinite volume limit.
At small $R$, the line hues close to the prediction for two free particles of mass $m_{\text{sc},1}$, 
but for increasing $R$ moves towards the dispersion relation of a single particle of mass $m_{\text{sc},2}$.
\subref{subfig:phi6_cross}: We zoom in on the 
spectrum in Fig. \ref{subfig:phi6_sp} showing an avoided crossing due to interactions.}
\label{fig:phi6_spectra}
\end{figure}

We have also tested the conjecture of Ref.~\cite{mussardo-2007} in the double well phase of the $\Phi^6$ LG theory with $g_2$, $g_4<0$ and $g_6>0$.
The zeroth order perturbative mass is easily computed to be:
\begin{equation}
 m_0^2 = -4g_2 + \frac{11}{3}\frac{g_4^2}{g_6} - \frac{4}{3}\frac{g_4}{g_6}\sqrt{g_4^2 - 3g_2g_6}  \,.
\end{equation}

As for the case of the $\phi^4$ interaction, we are limited to relatively small values of $g_4$ and $g_6$ in order
to obtain an appreciably sized scaling region in $R$.
In this regime we find again the same features that we saw with the $\phi^4$ potential, 
as can be seen in Fig.~\ref{subfig:phi6_sp}.
In this case, after passing from the small $R$ conformal region, 
a massive regime is reached characterized by a doubly degenerate ground state (green), a doubly degenerate 
single particle state, $B_1$ (blue), and a second doubly degenerate single particle state $B_2$ (red).
Again we find the unstable states of Eq.~(\ref{massphi4}) for $n\geq 3$ as well as multi-particle states with some $1/R$ dependence.

By studying the exponential splitting of the two ground states, we can again extract the kink mass 
$M$ (see Fig.~\ref{subfig:phi6_GS}) and, by means of Eq.~\eqref{massphi4}, give an estimate of the first two bound state masses 
$m_{\text{sc},1}$ and $m_{\text{sc},2}$.

In order to confirm the presence of a second bound state in the infinite volume limit, we again study the spectrum in the 
momentum-1 sector.  The result is shown in Fig.~\ref{subfig:phi6_mom1}.
The first level (blue solid line) is well fitted by the dispersion relation of a particle of mass $m_{\text{sc},1}$ (orange curve with triangles), 
the small differences coming most probably from unknown finite size corrections.
The second level (red solid line) is believed to correspond in the infinite volume limit to the second bound state $B_2$.
At finite $R$ we can see a transition between a two-particle state and a single particle one:
for small values of $m_{\text{sc},1} R$, the energy of the level is close to the one of two particles of type $B_1$ with mass $m_{\text{sc},1}$, 
one static and one with momentum $2\pi/R$ (light blue curve with circles).
For increasing values of $m_{\text{sc},1} R$, the energy gets closer and closer to the dispersion relation of a single particle of type $B_2$, 
with mass $m_{\text{sc},2}$ (green curve with squares).
Thus we conclude that at least at large $R$, there is a second bound state in the theory.
Finally, in Fig.~\ref{subfig:phi6_cross} we show the presence of an avoided crossing in the spectrum of the $\Phi^6$ theory.

\section{Conclusion and Discussion}
\label{conclusions}

In this paper we have studied the application of the TCSA to Landau-Ginzburg theories.
Because of the difficulties of using an uncompactified boson with its uncountable irreducible representations, 
we chose to approximate the LG theory by using a compactified boson.
However we always work at large compactification radius where the effects of compactification
are negligible at low energies.

As a first implementation of a Landau-Ginzburg theory, we demonstrated that this methodology 
can reproduce the spectrum of a free massive boson (a $\Phi^2$ theory).
Moving on to studying interacting LG theories, we first considered the unbroken phase of $\Phi^4$ LG theory.
We demonstrated that at least at weak coupling $g_4$, we could obtain good agreement between our numerics
and a two-loop perturbative computation.

Having run the methodology through these two tests, we then studied the low energy kink and neutral excitations 
of the $\Phi^4$ and $\Phi^6$ LG theories in their broken phase.  
We compared our numerics here with 
the semiclassical predictions made in Ref.~\cite{mussardo-2007} and found good agreement.
In this work, the study of the finite momentum sectors of the Hilbert space proved to be very useful
in establishing the presence of two bound states as predicted by semi-classics.

In future work it will be interesting to understand if we can represent the various conformal minimal models
in the language of LG theories, something originally envisioned in Ref.~\cite{zamlg}.  
However in order to do so one will need to understand the renormalized effective
potentials of a given LG theory.  A pure LG $\Phi^4$ will not simply be equivalent to a $c=1/2$ CFT because the
quantum renormalization of the $\Phi^4$ will lead to an effective $\Phi^2$ term, so spoiling scale invariance.  In order
to realize the $c=1/2$ theory, one would have at a minimum to find a bare LG potential which becomes critical
upon renormalization.  Whether this search for such a potential must be done numerically or can be performed
analytically in perturbation theory is however an open question.

\begin{acknowledgments}
The research herein was supported by the CMPMS Department, Brookhaven National Laboratory, in turn funded 
by the U.S. Department of Energy, Office of Basic Energy Sciences, under Contract No. DE-AC02-98CH10886 (RMK), by the National Science Foundation under grant no. PHY 1208521 (RMK), and by the Netherlands Organization for Scientific Research (NWO) and the Foundation for Fundamental Research on Matter (FOM) (GPB).  This research is also supported by the IRSES Grant QICFT. 
\end{acknowledgments}

\vfill\eject

\appendix

\section{Details of the two-loop perturbative computation of the mass in the unbroken phase of $\Phi^4$ LG}
\label{app:2loop}

In this Appendix we give some details on the computation of the two-loop contribution to the perturbative renormalization of 
the mass in the unbroken phase of $\Phi^4$ 
discussed in Section~\ref{sec:posg2g4} and shown in Figs.~\ref{subfig:phi4pos_IIo_set1} and \ref{subfig:phi4pos_IIo_set9}.

As discussed in the main text, we want to compute the second derivative of the effective action $\Gamma^{(2)}$, and solve the equation $\Gamma^{(2)}(p^2=-m_{\text{pt},1}^2)=0$.
We need to compute all one-particle irreducible diagrams with two external legs.
At one-loop, we showed that the only contribution is given by the tadpole diagram, computed in Eq.~\eqref{eq:tadpole}.
We also needed to consider the mass counterterm coming from the CFT prescription for normal ordering, leading to the result Eq.~\eqref{eq:Iorder}.
We now want to extend this analysis to second order in $g_4$.
There are two diagrams contributing to the mass renormalization at this order, the snowman and the Saturn diagrams~\cite{binney}.
Moreover, the mass counterterm must be inserted in the internal propagator of the one-loop tadpole of Eq.~\eqref{eq:tadpole}.
This will give a result of order $g_4^2$ and will cancel the divergence coming from the tadpole subdiagram present 
in the snowman.

The snowman diagram gives the following contribution:
\begin{equation}
\begin{split}
\begin{tikzpicture}[baseline=(vc.base),outer sep=5pt]
\coordinate[] (i) at (0,0);
\coordinate[] (v) at (1.25,0);
\coordinate[] (o) at (2.5,0);
\coordinate[] (vc) at (1.25,0.625);
\coordinate[] (vcup) at (1.25,1.25+0.625);
\draw[particle] (i) -- node[label_ext] {} (v);
\draw[particle] (v) -- node[label_ext] {} (o);
\draw[thick, postaction={decorate},%
decoration={markings,mark=at position 0.49 with {\arrowreversed[black,rotate=4,yshift=0.00cm]{triangle 45}},
                     mark=at position 0.98 with {\arrowreversed[black,rotate=8,yshift=0.00cm]{triangle 45}}}] (vc) circle (0.625cm);
\draw[thick, postaction={decorate},decoration={markings,mark=at position 0.23 with {\arrowreversed[black,rotate=12,yshift=0.01cm]{triangle 45}}}] (vcup) circle (0.625cm);
\node[] at ($(vcup)+(0,0.9)$) {$q_2$};
\node[] at ($(vc)-(0.9,0)$) {$q_1$};
\end{tikzpicture}
 &= 
 \frac{36}{\pi^2}m_0^2\left(\frac{g_4}{g_2}\right)^2\left(\frac{m_0 R}{2\pi}\right)^2
 \sum_{|n_1|<2N_{\text{tr}}} \int\frac{\ud q}{q^2+n_1^2+\left(\frac{m_0 R}{2\pi}\right)^2}
 \sum_{|n_2|<2N_{\text{tr}}} \int\frac{\ud q}{\left[q^2+n_2^2+\left(\frac{m_0 R}{2\pi}\right)^2\right]^2} \\
 &= 18m_0^2\left(\frac{g_4}{g_2}\right)^2\left(\frac{m_0 R}{2\pi}\right)^2
 \sum_{|n_1|<2N_{\text{tr}}} \frac{1}{\sqrt{n_1^2 + \left(\frac{m_0 R}{2\pi}\right)^2}}
 \sum_{|n_2|<2N_{\text{tr}}} \frac{1}{\left[n_2^2 + \left(\frac{m_0 R}{2\pi}\right)^2\right]^{3/2}} \,,
\end{split}
\end{equation}
and once we subtract the divergence we obtain
\begin{equation}
\begin{split}
\begin{tikzpicture}[baseline=(vc.base),outer sep=5pt]
\coordinate[] (i) at (0,0);
\coordinate[] (v) at (1.25,0);
\coordinate[] (o) at (2.5,0);
\coordinate[] (vc) at (1.25,0.625);
\coordinate[] (vcup) at (1.25,1.25+0.625);
\draw[particle] (i) -- node[label_ext] {} (v);
\draw[particle] (v) -- node[label_ext] {} (o);
\draw[thick, postaction={decorate},%
decoration={markings,mark=at position 0.49 with {\arrowreversed[black,rotate=4,yshift=0.00cm]{triangle 45}},
                     mark=at position 0.98 with {\arrowreversed[black,rotate=8,yshift=0.00cm]{triangle 45}}}] (vc) circle (0.625cm);
\draw[thick, postaction={decorate},decoration={markings,mark=at position 0.23 with {\arrowreversed[black,rotate=12,yshift=0.01cm]{triangle 45}}}] (vcup) circle (0.625cm);
\node[] at ($(vcup)+(0,0.9)$) {$q_2$};
\node[] at ($(vc)-(0.9,0)$) {$q_1$};
\end{tikzpicture}
&\hspace{2mm}+\hspace{2mm}
\begin{tikzpicture}[baseline=(vc.base),outer sep=5pt]
\coordinate[] (i) at (0,0);
\coordinate[] (v) at (1.25,0);
\coordinate[] (o) at (2.5,0);
\coordinate[] (vc) at (1.25,0.625);
\coordinate[] (vup) at (1.25,1.25);
\draw[particle] (i) -- node[label_ext] {} (v);
\draw[particle] (v) -- node[label_ext] {} (o);
\draw[thick, postaction={decorate},%
decoration={markings,mark=at position 0.49 with {\arrowreversed[black,rotate=4,yshift=0.00cm]{triangle 45}},
                     mark=at position 0.98 with {\arrowreversed[black,rotate=8,yshift=0.00cm]{triangle 45}}}] (vc) circle (0.625cm);
\draw[cross,fill=white,cross] (vup) circle (5pt);
\node[] at ($(vc)-(0.9,0)$) {$q$};
\end{tikzpicture} = \\
&\hspace{-1.5cm} 36m_0^2\left(\frac{g_4}{g_2}\right)^2\left(\frac{m_0 R}{2\pi}\right)^2
 \left\{\sum_{n_1=1}^{2N_{\text{tr}}}\left( \frac{1}{\sqrt{n_1^2 + \left(\frac{m_0 R}{2\pi}\right)^2}} - \frac{1}{n} \right) 
 + \frac{\pi}{m_0 R}\right\}
 \sum_{|n_2|<2N_{\text{tr}}} \frac{1}{\left[n_2^2 + \left(\frac{m_0 R}{2\pi}\right)^2\right]^{3/2}} \,.
\end{split}
\end{equation}

The Saturn diagram, unlike the snowman, is convergent and it is given at threshold by the following integral
\begin{equation}
\begin{split}
\begin{tikzpicture}[baseline=-1mm,outer sep=5pt]
\coordinate[] (i) at (0,0);
\coordinate[] (o) at (3,0);
\coordinate[] (vc) at (1.5,0);
\draw[thick, postaction={decorate},%
decoration={markings,mark=at position 0.2 with {\arrow[black]{triangle 45}},
		     mark=at position 0.55 with {\arrow[black]{triangle 45}},
		     mark=at position 0.9 with {\arrow[black]{triangle 45}}}] (i) --  (o);
\draw[thick, postaction={decorate},%
decoration={markings,mark=at position 0.24 with {\arrowreversed[black,rotate=4,yshift=0.00cm]{triangle 45}},
                     mark=at position 0.79 with {\arrow[black,rotate=-14,yshift=0.00cm]{triangle 45}}}] (vc) circle (0.625cm);
\node[] at ($(i)+(0.4,0.3)$) {$p$};
\node[] at ($(o)+(-0.5,0.3)$) {$p$};
\node[] at ($(vc)+(0,0.9)$) {$q_1$};
\node[] at ($(vc)-(0,0.9)$) {$q_2$};
\end{tikzpicture}
  =& \frac{24}{\pi^2}m_0^2\left(\frac{g_4}{g_2}\right)^2\left(\frac{m_0 R}{2\pi}\right)^2 \times \\
 &\hspace{-3.3cm}\times\sum_{|n_1|,|n_2|<2N_{\text{tr}}}\int\ud q_1 \ud q_2 
 \frac{1}{q_1^2+n_1^2+\left(\frac{m_0 R}{2\pi}\right)^2}
 \frac{1}{q_2^2+n_1^2+\left(\frac{m_0 R}{2\pi}\right)^2}
 \frac{1}{\left(\frac{pR}{2\pi}-q_1-q_2\right)^2+(n_1+n_2)^2+\left(\frac{m_0 R}{2\pi}\right)^2} \,.
\end{split}
\end{equation}
The two integrals can be performed and we finally obtain
\begin{equation}
\begin{split}
\begin{tikzpicture}[baseline=-1mm,outer sep=5pt]
\coordinate[] (i) at (0,0);
\coordinate[] (o) at (3,0);
\coordinate[] (vc) at (1.5,0);
\draw[thick, postaction={decorate},%
decoration={markings,mark=at position 0.2 with {\arrow[black]{triangle 45}},
		     mark=at position 0.55 with {\arrow[black]{triangle 45}},
		     mark=at position 0.9 with {\arrow[black]{triangle 45}}}] (i) --  (o);
\draw[thick, postaction={decorate},%
decoration={markings,mark=at position 0.24 with {\arrowreversed[black,rotate=4,yshift=0.00cm]{triangle 45}},
                     mark=at position 0.79 with {\arrow[black,rotate=-14,yshift=0.00cm]{triangle 45}}}] (vc) circle (0.625cm);
\node[] at ($(i)+(0.4,0.3)$) {$p$};
\node[] at ($(o)+(-0.5,0.3)$) {$p$};
\node[] at ($(vc)+(0,0.9)$) {$q_1$};
\node[] at ($(vc)-(0,0.9)$) {$q_2$};
\end{tikzpicture}
 =& 24m_0^2\left(\frac{g_4}{g_2}\right)^2\left(\frac{m_0 R}{2\pi}\right)^2 \times \\ &\hspace{-1cm}\times
 \sum_{|n_1|,|n_2|<2N_{\text{tr}}} \frac{f(n_1)+f(n_2)+f(n_1+n_2)}{f(n_1)f(n_2)f(n_1+n_2)}\frac{1}{\left[f(n_1)+f(n_2)+f(n_1+n_2)\right]^2-\left(\frac{pR}{2\pi}\right)^2} \,,
\end{split}
\end{equation}
where $f(n)=\sqrt{n^2+\left(m_0R/(2\pi)\right)^2}$.
We remark that for consistency at second order in $g_4$, we only need to
compute the Saturn diagram at $p^2=-m_0^2$.

\section{Finite-size behavior of the $2$-particle energy levels}
\label{app:2particle}

In this Appendix we are going to discuss how to understand the finite-size behavior of the $2$-particle energy levels for a 
non-integrable QFT such as the $\Phi^4$ LG theory, both in its symmetric and broken phase. The method a-priori applies only 
below $E_{\text{th}}$, where $E_{\text{th}}$ is the lowest energy threshold where inelastic production processes start to occur. 
In terms of the radius $R$ of the cylinder versus the energy levels $E_i(R)$ that are plotted, the condition $E < E_{\text{th}}$ translates 
into the equivalent condition $R > R_{\text{cr}}$, where $R_{\text{cr}}$ is the largest value of $R$ where an avoiding level crossing 
between the lowest 2-particle energy lines takes place.  Below this production threshold the computation actually mimics the one  applicable to purely integrable situation (see, for instance \cite{yurov-1990}) and then for $E < E_{\text{th}}$ ($R > R_{\text{cr}}$) there is only the elastic $2 \rightarrow 2$ particle processes to worry about.  This is described by the elastic part of the S-matrix amplitude
\begin{equation} \label{eq:S delta}
S_{\text{el}}(\theta) \equiv e^{i \delta(\theta)} \,,
\end{equation}
where $\delta(\theta)$ is the elastic phase-shift. In this range of $R$ we can then apply the Bethe ansatz quantization 
condition relative to the momenta of the two particles, labeled by their rapidities $\theta_1$ and $\theta_2$ respectively:
\begin{equation} \label{eq:BetheAnsatzQuantization}
\begin{aligned}
m R \sinh\theta_1 + \delta(\theta_1-\theta_2) &= 2 \pi N_1 \,, \\
m R \sinh\theta_2 + \delta(\theta_2-\theta_1) &= 2 \pi N_2 \,,
\end{aligned}
\end{equation}
where $m$ is the physical mass of the particle while $N_1$ and $N_2$ are the Bethe numbers.
We remark that the definition of the phase shift $\delta(\theta)$
depends on the choice of the branch cut
of the logarithm of Eq.~\eqref{eq:S delta}.
Different choices are of course equivalent and correspond to redefinitions of the Bethe numbers~\cite{thacker1981}.
In particular, if $\delta(\theta)$ is chosen to lie in the interval
$[-2\pi,0]$, it is continuous for all values of $\theta$ but it
reduces to a step function in the limit of vanishing interaction.
In this case the Bethe numbers are of a fermionic nature and must always be chosen such that $N_1\neq N_2$.
On the contrary, if the phase shift lies between $-\pi$ and $\pi$, it
possesses a finite discontinuity at $\theta=0$ 
for any finite value of the interaction parameter $g_4$.
Moreover, it vanishes everywhere when $g_4\rightarrow 0$.
In this case $\delta(\theta)$ is the true interaction phase shift and the
Bethe numbers have a bosonic nature and can 
be chosen without any restriction.
We follow this second prescription throughout the paper.

The energy of the state corresponding to the quantization of Eq.~\eqref{eq:BetheAnsatzQuantization} is simply
\begin{equation}
E(R) = m (\cosh\theta_1 + \cosh\theta_2) \,,
\label{eq:freeexpression}
\end{equation}
where the two rapidities $\theta_{1,2}(R)$ have acquired an $R$-dependence as solutions of the Eqs. 
\eqref{eq:BetheAnsatzQuantization}. Since $\delta(\theta) = - \delta(-\theta)$, adding the two equations written above we have
\begin{equation} \label{eq:Bethe Eq in CM}
m R (\sinh\theta_1 +\sinh\theta_2) \,=\, 2\pi (N_1 + N_2) \,,
\end{equation}
which is nothing more than the conservation of the total momentum on the circle.
In the momentum-0 sector ($N_1=-N_2$ and $\theta_1= - \theta_2$), the Bethe ansatz equations reduce to 
\begin{equation}
m R \sinh\theta_1 + \delta(2 \theta_1) \,=\, 2\pi N_1 \,.
\label{centerofmasseq}
\end{equation}

To proceed further with this approach, we need to derive an expression for the phase-shift, $\delta(\theta)$. 
We will derive this quantity in the lowest order approximations in the coupling constant $g_4$ of the $\Phi^4$ LG theory.
Since we are interested in scattering processes, we will perform the computation in real time.
The Lagrangian is then:
\begin{equation} \label{eq:Lagrangian Mink}
\mathcal{L}(\phi) = \frac{1}{4\pi} \Big( \frac{1}{2}(\partial\phi)^2 - U(\phi) \Big)\,, \qquad\quad
U(\phi) = \frac{1}{2}\left(g_2\,\phi^2 + g_4\, \phi^4\right) \,.
\end{equation}
In order to have the usual Feynman rules with the field normalization equal to one, we can rescale the field by a constant factor.
Substituting $\phi\rightarrow\sqrt{4\pi}\phi$ in Eq.~\eqref{eq:Lagrangian Mink}, we obtain
\begin{equation} \label{eq:Lagrangian Mink new phi}
\mathcal{L}(\phi) = \frac{1}{2}(\partial\phi)^2 - U_{\text{res}}(\phi) \,, \qquad\quad
U_{\text{res}}(\phi) = 
\frac{g_2}{2}\phi^2 + 2\pi g_4\,\phi^4 \,.
\end{equation}
From this last expression we extract the Feynman rules that will be used in the following computation of the $S$-matrix at lowest order in perturbation theory.
We first discuss the case of the unbroken phase and then turn to the broken phase. 

\vskip 10pt
\noindent {\bf Unbroken phase:} In this phase $g_2>0$ and the potential has a single minimum in $\phi^{(0)}=0$ around which we can do perturbation theory.
At the lowest order in $g_4$, the physical mass of the particle is $m = m_0$, while the higher order corrections in $g_4$ 
shift its value to the one given in the text:
\begin{equation}
m(g_4) = m_0 + {\cal O}(g_4) \dots
\end{equation}
At the lowest order in $g_4$, the elastic transmission amplitude, $T$, 
is given by the $\phi^4$-vertex, whose Feynman diagram expression is 
\begin{equation} \label{eq:Smatrix unbroken Iorder}
T_{\text{el}} \simeq \;
\begin{tikzpicture}[baseline=(c.base),outer sep=5pt]
\coordinate[] (bl) at (0,0);
\coordinate[] (br) at (1.5,0);
\coordinate[] (tl) at (0,1.5);
\coordinate[] (tr) at (1.5,1.5);
\coordinate[] (c)  at (0.75,0.75);
\draw[thick, postaction={decorate},decoration={markings,mark=at position 0.6 with {\arrow[black]{triangle 45}}}] (bl) --  (c);
\draw[thick, postaction={decorate},decoration={markings,mark=at position 0.75 with {\arrow[black]{triangle 45}}}] (c)  --  (br);
\draw[thick, postaction={decorate},decoration={markings,mark=at position 0.6 with {\arrow[black]{triangle 45}}}] (tl) --  (c);
\draw[thick, postaction={decorate},decoration={markings,mark=at position 0.75 with {\arrow[black]{triangle 45}}}] (c)  --  (tr);
\end{tikzpicture}
\;= - 48\pi i g_4.
\end{equation}
Now using the formula~\cite{mussardo-book}
\begin{equation}
S(s) = 4 m^2 \sinh\theta \,S(\theta) \,,
\end{equation}
linking the elastic part of the 2-body $S$-matrix given in terms of the $s$-Mandelstam variable to this same quantity given in terms of
the rapidity difference $\theta=\theta_1-\theta_2$, we arrive at an expression for $T$ in terms of $\theta$:
\begin{equation}
T(\theta) \simeq - \frac{48\pi i\,g_4}{4m^2 \sinh\theta} = -12\pi i\frac{g_4/m^2}{\sinh\theta} \,.
\label{pertSSS}
\end{equation}
The general solution of the unitarity and crossing symmetry equations for the elastic $S$-matrix of a neutral particle is given 
by~\cite{Zam,mussardo-book}
\begin{equation}
S_{\alpha}(\theta) = \frac{\tanh\frac{1}{2} (\theta - i \pi \alpha)}{\tanh\frac{1}{2} (\theta + i \pi \alpha)} \, .
\end{equation}
For $\alpha \sim 0$ this can be expanded as 
\begin{equation}
S_{\alpha}(\theta) \simeq 1 - \frac{2 \pi i \alpha}{\sinh\theta} + \dots \,.
\end{equation}
If we compare the second term of this expression (i.e.\ the transmission amplitude) with Eq.\eqref{eq:Smatrix unbroken Iorder}, 
we can identify the parameter 
$\alpha$, i.e.
\begin{equation}
\alpha = 6\frac{g_4}{m^2} .
\end{equation}
In this way, we have for a final expression for the S-matrix:
\begin{equation}
S_{\text{el}}(\theta) = \frac{\tanh\frac{1}{2} \left(\theta - i 6\pi \frac{g_4}{m^2}\right)}{\tanh\frac{1}{2} \left(\theta + i 6\pi \frac{g_4}{m^2}\right)} \,.
\end{equation}
Notice that, even though we compute the $T$-matrix to first order in $g_4$, 
our final expression for the S-matrix contains higher order contributions, both from the physical (renormalized) mass $m$ and from 
insisting that the S-matrix takes a form that obeys unitarity and crossing.

\vskip 10pt
\noindent {\bf Broken phase:} In order to compute the elastic $S$-matrix
in this phase, it is necessary first to shift the field with respect to its vacuum expectation value: 
\begin{equation}
\phi_\pm^{(0)}(x) = \pm\sqrt{\frac{-g_2}{8\pi g_4}} + \eta(x)
\end{equation}
and then expand the potential in terms of the new fluctuating field $\eta(x)$,
\begin{equation}
U_{\text{res}}(\phi_{\pm}^{(0)} + \eta) = - g_2 \,\eta^2 \pm \sqrt{-8\pi g_2 g_4}\,\eta^3 + 2\pi g_4\, \eta^4 \,.
\end{equation}
The lowest order value of the particle mass is now $m_0\equiv \sqrt{-2g_2}$.
The Feynman rules for this theory can be easily extracted and have both 3-body and 4-body interactions.
At order $g_4$, the 2-body transmission scattering amplitude is given, in addition to the 4-body vertex, 
by the sum of the $s$-, $t$- and $u$-channel diagrams built with two 3-leg vertices, each of them being proportional 
to $\sqrt{g_4}$.

\begin{figure}
\centering
\subfigure[]{\includegraphics[width=0.48\textwidth]{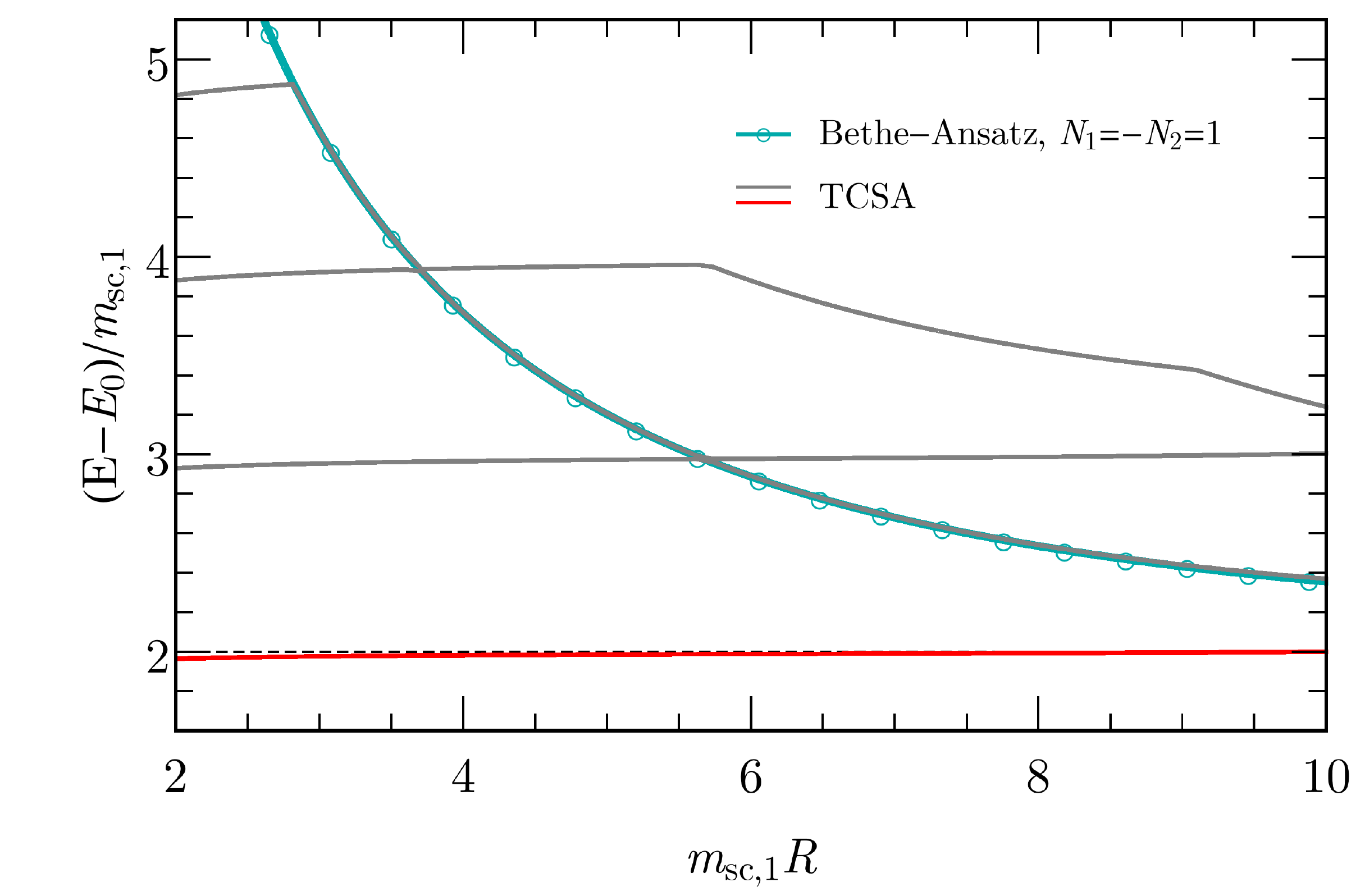}\label{subfig:phi4_2pN1_set11}}\hspace{1mm}
\subfigure[]{\includegraphics[width=0.48\textwidth]{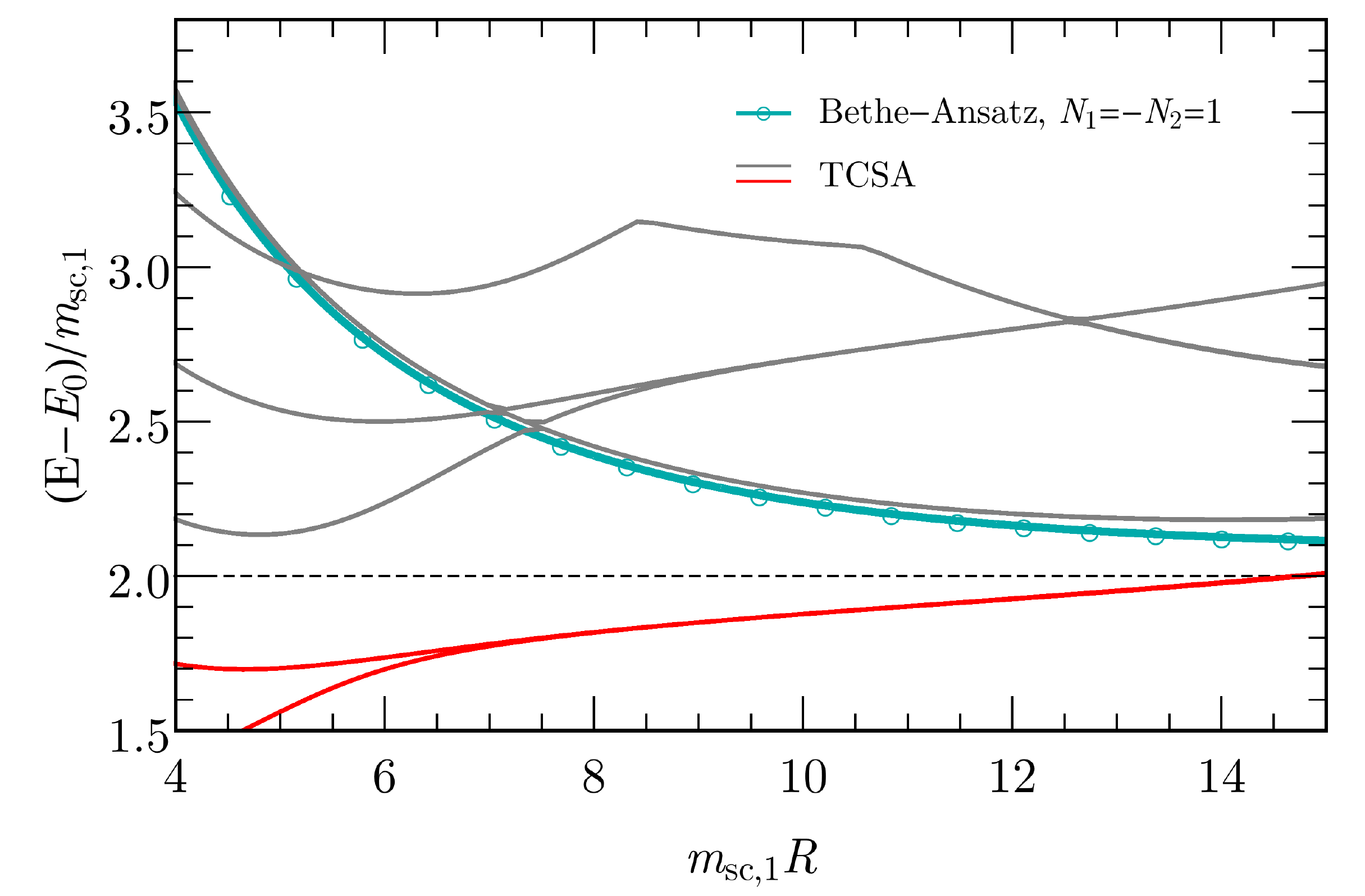} \label{subfig:phi4_2pN1_set6}}
\caption{Two examples of the energy levels predicted by the Bethe ansatz quantization procedure 
for the low lying momentum-0 energy levels of $\Phi^4$ LG theories with the 
same parameters as in Fig.~\ref{fig:phi4_spectra}, namely \subref{subfig:phi4_mom1_set11} 
$g_2=-0.1$, $g_4=6\times 10^{-5}$, $\beta=0.06$ and $N_{\text{tr}}=6$ and
\subref{subfig:phi4_mom1_set6} $g_2=-0.1$, $g_4=1.2\times 10^{-3}$, $\beta=0.2$ and $N_{\text{tr}}=7$.
The ground state energy belonging to the momentum-0 sector has been subtracted.
All energies are normalized with respect to $m_{\text{sc},1}$ and are plotted as function of $m_{\text{sc},1}R$.
The Bethe numbers used here are $N_1=1$ and $N_2=-1$.  The match with the data is very good 
far from the avoided crossings and values of $m_{\text{sc},1}R$ on the order of $14$ (beyond this value,
one can see truncation effects arise).}
\label{fig:phi4_2pN1}
\end{figure}

At the threshold (i.e.\ $p_1=p_2=0$ where $p_i$ is the space component of the momenta of the incoming particles), the 
contributions to the $T$-matrix are
\begin{equation}
\begin{aligned}
T &\simeq
\begin{tikzpicture}[baseline=(c),outer sep=5pt]
\coordinate[] (bl) at (0,0);
\coordinate[] (br) at (1.5,0);
\coordinate[] (tl) at (0,1.5);
\coordinate[] (tr) at (1.5,1.5);
\coordinate[] (c)  at (0.75,0.75);
\draw[thick, postaction={decorate},decoration={markings,mark=at position 0.6 with  {\arrow[black]{triangle 45}}}] (bl) --  (c);
\draw[thick, postaction={decorate},decoration={markings,mark=at position 0.75 with {\arrow[black]{triangle 45}}}] (c)  --  (br);
\draw[thick, postaction={decorate},decoration={markings,mark=at position 0.6 with  {\arrow[black]{triangle 45}}}] (tl) --  (c);
\draw[thick, postaction={decorate},decoration={markings,mark=at position 0.75 with {\arrow[black]{triangle 45}}}] (c)  --  (tr);
\end{tikzpicture}
\;+\;
\begin{tikzpicture}[baseline=(c1),outer sep=5pt]
\coordinate[] (bl) at (0,0);
\coordinate[] (br) at (2.5,0);
\coordinate[] (tl) at (0,1.5);
\coordinate[] (tr) at (2.5,1.5);
\coordinate[] (c1)  at (0.6,0.75);
\coordinate[] (c2)  at (1.9,0.75);
\draw[thick, postaction={decorate},decoration={markings,mark=at position 0.6 with  {\arrow[black]{triangle 45}}}] (bl) -- (c1);
\draw[thick, postaction={decorate},decoration={markings,mark=at position 0.75 with {\arrow[black]{triangle 45}}}] (c2) -- (br);
\draw[thick, postaction={decorate},decoration={markings,mark=at position 0.6 with  {\arrow[black]{triangle 45}}}] (tl) -- (c1);
\draw[thick, postaction={decorate},decoration={markings,mark=at position 0.75 with {\arrow[black]{triangle 45}}}] (c2) -- (tr);
\draw[thick, postaction={decorate},decoration={markings,mark=at position 0.65 with {\arrow[black]{triangle 45}}}] (c1) -- (c2);
\end{tikzpicture}
\;+\;
\begin{tikzpicture}[baseline=(cm),outer sep=5pt]
\coordinate[] (bl) at (0,0);
\coordinate[] (br) at (1.5,0);
\coordinate[] (tl) at (0,2.5);
\coordinate[] (tr) at (1.5,2.5);
\coordinate[] (c1)  at (0.75,0.6);
\coordinate[] (c2)  at (0.75,1.9);
\coordinate[] (cm)  at (0.75,1.25);
\draw[thick, postaction={decorate},decoration={markings,mark=at position 0.6 with  {\arrow[black]{triangle 45}}}] (bl) -- (c1);
\draw[thick, postaction={decorate},decoration={markings,mark=at position 0.75 with {\arrow[black]{triangle 45}}}] (c1) -- (br);
\draw[thick, postaction={decorate},decoration={markings,mark=at position 0.6 with  {\arrow[black]{triangle 45}}}] (tl) -- (c2);
\draw[thick, postaction={decorate},decoration={markings,mark=at position 0.75 with {\arrow[black]{triangle 45}}}] (c2) -- (tr);
\draw[thick, postaction={decorate},decoration={markings,mark=at position 0.65 with {\arrow[black]{triangle 45}}}] (c1) -- (c2);
\end{tikzpicture}
\;+\;
\begin{tikzpicture}[baseline=(cm),outer sep=5pt]
\coordinate[] (bl) at (0,0);
\coordinate[] (br) at (1.5,0);
\coordinate[] (tl) at (0,2.5);
\coordinate[] (tr) at (1.5,2.5);
\coordinate[] (c2)  at (0.75,1.9);
\coordinate[] (cm)  at (0.75,1.25);
\coordinate[] (cm)  at (0.75,1.25);
\draw[thick, postaction={decorate},decoration={markings,mark=at position 0.6 with  {\arrow[black]{triangle 45}}}] (bl) -- (c1);
\draw[thick, postaction={decorate},decoration={markings,mark=at position 0.75 with {\arrow[black]{triangle 45}}}] (c1) -- (tr);
\draw[thick, postaction={decorate},decoration={markings,mark=at position 0.6 with  {\arrow[black]{triangle 45}}}] (tl) -- (c2);
\draw[thick, postaction={decorate},decoration={markings,mark=at position 0.75 with {\arrow[black]{triangle 45}}}] (c2) -- (br);
\draw[thick, postaction={decorate},decoration={markings,mark=at position 0.65 with {\arrow[black]{triangle 45}}}] (c1) -- (c2);
\end{tikzpicture} \\[2mm]
&= \; - 48 \pi i g_4 \qquad\quad\; - 48i\pi g_4 \qquad\quad\; + 144i\pi g_4 \qquad + 144i\pi g_4 \,,
\end{aligned}
\end{equation}
and therefore, near threshold, we have
\begin{equation}
T \simeq 192 i \pi\, g_4 \,.
\end{equation}
Repeating the same steps done before to identify the parameter $\alpha$ entering the exact expression of the 
elastic $2$-body $S$-matrix, we end up (in terms of the physical mass $m$) with
\begin{equation}
\alpha = - 24\frac{g_4}{m^2}.
\end{equation}
Notice that in this case $\alpha$ is negative and the corresponding elastic $S$-matrix entering the Bethe-Ansatz equation 
has a pole at $\theta = i 24\pi \frac{g_4}{m^2}$
\begin{equation} \label{eq:Smatrix broken}
S_{\text{el}}(\theta) = \frac{\tanh\frac{1}{2} \left(\theta + i 24\pi \frac{g_4}{m^2}\right)}{\tanh\frac{1}{2} \left(\theta - i 24\pi \frac{g_4}{m^2}\right)} \,.
\end{equation}
This pole corresponds to the first bound state that exists in the broken phase. It is easy to see that the value of the mass of the bound state extracted from this $S$-matrix coincides (at the lowest non-trivial order) with the general expression of the mass-spectrum provided by the semi-classical analysis done in the text. 

Finally we show in Fig.~\ref{fig:phi4_2pN1} that the Bethe ansatz quantization conditions together with above $S$-matrix
can describe some of the higher energy states in the momentum-0 sector of
the broken phase.
From these plots it is clear that even if we only expect the purely elastic $S$-matrix 
to give an accurate description below threshold, the solution of the Bethe equations 
still describes the data well above this, at least away from avoided level crossings.

\bibliography{./biblio}

\end{document}